\documentclass[pdflatex,sn-mathphys-num,hyphens]{sn-jnl}

\usepackage{amssymb}
\usepackage{amsmath}
\usepackage{tabularx}
\usepackage{multirow}
\usepackage{soul}
\usepackage{booktabs}
\usepackage{url}
\usepackage{siunitx}
\usepackage{subcaption}
\usepackage{xcolor}
\usepackage{adjustbox}
\usepackage{fontawesome5}
\usepackage[normalem]{ulem}

\usepackage[T1]{fontenc}

\definecolor{mygreen}{HTML}{1a9850}
\definecolor{myred}{HTML}{fb8072}

\newcommand{\dbField}[1]{{\small\texttt{#1}}}

\newcommand{\rev}[1]{\textcolor{black}{#1}}

\theoremstyle{thmstyleone}%

\theoremstyle{thmstyletwo}%

\theoremstyle{thmstylethree}%

\begin{document}

\title{When Transparency Falls Short: Auditing Platform Moderation During a High-Stakes Election}

\author[1,2]{\fnm{Benedetta} \sur{Tessa}}\email{benedetta.tessa@phd.unipi.it}
\equalcont{These authors contributed equally to this work.}

\author*[3]{\fnm{Gautam Kishore} \sur{Shahi}}\email{gautam.shahi@uni-due.de}
\equalcont{These authors contributed equally to this work.}

\author[2]{\fnm{Amaury} \sur{Trujillo}}\email{amaury.trujillo@iit.cnr.it}

\author[2]{\fnm{Stefano} \sur{Cresci}}\email{stefano.cresci@iit.cnr.it}

\affil[1]{\orgdiv{Department of Computer Science}, \orgname{University of Pisa}, \orgaddress{\country{Italy}}}

\affil[2]{\orgname{IIT-CNR}, \orgaddress{\city{Pisa}, \country{Italy}}}

\affil[3]{\orgname{University of Duisburg-Essen}, \orgaddress{\country{Germany}}}

\abstract{
During major political events, social media platforms encounter increased systemic risks. However, it is still unclear if and how they adjust their moderation practices in response.
The Digital Services Act Transparency Database provides---for the first time---an opportunity to systematically examine content moderation at scale, allowing researchers and policymakers to evaluate platforms' compliance and effectiveness, especially at high-stakes times. 
Here we analyze 1.58 billion self-reported moderation actions by the eight largest social media platforms in Europe over an eight-month period surrounding the 2024 European Parliament elections.
We found that platforms did not exhibit meaningful signs of adaptation in moderation strategies as their self-reported enforcement patterns did not change significantly around the elections. This raises questions about whether platforms made any concrete adjustments, or whether the structure of the database may have masked them.
On top of that, we reveal that initial concerns regarding platforms' transparency and accountability still persist one year after the launch of the Transparency Database. \textcolor{black}{Our findings highlight the limits of current co-regulatory and self-regulatory approaches}, and point to the need for stronger enforcement and better data access mechanisms to ensure that online platforms meet their responsibilities in protecting the democratic processes.
}

\maketitle

\section{Introduction}
\label{sec:intro}
\rev{Social media plays a significant role in election campaigns, influencing public opinion, voting behavior, and ultimately election outcomes \cite{stier2026online}. Previous elections, including the 2019 European Parliament election \cite{shahi2024agenda} and the 2024 United States presidential election \cite{kuberski2025prevalence}, have demonstrated the impact of unmoderated and misleading content disseminated through social media platforms. In response to these challenges, the European Union (EU) has consolidated its efforts to improve online platform accountability by introducing the Digital Services Act (DSA) \cite{drolsbach2024content}. Its co-regulatory approach, in which public and private actors share risk mitigation responsibilities~\cite{shattock2021self}, is based on a combination of voluntary self-regulation and statutory regulation. For instance, the largest platforms are greatly encouraged to undertake self-initiatives to mitigate systemic risks and to adopt voluntary codes of conduct, while at the same time they are obliged to publish detailed transparency data and give platform access to vetted researchers. The 2024 European Parliament election is the first major electoral event under this new regulatory framework, thus providing a litmus test on the effectiveness of its transparency mechanisms in the face of a democratic systemic risk.}

The European Parliament elections constitute a cornerstone of the EU political landscape. Held every five years, they give citizens the opportunity to choose their representatives in the Parliament, which in turn selects the President of the European Commission.
The latest elections were held on 6--9 June 2024 across all EU countries, in a year that also saw a large number of national elections held worldwide. The President was then elected on 18 July 2024 through a secret ballot.
National and supra-national elections have an extensive impact on social media platforms, which play a key role in hosting political campaigns and spreading election-related information~\cite{rho2020political,papakyriakopoulos2023upvotes,shahi2024agenda}. For instance, politicians frequently use their social media to engage with citizens and to encourage political learning and electoral participation~\cite{cresci2014criticism,fatema2022social,kim2022observation,bene2022keep}.
At the same time, online electoral discourse can be exposed to malicious activities aimed at influencing public opinion before elections~\cite{chen2022social}. Such activities include information manipulation~\cite{cinelli2020limited,tardelli2020characterizing,matatov2022stop,mazza2022investigating}, the use of deepfakes~\cite{haq2024,diakopoulos2021anticipating}, targeted harassment~\cite{hua2020characterizing}, and opaque or unfair political advertising~\cite{bar2024systematic}. 
These phenomena are only a few of the risks arising from a period of heightened political activity in which significant political and economic interests intersect. One of the main tools that social media platforms have to mitigate these risks is content moderation~\cite{grimmelmann2015virtues,gillespie2018custodians}. During elections, platforms may adjust their moderation strategies to better address the challenges that they face during these high-risk periods.

Content moderation has become increasingly more important not only for platform users but also for EU regulators. In October 2022, the EU introduced the DSA to regulate online platforms and promote a more transparent, inclusive, and safe digital environment~\cite{eu2020DSA}.
Among several transparency-related measures, the DSA mandates large online platforms to report all their moderation actions within the EU by submitting clear, detailed, and timely statements of reasons (SoRs) to the DSA Transparency Database\footnote{\url{https://transparency.dsa.ec.europa.eu/}} (\texttt{DSA-TDB})—an open, centralized repository managed by the European Commission~\cite{trujillo2023dsa,kaushal2024automated}.
Launched in September 2023, this database represents a novel and groundbreaking tool for ensuring transparency and accountability in content moderation practices.
Several early studies have examined the initial data that platforms submitted to the database during its first months of operation~\cite{trujillo2023dsa,dergacheva2023one,aspromonte-etal-2024-llms,drolsbach2024content}. Interestingly, these studies revealed various compliance shortcomings and identified significant structural issues within the database, all of which may limit its overall reliability and usefulness~\cite{trujillo2023dsa,kaushal2024automated}.

In this study---an extension of the work carried out in \cite{shahi2025year}---we present one of the most comprehensive analysis of the \texttt{DSA-TDB} to date, conducted nearly one year after its initial release. We examine an eight-month period surrounding the 2024 EU elections and analyze the self-reported moderation actions of the eight largest social media platforms operating in the EU.
By examining 1.58 billion SoRs during a politically sensitive period, we aim to determine whether the \texttt{DSA-TDB} has fulfilled its transparency objective and to assess if and how the reported moderation practices of major social media platforms changed in response to heightened integrity risks.
These extensive analyses also allow to assess whether the aforementioned shortcomings have been resolved.
Herein we address the following research questions:

   \noindent   \textbf{RQ1:}  \textit{How did self-reported moderation practices change during the European electoral period?} \\ %
 To answer this RQ, we collected and analysed 1.58 Billion SoRs, and  we aim to identify significant changes in content moderation practices before and after the 2024 EU elections, including shifts in the volume and types of content being moderated. \\
\textbf{RQ2:}     \textit{How has the reliability and consistency of the database evolved since its initial release?} \\
Early analyses revealed major issues that restricted the database’s usefulness as a transparency tool. In this study, we re-evaluate the issues identified in ~\cite{trujillo2023dsa} by replicating and extending its analysis using a larger and more recent dataset, evaluating both progress made and ongoing challenges in ensuring meaningful platform transparency.

\rev{Our findings are relevant and multifold as they examine whether the \texttt{DSA-TDB} can support the co-regulation of platform moderation during a major European Union election. Electoral periods often lead platforms to adapt their enforcement practices in response to increased risks of information manipulation, harmful content, and coordinated abuse. However, assessing platform responses through the \texttt{DSA-TDB} requires examining whether the database provides data that are sufficiently systematic, complete, and comparable for this purpose. In practice, the ability to evaluate platform responses during elections depends not only on the availability of transparency data, but also on whether such data are reported consistently across platforms and contain enough contextual information to distinguish meaningful changes in moderation practices from other reporting dynamics. By focusing on the 2024 European Parliament elections, our analysis shows what the \texttt{DSA-TDB} can and cannot reveal about platform responses during politically sensitive events.}

\rev{This work also contributes to research at the intersection of social media, elections, content moderation, and regulatory transparency \cite{kaushal2024automated, gillespie2018custodians,marsden2011internet}. It extends work on online political communication and electoral integrity, but also contributes to the emerging literature on the DSA \cite{groesch2025big} and its Transparency Database by reassessing the database’s consistency, completeness, and practical usefulness nearly one year after its launch \cite{meyer2025transparency}.
}

\section{Related Work}
\label{sec:relwork}

\rev{
In this section, we briefly describe the literature concerning the main approaches to online content moderation, the influence of social media on elections, the model of co-regulation that the DSA follows to share the responsibility of systemic risk mitigation among government and online platforms, and the \texttt{DSA-TBD} as a mechanism to implement such approach. This overview thus provides both the background and conceptual framework of our study on the compliance and effectiveness of the \texttt{DSA-TDB}, with respect to the democratic systemic risk during the 2024 EU Parliament Elections.
}

\subsection{Online Content Moderation}

\rev{Content moderation is fundamental for maintaining the well-being of online communities by preventing the spread of hate speech, harmful or illegal content, upholding community norms, and promoting fairness and respect among users~\cite{gillespie2018custodians,gorwa2020algorithmic,trujillo2023dsa}.
There are two broad moderation strategies: \emph{hard} and \emph{soft}.
On one hand, the \emph{hard} strategy, sometimes called \textit{deplatforming}, implies the removal of content, accounts, or even entire communities. It is the most frequently used strategy and as such has been the center of various studies, particularly regarding its efficacy and consequences~\cite{trujillo2021echo,jhaver2021evaluating,horta2024deplatforming,cima2024great}.
Such strategy, however, raises concerns about limitations to the right of free speech~\cite{jhaver2023bans,zannettou2021won}.
On the other hand, the \textit{soft} strategy is less punitive, as the goal is to warn users about their content or accounts rather than removing them. An illustrative example is Reddit's \emph{quarantine}, which reduces the visibility of an entire community~\cite{chandrasekharan2022quarantined,trujillo2022make,trujillo2023one}. Another example is attaching warning labels to posts to make users aware about their possible harm~\cite{pennycook2020implied,zannettou2021won}.}

\rev{Both hard and soft moderation interventions can lead to undesired and unintended side effects, however. For example, users may become less active and/or more toxic, or even migrate to other more permissive platforms, which only moves the problem without actually solving it~\cite{trujillo2023one,trujillo2021echo,horta2024deplatforming}. In trying to mitigate these downsides, recent efforts seek to anticipate the outcomes of a moderation intervention. For example, by using machine learning models to predict the levels of user activity, toxicity, and participation diversity after a community ban on Reddit~\cite{tessa2024beyond,tessa2025quantifying}, by predicting the ban of a whole Reddit community~\cite{habib2022proactive}, or predicting which users would evade a Wikipedia ban by creating another account~\cite{niverthi2022characterizing}.
Still, the balance between enforcing content moderation (either hard or soft actions) and infringing on freedom of expression remains a significant challenge for online platforms, particularly during elections and other politically sensitive periods.}

\subsection{Role of Social Media in Elections Interference}

\rev{Social media platforms have transformed political communication by enabling candidates, political parties, journalists, and citizens to exchange information rapidly and at an unprecedented scale \cite{shahi2024agenda}. Because of this, online platforms have also become powerful tools for election interference. Malicious actors such as foreign governments, political organizations, coordinated influence campaigns, and automated bot networks exploit social media to manipulate public opinion, amplify political polarization, suppress voter participation, and undermine trust in democratic institutions \cite{bradshaw2018challenging, mont2024electoral}.}

\rev{One of the primary interference mechanisms is the dissemination of false information (misinformation), sometimes done deliberately with the intent to deceive (disinformation). During election campaigns, misleading narratives, fabricated news articles, manipulated images and videos, and conspiracy theories can spread rapidly through algorithmic recommendation systems and user sharing behavior. For instance, research by Vosoughi et al.~\cite{vosoughi2018spread} demonstrated that false news spreads significantly faster and reach more people than truthful information on social media, making electoral misinformation particularly difficult to contain.
A related concept is \emph{computational propaganda}, defined by Musfta et al.~\cite{mustafa2025conceptualizing} as the strategic use of algorithms, automation, and human curation to manipulate public opinion across digital platforms, which may distort public discourse and influence political agendas on social media and whose authors might be  local, national, or foreign.
For example, investigations into the 2024 United States presidential election revealed foreign interference efforts by the Internet Research Agency to conduct coordinated disinformation campaigns across multiple social media platforms \cite{yan2025origin}. Similar influence operations have since been documented in elections across Europe \cite{casero2025spreading}, Asia \cite{shah2026making}, and Latin America \cite{hale2024analyzing}, illustrating the global nature of online election interference.
Recent advances in generative artificial intelligence have exacerbated the issue, underlining the need for multimodal detection systems capable of identifying manipulated and misleading text, images, audio, and video across social media platforms \cite{shahi2025multimodal}.}

\rev{Additionally, social media advertising enables election interference through micro-targeting by using detailed user profiles to deliver highly personalized messages tailored to individual demographics, interests, and psychological characteristics. While targeted political communication can improve campaign efficiency, it also raises concerns regarding transparency, accountability, voter manipulation, and unequal exposure to political information. The controversy surrounding Cambridge Analytica highlighted how personal data collected from millions of users could be leveraged for political advertising and behavioral targeting without informed consent \cite{heawood2018pseudo}.
A related challenge is the formation of echo chambers and filter bubbles. Recommendation algorithms tend to prioritize content that aligns with users' existing preferences, reinforcing ideological homogeneity and reducing exposure to diverse political viewpoints. Such environments facilitate the rapid spread of partisan misinformation while increasing political polarization and difficult to tackle~\cite{marino2023seven}.}

\rev{Due to the increasing prevalence of the above issues in the last few years, governments and online platforms across the world have taken different initiatives to regulate and govern social media to mitigate its negative effects to the democratic process. In this regard, Marsden et al.~\cite{marsden2020platform} have proposed \emph{co-regulation} as the best approach to oversee and control disinformation in the context of democratic elections.}

\subsection{Social Media Co-regulation via the DSA}

\rev{In regulatory theory, \emph{co-regulation} refers to a flexible regulatory pluralism in which the responsibility of social control is shared among organizations of the public and private sectors, with the rationale being that multiple policy instruments and a broader range of actors will result in better regulation~\cite{gunningham2017smart}. Being midway between voluntaristic self-regulation and statutory regulation, co-regulation is a kind of \emph{enforced self-regulation}: it draws on the expertise of engaged actors within regulated organizations to make sure risks are controlled properly, whilst maintaining external regulatory oversight to ensure the integrity of the process~\cite{haines2017regulation}.}

\rev{Co-regulation has been heralded by experts as the appropriate regulatory against online disinformation, by encouraging independence and flexibility of online platforms in tackling the issue using innovative means (e.g., via the massive deployment of artificial intelligence)~\cite{marsden2020platform}. Indeed, given that within the EU co-regulation has been for many years the main guiding model used to regulate the Internet~\cite{marsden2011internet}, it has also been adopted for the design of the DSA to tackle these and other systemic risks~\cite{shattock2021self}. For instance, Section 6 of the DSA explicitly states that the EC shall promote voluntary standards and codes of conduct among online platforms for the proper application of the regulation. Hence, the idea is to foster new and existing self-regulating initiatives by platforms, such as internal content moderation policies \cite{prollochs2022community}, fact-checking partnerships \cite{waller2026understanding}, political advertising transparency measures, and coordinated inauthentic behavior detection systems. At the same time, the DSA establishes statutory obligations that platforms must fulfill via specific mechanisms, such as sending detailed moderation actions to the \texttt{DSA-TDB}, which can then be used to audit and verify the implementation of the aforementioned voluntary initiatives.}

\subsection{The DSA Transparency Database}

\rev{Since its launch, the \texttt{DSA-TDB} has been the center of several studies examining its usefulness and effectiveness. Albeit these initial studies were based on limited data, they revealed variability in content moderation practices, including differences in the types of content moderated, the implementation of visibility restrictions, and the reported use of automated tools, as well as some inconsistencies in the data~\cite{kaushal2024automated,drolsbach2024content,dergacheva2023one}. A later and more complete work focused on social media~\cite{trujillo2023dsa} found that while platforms formally comply with the DSA requirements, many optional yet informative details are omitted, limiting the database's utility; in addition, there were various and significant inconsistencies with the platforms’ own transparency reports.}

\rev{Beyond the evaluations of the database’s quality, other works have explored ways to improve its usability. For example, \citep{aspromonte-etal-2024-llms} used a multi-agent system based on large language models (LLMs) to link SoRs to the relevant sections of platforms’ Terms of Service. Their results show that LLMs can offer helpful context, making moderation decisions easier to understand and potentially encouraging greater engagement with the DSA. Following this line of research, \cite{esser-spanakis-2025-linking} studied how statements of reasons can be linked to the most relevant clauses in platform policy documents. They also introduced a small dataset of SoRs by TikTok and evaluated a set of retrieval and language models to connect moderation decisions to policy text, showing that such linking can improve the interpretability of platform enforcement and support fairness assessments.}

\rev{These previous works approached the \texttt{DSA-TDB} and related transparency mechanisms from different perspectives: compliance with reporting obligations,  informativeness of the reported data, meeting the minimum schema requirements, or the extent to which the reported information enables meaningful scrutiny of moderation practices. Herein, however, we assess whether the initial concerns about consistency and completeness have been addressed in the context of a major political for which platforms may have greatly adjusted their moderation practices. By the same token, we also asses the effectiveness (from a co-regulatory perspective) of this statutory mechanism in verifying the mitigation of the systemic risk to the electoral process following the platforms' voluntary initiatives.}

\section{Approach}
\label{sec:approach}

We focus on the eight largest social media platforms in the EU, i.e., those defined as very large online platforms (VLOPs) by the DSA~\cite{eu2020DSA}. VLOPs are platforms with more than 45M users within the EU---namely: Facebook, Instagram, LinkedIn, Pinterest, Snapchat, TikTok, X (formerly Twitter), and YouTube. To better understand how each platform prepared for the 2024 European elections, in this section we first present a precursory analysis of the corresponding self-reported systemic risk assessment regarding the electoral process. We then describe the time frames and datasets used for our analyses on potential moderation changes around the EU elections (detailed in \S\ref{sec:results-rq1}) and changes in data reliability and consistency since initial release of the \texttt{DSA-TDB} (detailed in \S\ref{sec:results-rq2}).
 
\subsection{\textcolor{black}{Self-assessed Risk Management}} %
\label{sec:self-assessed-risk}

{\color{black}
Pursuant to Article 34 and Article 42 of the DSA, very large online platforms must release annual DSA Systemic Risk Assessment Reports~\cite{eu2020DSA}. \rev{This is another co-regulatory example of a DSA statutory obligation designed to ease the assessment the platforms' voluntary initiatives to mitigate risks.} 
Based on the reports released between August 2024 and January 2025 (see Appendix~\ref{sec:appendix-dsa-systemic-risk-reports}), we classified the eight selected platforms into the following three relative categories, according to the impact of the 2024 electoral process in the analyzed risk assessment and mitigation measures.

\subsubsection{Lower Impact}
Both Pinterest and LinkedIn stated multiple times in their reports that their user base does not seek to engage with political content on their platforms due to their leisure and professional natures, respectively. Therefore, even if these platforms implemented new mitigation measures around the elections, these were not as exhaustive as done by other platforms. In addition, both platforms already had in place policies that significantly limited political content and ads.

\subsubsection{Middle Impact}
Snapchat, TikTok, and X were each impacted differently by the elections, but in general this impact was moderate compared to other platforms. Within this category, Snapchat suffered the least impact. Its reach is moderate and even if it allows political content and ads, there are policies in place to limit their visibility to certain in-app features. TikTok, on the other hand, has an extensive reach, but it severely limits political content on its platform---having created new automatic mitigation measures specifically for the elections---and does not allow political advertising. X has a EU user base comparable to that of Snapchat, but it includes more public figures and official institutions. Curiously, X allows political ads in several jurisdictions, but not within the EU. In addition, X has been under heavier scrutiny due to DSA-related violations~\cite{eu_proceedings_ag_x}, hence its pressure around the elections was higher than that of the other two platforms in this category.

\subsubsection{Higher Impact}
Facebook, Instagram, and YouTube were the three platforms that implemented the most mitigation measures due to their higher inherent risks for elections. This is because all three platforms have a very extensive reach, widely allow political content, and they also allowed political advertising within the EU. In addition, the risk management strategies of all three platforms were conducted within the much larger approach of their respective behemoth parent companies, Meta and Google.}

\begin{table*}[t]
\centering
\begin{tabular}{cccrl}
\toprule
\textbf{phase} & \textbf{start date} & \textbf{end date} & \textbf{\# days} & \textbf{description} \\
\midrule
1 & Mar 01 & Jun 05  & 97 & Pre-electoral \\
2 & Jun 10 & Jul 17  & 38 & Inter-electoral\\
3 & Jul 19 & Oct 31 & 105 & Post-electoral\\
\bottomrule
\end{tabular}
\caption{{\color{black}Overview of the three phases of interest around the European elections}}
\label{tab:phases}
\end{table*}

\subsection{Time Frames}
\label{sec:time-frames}
For RQ1 (moderation changes around the 2024 EU elections), we first defined the time frame for data collection, which spans from 1 March to 31 October. %
This period covers approximately 14 weeks before and 20 weeks after the election, consistent with the timeline outlined in the EU harmonised electoral framework~\cite{eprs2023briefing751463}. It also allows us to capture potential shifts in moderation practices across both pre- and post-electoral phases.
The post-election period is intentionally longer to take into consideration possible delayed enforcements and post-hoc moderation, which are plausible for such large scale events in which the online activity notably increases. To further analyze how moderation practices evolved over time around the electoral period, we further divided this time frame into three phases encompassing the pre-, intra-, and post-electoral temporal windows, as reported in Table~\ref{tab:phases}. %

For RQ2 (data reliability and consistency compared to the initial period), we also collected data from the initial time frame introduced in \cite{trujillo2023dsa}, comprising the SoRs submitted during the first 100 days of the database (from September 25, 2023, to January 2, 2024). 
Hencefort, we will refer to this time frame as \textit{initial} period.

\subsection{Data}
\label{sec:datasets}
{\color{black}
To answer RQ1, we downloaded 1.58B SoRs from the publicly-available \texttt{DSA-TDB}.\footnote{\url{https://transparency.dsa.ec.europa.eu/explore-data/download}} These SoRs cover all self-reported moderation actions that the eight considered VLOPs took in the EU between 1 March to 31 October. %
In detail, our data includes 646.1M SoRs from TikTok, 300.8M from Instagram, 260.2M from Facebook, 81.9M from Pinterest, 36.3M from YouTube, 2.3M from Snapchat, 628K from X, and 293K from LinkedIn.

To answer RQ2, instead, we collected 353M SoRs comprising 184.8M SoRs from TikTok, 60.3M from Pinterest, 79.1M from Facebook, 19.2M from YouTube, 8.1M from Instagram, 1.1M from Snapchat, 0.5M from X, and 0.04M from LinkedIn.
As explained in the \texttt{DSA-TDB}'s official documentation, each SoR is composed of multiple fields.\footnote{\url{https://transparency.dsa.ec.europa.eu/page/documentation}} All those utilized in this study are described in Appendix Section~\ref{sec:appendix-description} and Table~\ref{tab:dsa-fields}.

}

\subsection{Methods}
\label{sec:methods}
{\color{black}
To address our research objectives, we tackle two complementary questions using distinct methodological approaches. RQ1 focuses on detecting potential shifts in content moderation practices before, during, and after the elections. For this, we rely on time series analyses, first by discovering any possible patterns or trend in number of moderation actions and delays. To deepen our analysis, we also perform time series decomposition to compute trend-specific metrics such as slope and strength index.
These additional analysis have been carried out not on the full observation period, but on three separate time phases following the definition provided in Table \ref{tab:phases}. To avoid possible spurious fluctuations, we excluded the electoral days (6--9 June) and the presidential election day (18 July). 
To complement and contextualize these results, we also perform anomaly detection and change point detection analyses.

Instead, RQ2 investigates the quality and reliability of the data submitted to the \texttt{DSA-TDB}, reassessing earlier concerns around incomplete, vague, or implausible reporting. Here, we combine descriptive statistics and cross-platform comparisons to evaluate whether previously observed inconsistencies persist. In particular, we repeat the same analyses carried out with the SoRs submitted during the initial period with the ones from the last 100 days of our observation period (573M SoRs from July 24 to October 31, 2024). 
We will refer to the last 100 days as \textit{latest} period.
}

\begin{figure*}[t]
\centering
    \begin{subfigure}{1\textwidth}%
      \includegraphics[width=\textwidth]{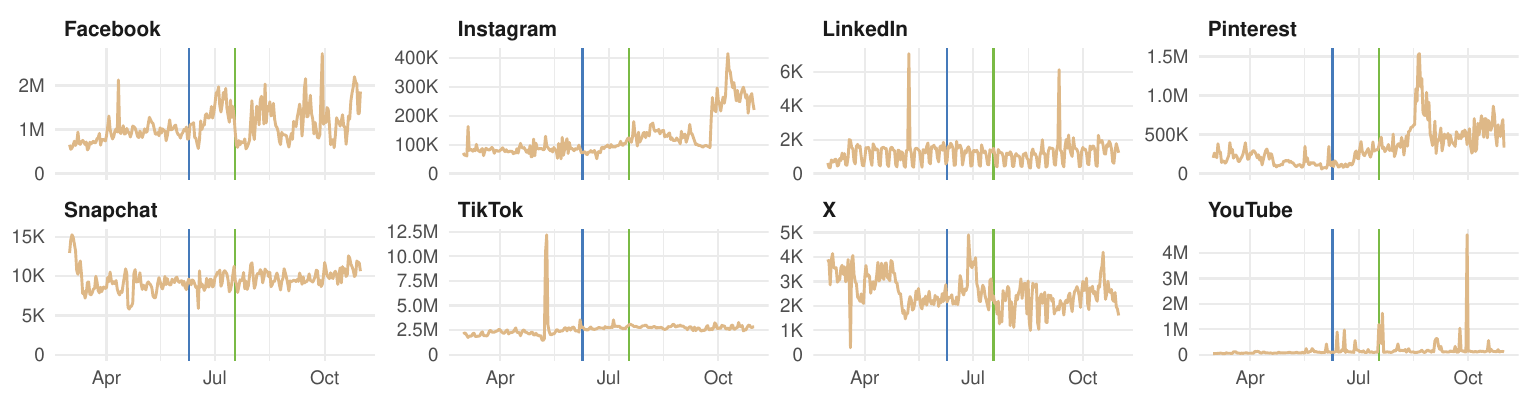}%
      \caption{Number of SoRs.}
      \label{fig:sor_count_time_series}%
    \end{subfigure}%
    \\
    \begin{subfigure}{1\textwidth}%
      \includegraphics[width=\textwidth]{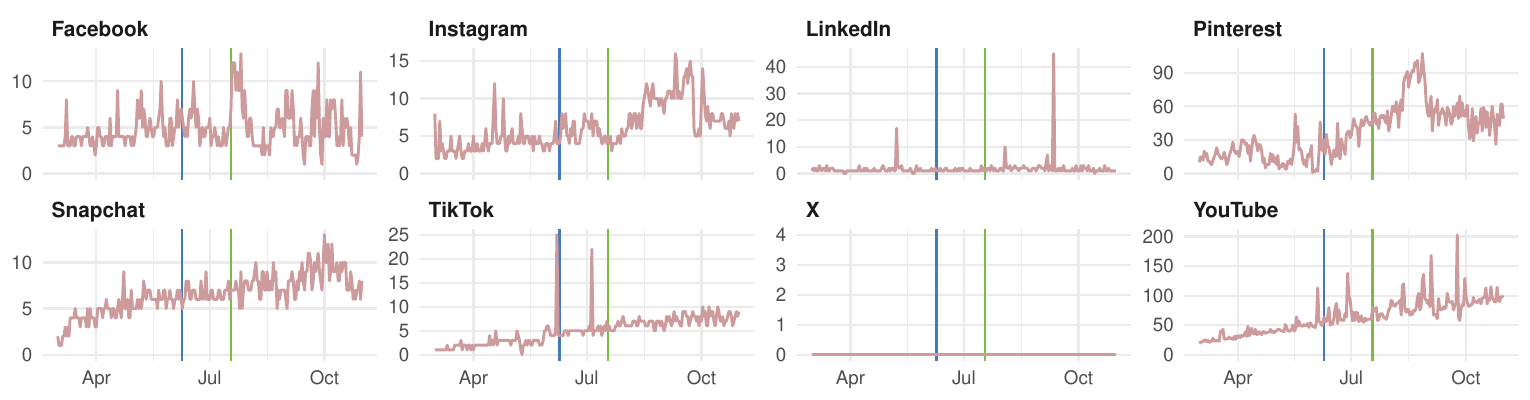}%
      \caption{Moderation delay (days).}
      \label{fig:avg_delay_time_series}%
    \end{subfigure}%
\caption{Daily time series of (a) the number of submitted SoRs and (b) their average delay from content creation to applied decision. The blue vertical line indicates the Parliament election while the green one indicates the President election.}
\label{fig:timeline}
\end{figure*}

\section{Moderation Efforts During The Elections}
\label{sec:results-rq1}

\subsection{Moderation Patterns Overview}
In order to answer to RQ1 and understand whether platforms adapted their moderation practices, we begun by assessing if platforms changed the volume or timeliness of their moderation actions in response to the heightened integrity risks surrounding the 2024 EU elections. 
An increase in moderation activity---or even a reduction in moderation delays---could point to heightened moderation efforts, particularly with respect to each platform's risk self-assessment and mitigation measures. On the other hand, stable patterns could suggest that platforms did not adjust their practices to address the risks.
To carry out this analysis we computed the time series of the daily number of SoRs submitted by each platform. Furthermore, we also computed the daily average moderation delay for each SoR, defined as the number of days between the content’s creation and the corresponding moderation action.
Figure~\ref{fig:timeline} shows both the volume (orange, Figure~\ref{fig:sor_count_time_series}) and moderation delay (red, Figure~\ref{fig:avg_delay_time_series}) time series, allowing us to assess whether moderation practices changed significantly during the electoral period compared to routine periods. 

Figure~\ref{fig:timeline} reveals high heterogeneity across all platforms both in terms of volume and delay. For example, while some platforms moderate millions of content daily, others only a few thousands.
Some platforms---such as X---showed steady moderation delays, while others---like LinkedIn---showed regular fluctuations in moderation volume. On top of that, some platforms often present abrupt variations which could be a sign of event-driven moderation policies. 
The same differences can be found when analyzing the platform’s number of active users~\cite{trujillo2023dsa}. 
Nevertheless, the time series showed no meaningful signs of adaptation during the electoral period, represented in figure as vertical blue and green lines corresponding to the European Parliament and Presidential elections, respectively.
This suggests that, overall, moderation activity remained relatively stable before, during, and after the elections. 

\begin{figure}[t]
    \includegraphics[width=\linewidth]{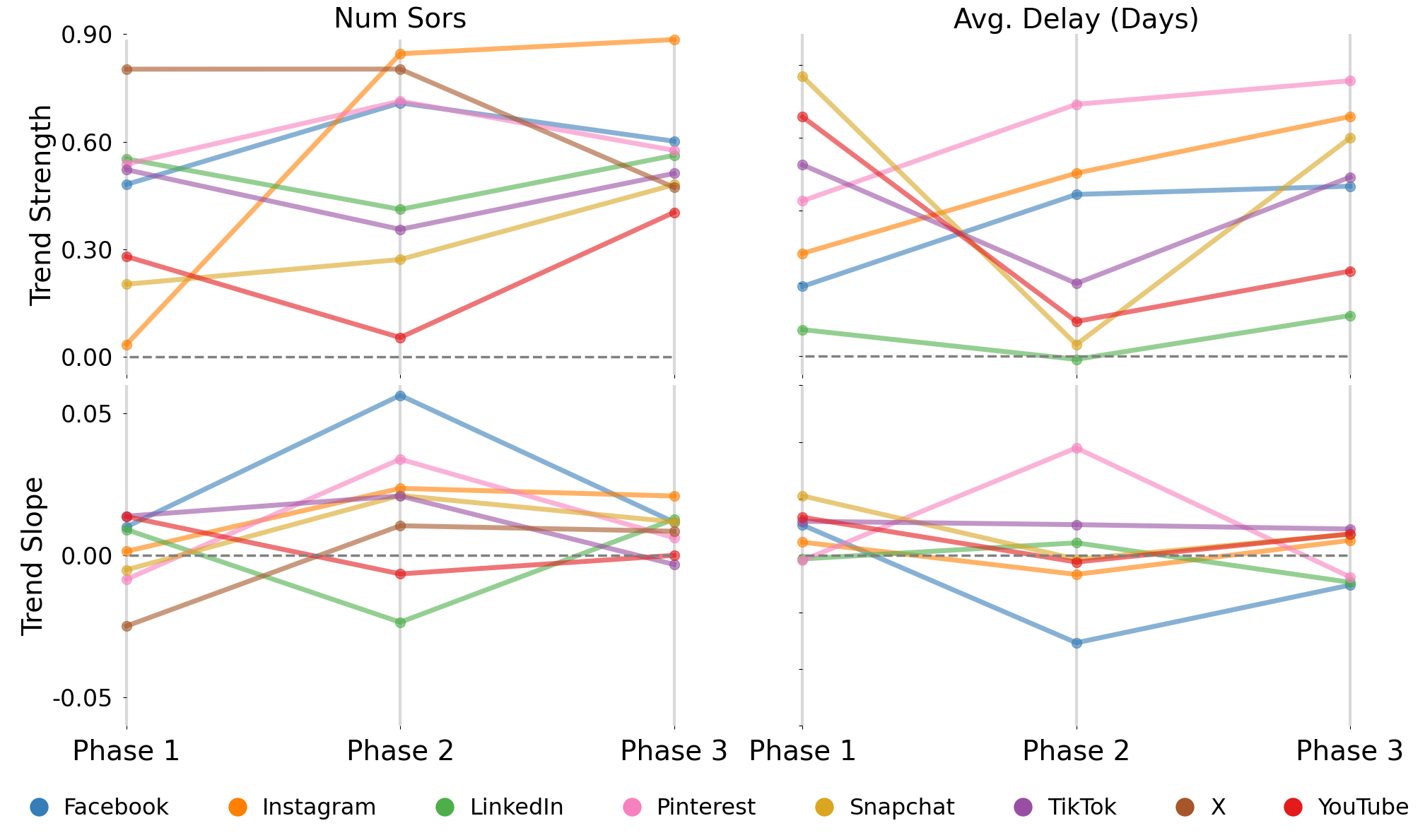}
    \caption{Parallel coordinates across the three phases, showing the patterns of the number of SoRs and delay. 
    For each platform, we plot, for each Phase, the Trend Strength Index and the trend slope for the number of SoRs and average delay. Higher slope values indicate stronger increasing trends, while higher Trend Strength Index values are a sign of more pronounced and stable trends.
    In general, changes across phases remain limited
    , with slightly stronger trends during the electoral period for some platforms and largely stable delay dynamics.}
    \label{fig:parallel-coordinates}
\end{figure}

{\color{black}
To get a deeper understanding of the trends shown in Figure~\ref{fig:timeline}, we carried out additional and more targeted analyses on the time series obtained from the division in phases.
This division into phases supports a more nuanced and detailed analysis of how moderation practices might have shifted and evolved. For instance, it allowed to thoroughly assess possible shifts in moderation practices before and after the electoral period, taking into account both long- and short-term effects of the elections. For each time series, we first removed the outliers and standardized them. Then, we obtained the trend, seasonality, and residuals via decomposition. We were then able to derive not only the trend slope but also the Trend Strength Index, defined as 
\[
\text{Trend Strength Index} = 1 - \frac{\operatorname{Var}(\text{Residuals})}{\operatorname{Var}(\text{Trend + Residuals})}
\]
which indicates how dominant the trend component is over the other components.

\subsubsection{Changes In Moderation Trends}
To understand how moderation activity changes over time, we computed the trend slope for each time series. The slope indicates whether the number of SoRs or the average delay tend to increase or decrease over time, highlighting possible periods of heightened moderation.
Figure~\ref{fig:parallel-coordinates} shows a parallel-coordinates plot of the trend slope and the Trend Strength Index computed for each phase, both for the number of SoRs and the average delay. In particular, the bottom panels represents the trend slopes, with the left panel showing the slopes for the number of SoRs and the right panel showing the slopes for the average delay. For the number of SoRs, we note that almost all platforms increase their trend slope going from Phase 1 to Phase 2, with a consequent decrease in Phase 3. While slopes in Phase 1 can be both positive and negative, they are mostly positive in Phases 2 and 3, which may indicate increased moderation efforts during and after the elections. 
All but two platforms increased their slope in Phase 2 with respect to Phase 1, indicating stronger moderation efforts, especially Facebook. Instead, LinkedIn and YouTube exhibited negative slopes and minimal changes, pointing to limited or delayed adjustments. Overall, the changes were modest, and most platforms maintained similar slopes across phases.
For the average moderation delay, slopes remained relatively stable across most platforms. Some exceptions were observed in Phase 2. Pinterest showed a marked increase in delay, whereas Facebook showed a notable decrease. These patterns reflect the volume trends: the increase in Pinterest’s moderation activity led to longer delays, while Facebook managed the additional workload efficiently---likely thanks to its automated processes~\cite{10.1145/3543507.3583275}. 

\subsubsection{Stability Of Moderation Trends}
While the trend slope captures the direction and magnitude of change over time, it does not consider the stability of such change. The Trend Strength Index (TSI) addresses this limitation by quantifying how dominant the trend component is relative to residual variability. High values indicate that the observed temporal evolution of the trend is consistent and stable, whereas low values are caused by more unstable volatile patterns.
The upper-most part of Figure~\ref{fig:parallel-coordinates} shows the TSI for both the number of SoRs and moderation delay.
For the number of SoRs, Instagram, X, Pinterest, and Facebook exhibited notably high TSI values in Phase 2 whereas the other platforms have low to moderate values, with YouTube showing an extremely low TSI. 
Overall, these patterns suggested that while all platforms slightly increased their moderation activity during the electoral period, only a subset of platforms did it consistently and significantly.
Instead, for the average moderation delay, the TSI values were generally lower and more variable across platforms. Pinterest showed high TSI in Phase 2, consistent with a stable increase in delay, whereas Facebook displayed moderately low TSI, suggesting that its decrease in delay was less stable. These results further reinforce that, although the magnitude of moderation changes remained modest across platforms, the stability of such changes varied substantially, highlighting the heterogeneity in how platforms maintained their moderation patterns over time.

\subsubsection{Cross-platform Temporal Dynamics}
Trend slope and TSI capture changes in magnitude and stability. However, they do not consider differences in the timing of such changes across platforms. In fact, platforms may have adjusted their moderation practices, but not necessarily all at the same time.
To compare moderation dynamics across platforms while accounting for temporal misalignments, we used Dynamic Time Warping (DTW)---a well-known distance measure for comparing temporal trajectories.
Figure~\ref{fig:dtw-sors} in Appendix Section~\ref{sec:appendix-dtw} shows the pairwise DTW distances computed for the number of SoRs. Relatively low distances emerged in Phase 1, indicating broadly similar moderation dynamics before the elections. In Phase 2, distances increased consistently across platforms, suggesting that moderation trajectories became more heterogeneous during the electoral period. In Phase 3, distances decreased for some platforms but increased for others, most notably Pinterest and TikTok, suggesting divergent post-election behaviour.

When analyzing the DTW distances for the delay shown in Figure~\ref{fig:dtw-delay} in Appendix Section~\ref{sec:appendix-dtw}, we observe that they remain generally low across platforms. However, in Phase 3 Facebook stands out with noticeably higher distances compared to all other platforms. 
Taken together, these results highlight a dual divergence: platforms differ not only in the volume of SoRs they submit to the database but also in the timeliness of their moderation actions. This reinforces the idea that, rather than managing systemic risks in a consistent way, platforms responded differently depending on how imminent or significant they perceived the risks to be.

\begin{figure*}[t]
    \centering
    \includegraphics[width=\columnwidth]{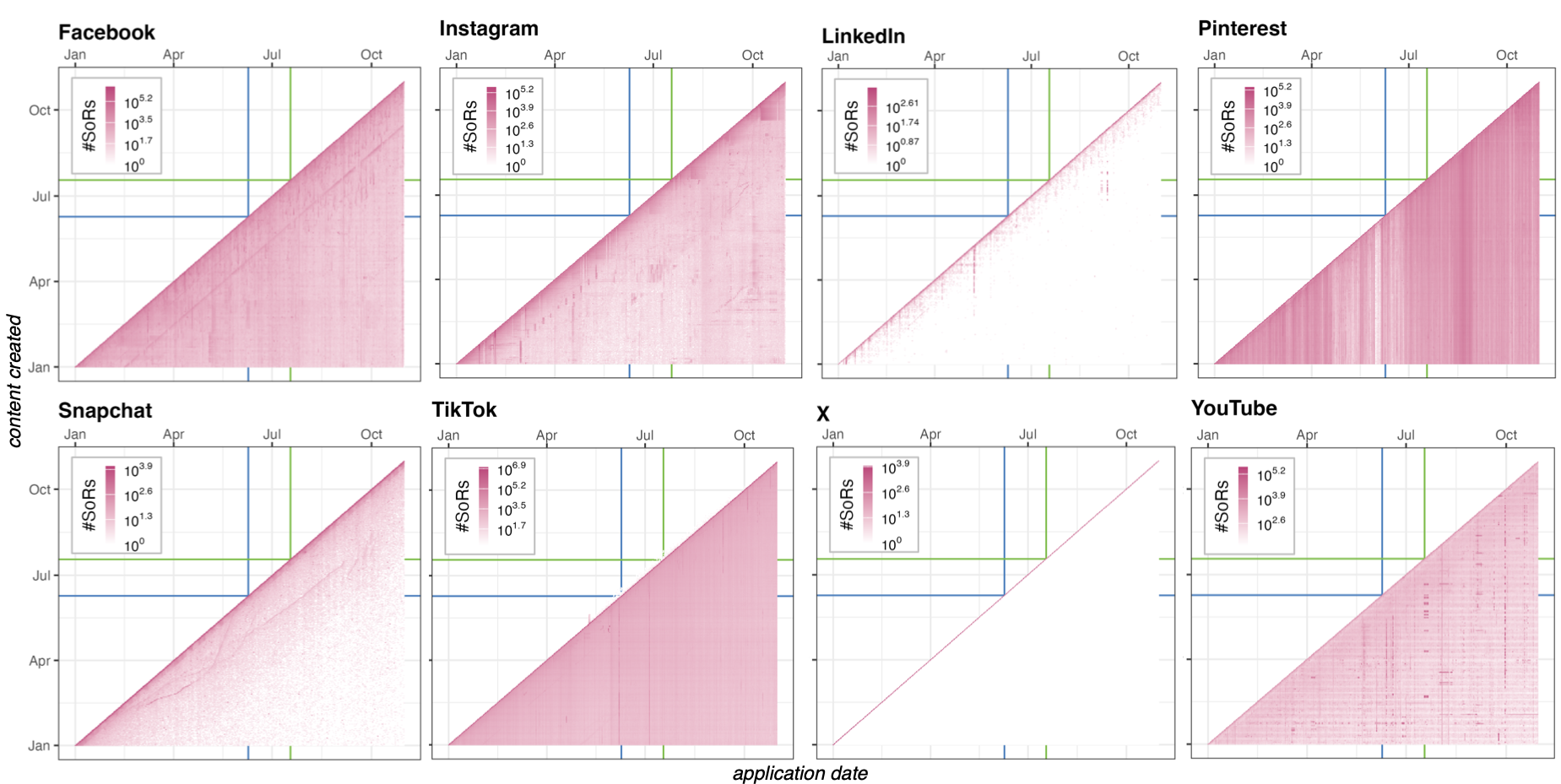}
    \caption{Analysis of moderation delays. For each platform, the heatmaps show the relationship, on a logarithmic scale, between the date when content was moderated (\textit{x} axis) and the date when the same content was published (\textit{y} axis). Blue lines indicate the Parliament elections days, while green lines indicate the President election day.}
    \label{fig:delay_analysis}
\end{figure*}
}

\subsection{Delays In Moderation Actions}
The analysis of moderation trends revealed no meaningful changes in terms of volume during the observed period. %
However, the time series of average daily moderation delays (Figure~\ref{fig:avg_delay_time_series}) shows that platforms differ in how quickly they moderate content, with some platforms displaying noticeable increases in delay during the Phase 3.
For example, as also shown in Figure~\ref{fig:parallel-coordinates}, YouTube, Pinterest, Snapchat, Facebook, and Instagram increased their moderation delay, which was particularly evident from late August to early September 2024. %
This trend suggests that platforms may not have moderated election-period content until after the elections concluded, which could explain both the increased delays in the post-electoral period (Phase 3) and the lack of changes in moderation volume during the election (Phase 2). To further explore this hypothesis, we performed a more detailed analysis of moderation delays to determine whether content posted during the elections was moderated at a later time.

The heatmaps in Figure~\ref{fig:delay_analysis} show the relationship between these dates on a logarithmic color scale. Points along the main diagonal represent no moderation delay while points below the diagonal indicate moderation of older content. In this visualization, no point should appear above the main diagonal, as that would imply content moderated before it was published. Figure~\ref{fig:delay_analysis} shows once again how different platforms display very different moderation patterns. For LinkedIn the moderation delay is minimal, and for X it is exactly zero. Instead, YouTube, Instagram, and Snapchat primarily focus on recent content despite a few fluctuations. In contrast, Facebook and TikTok show a more uniform distribution, with moderation delays spread more evenly over time. Lastly, Pinterest displays a moderation pattern largely independent of publication date, as reflected by the vertical lines in its heatmap. Other patterns include the dark diagonal lines visible in the heatmaps for Facebook and Snapchat. Facebook, in particular, exhibits a consistent one-month lag for certain moderation actions, which may reflect either batch reviews or scheduled automated moderation processes. Similar, though less pronounced, patterns appear in Instagram and YouTube.

Overall, this analysis allowed us to test our initial hypothesis. If platforms had significantly moderated content after the electoral period, we would expect to see a distinct pattern in the heatmaps. In particular, we would see a multitude of moderation actions from early September onward, targeting content published around the Parliament and Presidential elections. However, there is no sign of such pattern in Figure~\ref{fig:delay_analysis}, which rejects the hypothesis.

\subsection{External Event's Influence On Platform Moderation}

While overall moderation trends around the elections showed little change, certain related events or other major political occurrences may have influenced users and platform behaviors. To investigate this, we applied Pruned Exact Linear Time (PELT) to the clean and standardized time series to detect change points. PELT is a state-of-the-art change point detection technique that provides optimal segmentation with linear computational complexity under mild assumptions, making it particularly suitable for large-scale time series analysis~\cite{killick2012optimal,truong2020selective}. This allowed identifying specific dates when moderation patterns shifted significantly. After identifying these points, we investigated the main events occurred on or just before those dates, to understand potential triggers for the observed changes. For the number of SoRs time series, as shown in Appendix Figure~\ref{fig:cp-sors}, two dates emerged as the most recurring across all time series: 29 June and 19 July.

The former (29 June) is identified across five platforms---that is, Instagram, Pinterest, Snapchat, TikTok, and X. This date follows shortly after the European Parliament elections and coincides with the day the EU adopted a set of new sanctions towards Belarus due to its involvement with Russia in the Russo-Ukrainian war~\cite{xu2025social}, likely causing an increase of conflict-related content.\footnote{\url{https://finance.ec.europa.eu/news/eu-adopts-new-set-sanctions-against-belarus-2024-06-29_en}}
On the other hand, the latter (19 July) is detected for Facebook, Instagram, and LinkedIn. On this day, a major IT outage caused by a faulty CrowdStrike update affected millions of Windows systems worldwide.\footnote{\url{https://www.theguardian.com/business/live/2024/jul/19/retail-sales-great-britain-slump-12-government-borrowing-june-figure-lowest-2019-horizon-business-live}} The incident caused widespread concern and discussion, potentially causing increased online activity and subsequent moderation.

Following the same rationale, we also examined the delay time series. Results are shown in Appendix Figure~\ref{fig:cp-delays}. The most frequent change points occurred on 10 April (for Instagram, TikTok, and X), 29 July (for Instagram, LinkedIn, and X), and 3 August (for Facebook, X, and YouTube). The first date coincides with the European Parliament vote on migration and asylum reforms.\footnote{\url{https://www.europarl.europa.eu/news/en/press-room/20240408IPR20290/meps-approve-the-new-migration-and-asylum-pact}} Instead, the second and third dates fall within periods of intensified conflict in Gaza and the West Bank.\footnote{\url{https://www.un.org/unispal/document/humanitarian-situation-update-312-west-bank/}} This could have caused an increase in the number of moderated content leading to increased moderation delays.

Overall, these observations suggest that platform moderation is influenced not only by internal policies but also by major external world events. Periods of geopolitical tension, such as the European Parliament vote on migration reforms and the escalation of the Israel–Palestine conflict, might have affected both the volume of moderated content and the delays in moderation. However, there is still no evidence of notable changes directly associated with the European Parliament elections themselves.

\begin{figure}[t]
    \centering
    \includegraphics[width=0.7\linewidth]{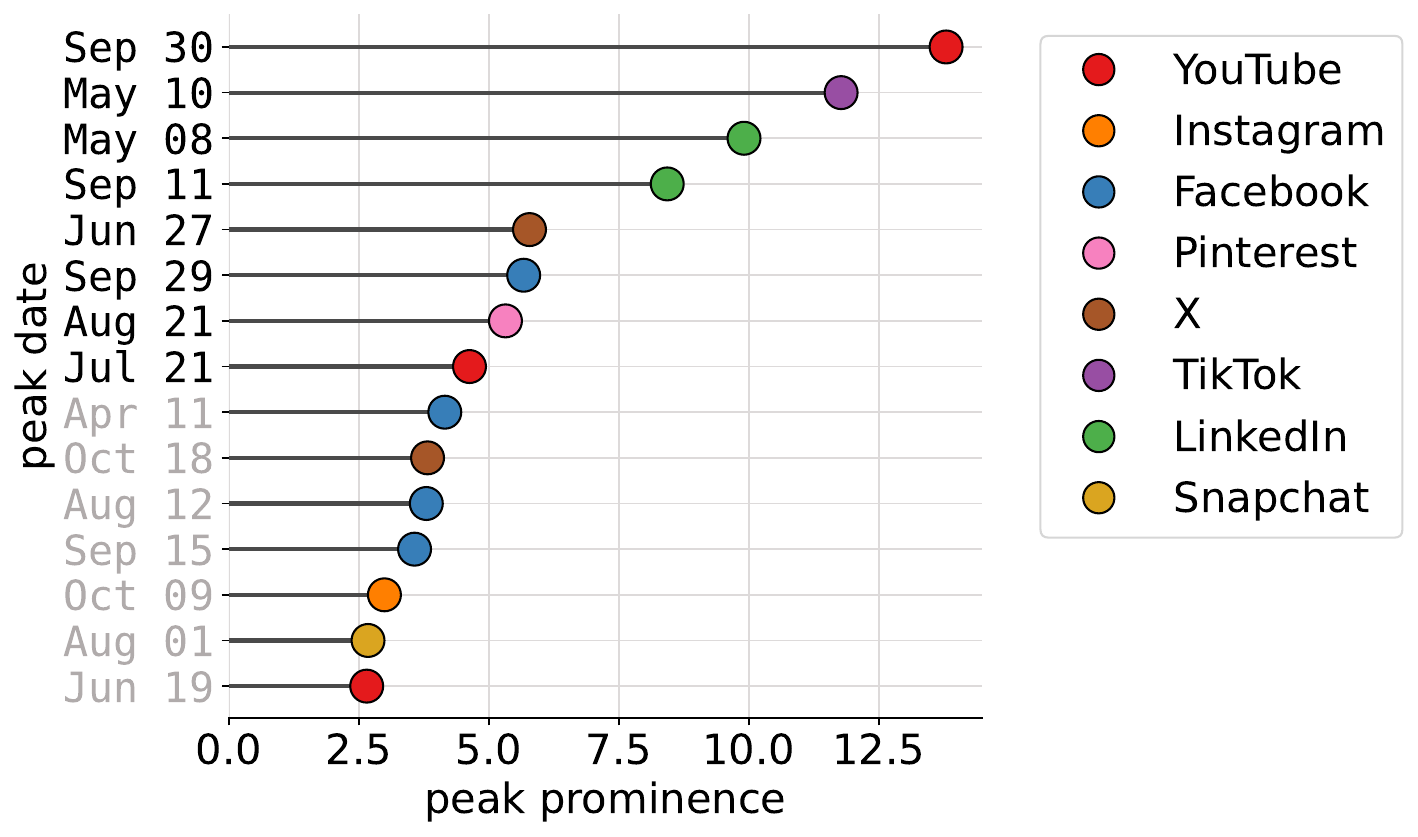}
    \caption{Top 15 global peaks sorted by decreasing prominence score. Here, we focus on the eight most prominent ones, with the corresponding date colored in black.}
    \label{fig:all-top-peaks}
\end{figure}

\subsection{Anomalies In Moderation Actions}

While the previous time series and trend analysis did not reveal clear signs of adaptation, we did notice various sharp peaks in moderation volume. This might be a signal of localized moderation interventions.
For example, these anomalous peaks could point to mass election-related moderation that did not emerge in a broader trend analysis~\cite{decook2022r}.

To test this hypothesis, we compared the attributes of SoRs associated with notable moderation peaks to those submitted on surrounding routine days, aiming to identify possible differences. For this analysis, we consider routine days those occurring within the two weeks before and after the peak. Out of all SoR attributes and values, we focused on those that could indicate election-related tampering (e.g., the predefined category \dbField{negative\_effects\_on\_civic\_discourse\_or\_elections}). We also analyzed the type of infringement (\dbField{category}), the type of moderated content (\dbField{content\_type}), the specific reason for incompatibility with platform policies (\dbField{incompatible\_ground}), and the use of automation in moderation, distinguishing between \dbField{automated\_detection} and \dbField{automated\_decision}. 
For clarity, we report only the attributes most frequently used by each platform, as well as for those revealing clear differences. For each attribute, we provide additional details in Appendix Section~\ref{sec:appendix-values}. For each considered peak, we plot the differences using diverging bar charts: left-aligned blue bars represent SoRs from routine (non-peak) days, while right-aligned red bars correspond to SoRs associated with the peak.
Each platform exhibited several peaks, but our analysis focuses on the most prominent ones. Figure~\ref{fig:all-top-peaks} reports the 15 peaks with the highest prominence, defined as the vertical distance between the peak and its lowest contour line. From this set, we selected the eight most prominent peaks for detailed analysis.

\begin{figure*}[h!]
\centering
    \begin{subfigure}{0.45\textwidth}%
        \includegraphics[width=\textwidth]{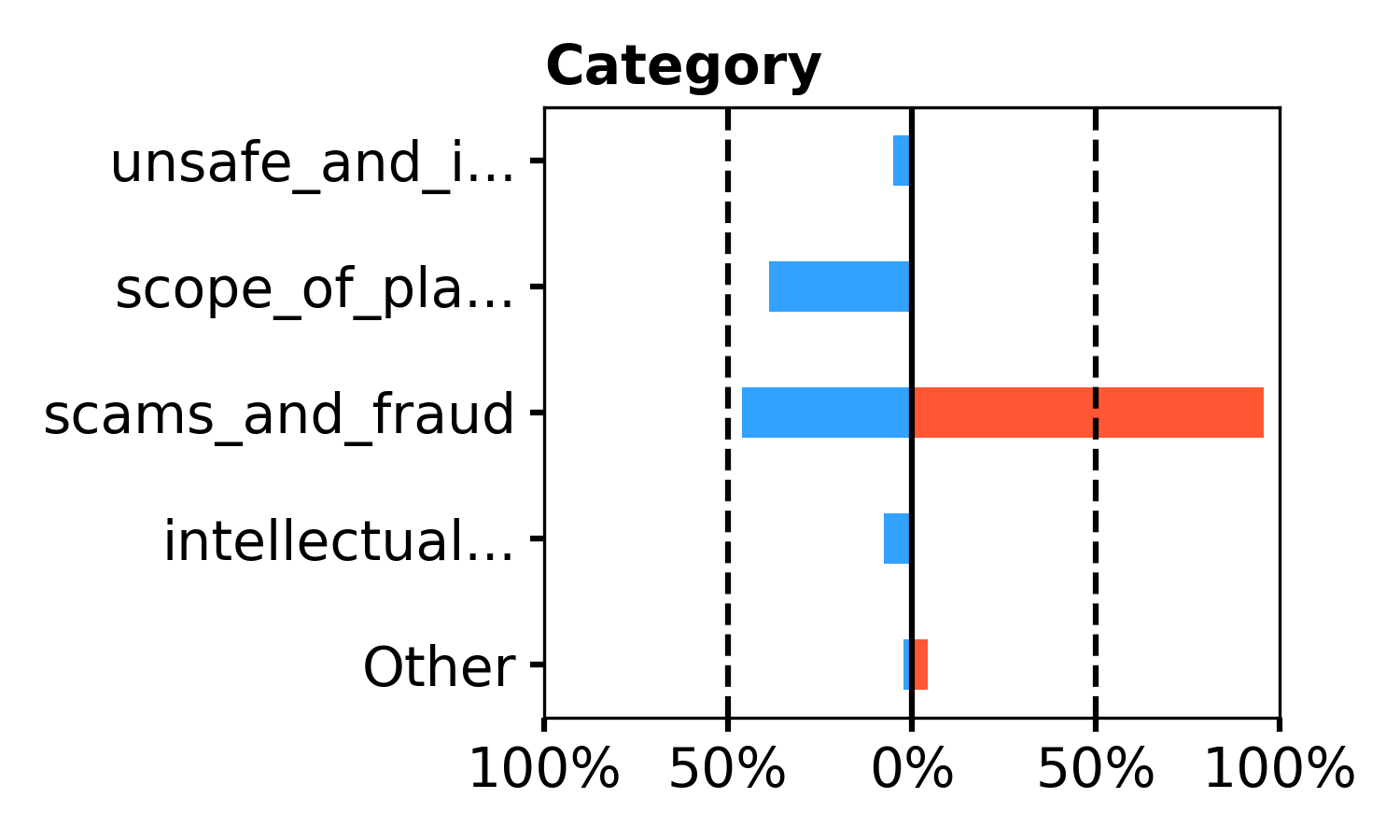}%
    \end{subfigure}%
    \begin{subfigure}{0.45\textwidth}%
        \includegraphics[width=\textwidth]{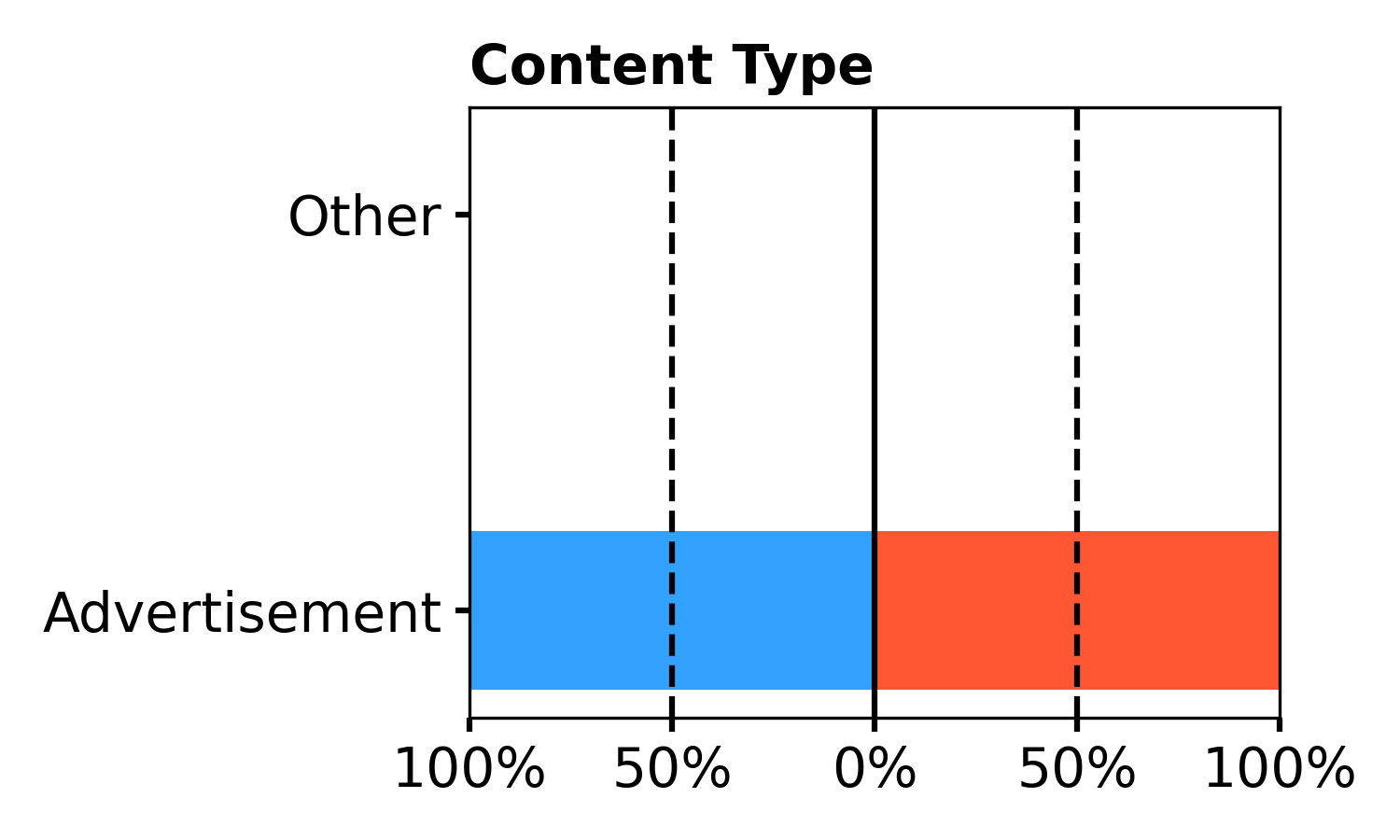}%
    \end{subfigure}%
    \\
    \begin{subfigure}{0.45\textwidth}%
        \includegraphics[width=\textwidth]{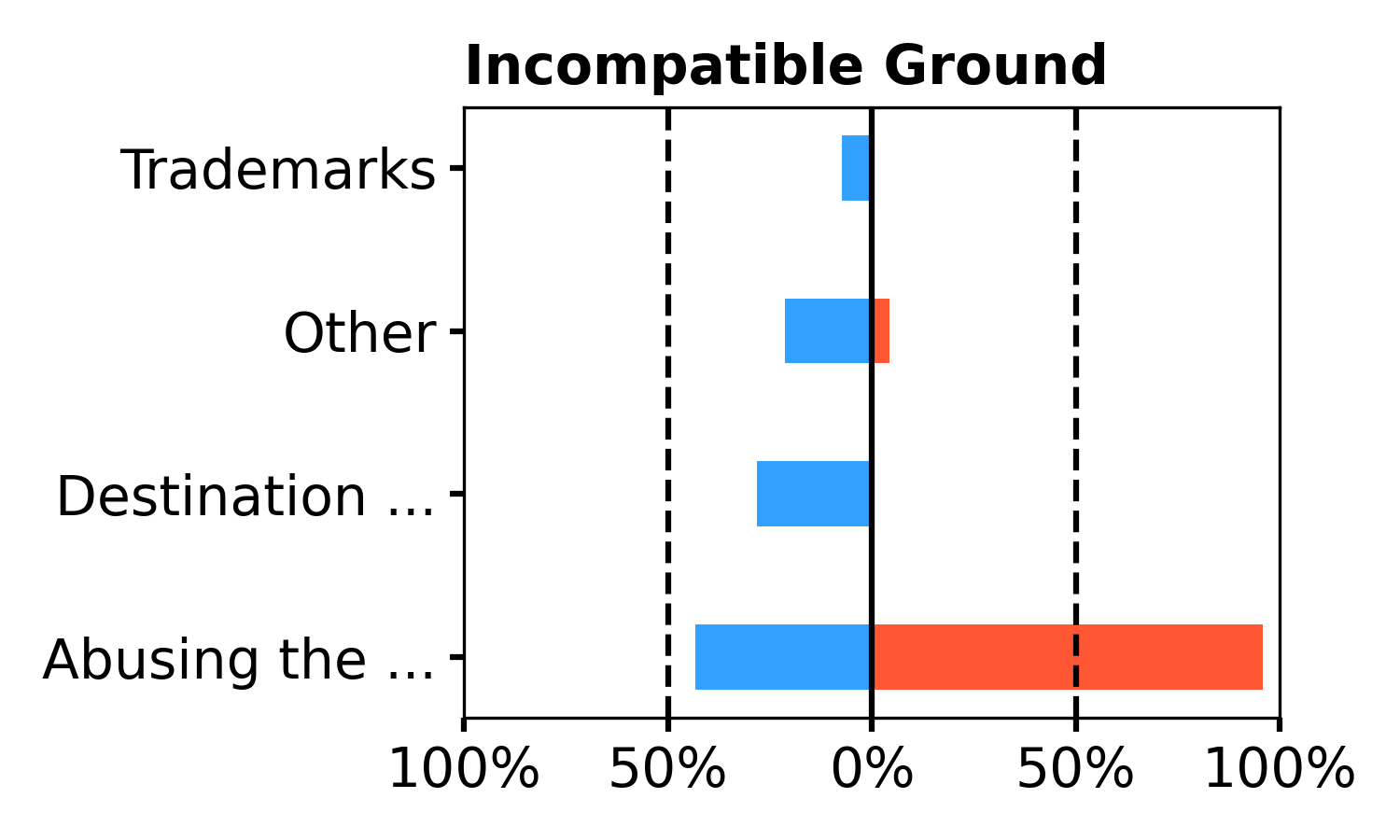}%
    \end{subfigure}%
    \begin{subfigure}{0.45\textwidth}%
        \includegraphics[width=\textwidth]{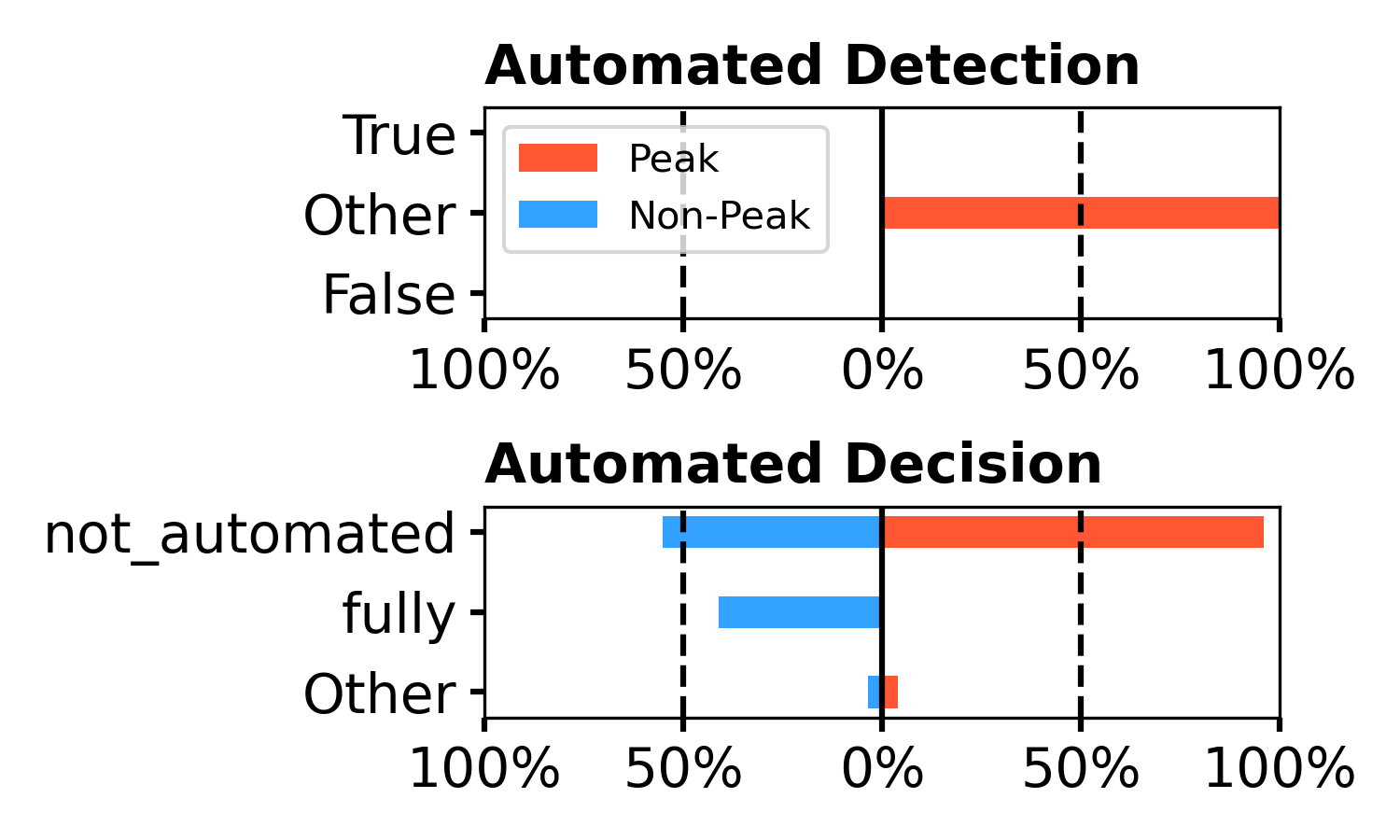}%
    \end{subfigure}%
    \caption{Comparison between the SoRs (right-aligned, red-colored) that caused the moderation anomaly reported for \faYoutube~\textbf{YouTube} during \textbf{September 30, 2024} and the SoRs (left-aligned, blue-colored) from the surrounding routine days.}
    \label{fig:anomaly-youtube-peak01}
\end{figure*}

\noindent\textbf{\faYoutube~YouTube---September 30, 2024.} Figure~\ref{fig:anomaly-youtube-peak01} reveals that the anomaly derives from an excessive posting of fraudulent advertisement, responsible for abusing the ad network. This increase suggests coordinated and systematic, rather than isolated, violations. However, there is no sign of those infractions being correlated to the European elections. Moreover, while the detection of the content is partially automated, the decision to moderate is fully manual, as opposed to the routine days.

\begin{figure*}[h!]
\centering
    \begin{subfigure}{0.45\textwidth}%
        \includegraphics[width=\textwidth]{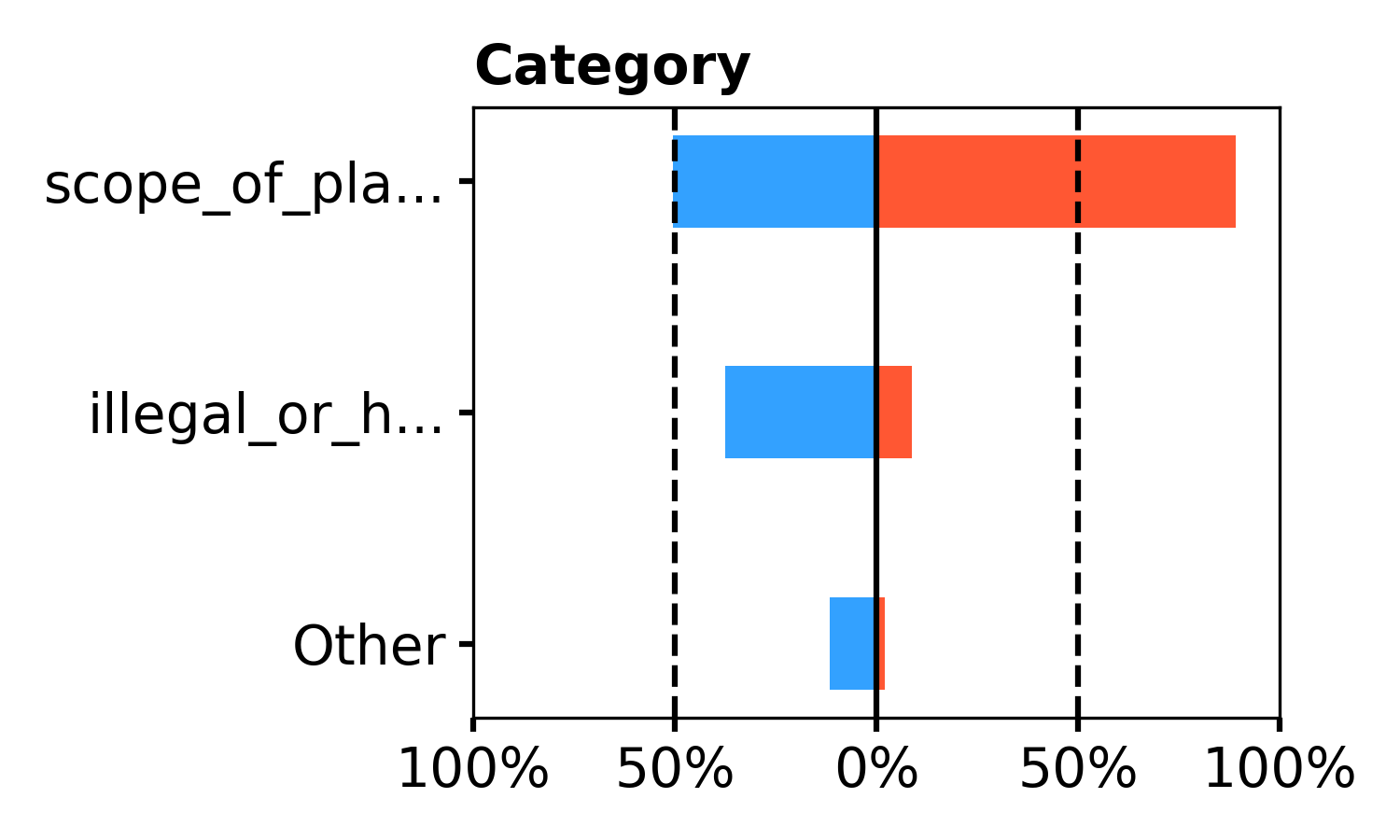}%
    \end{subfigure}%
    \begin{subfigure}{0.45\textwidth}%
        \includegraphics[width=\textwidth]{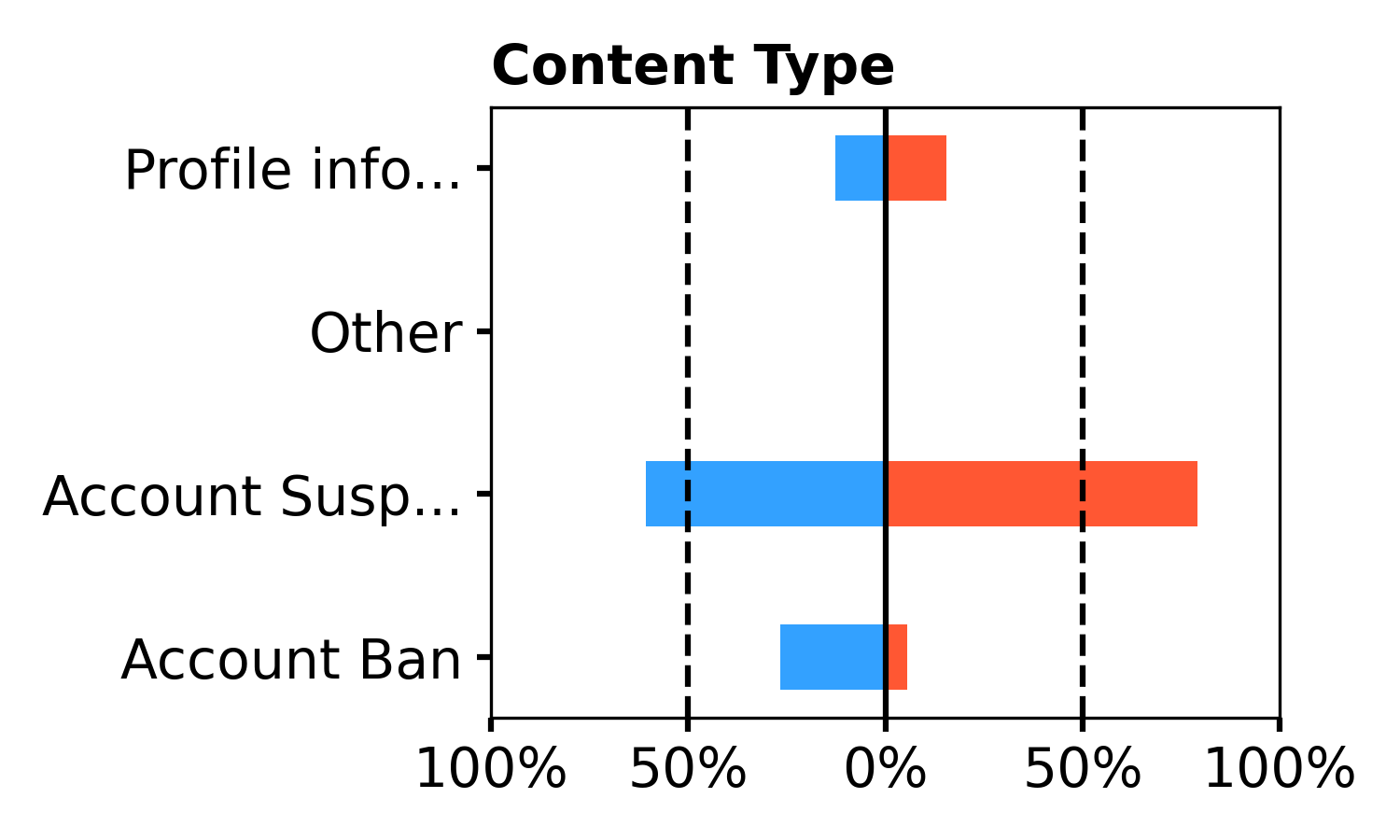}%
    \end{subfigure}%
    \\
    \begin{subfigure}{0.45\textwidth}%
        \includegraphics[width=\textwidth]{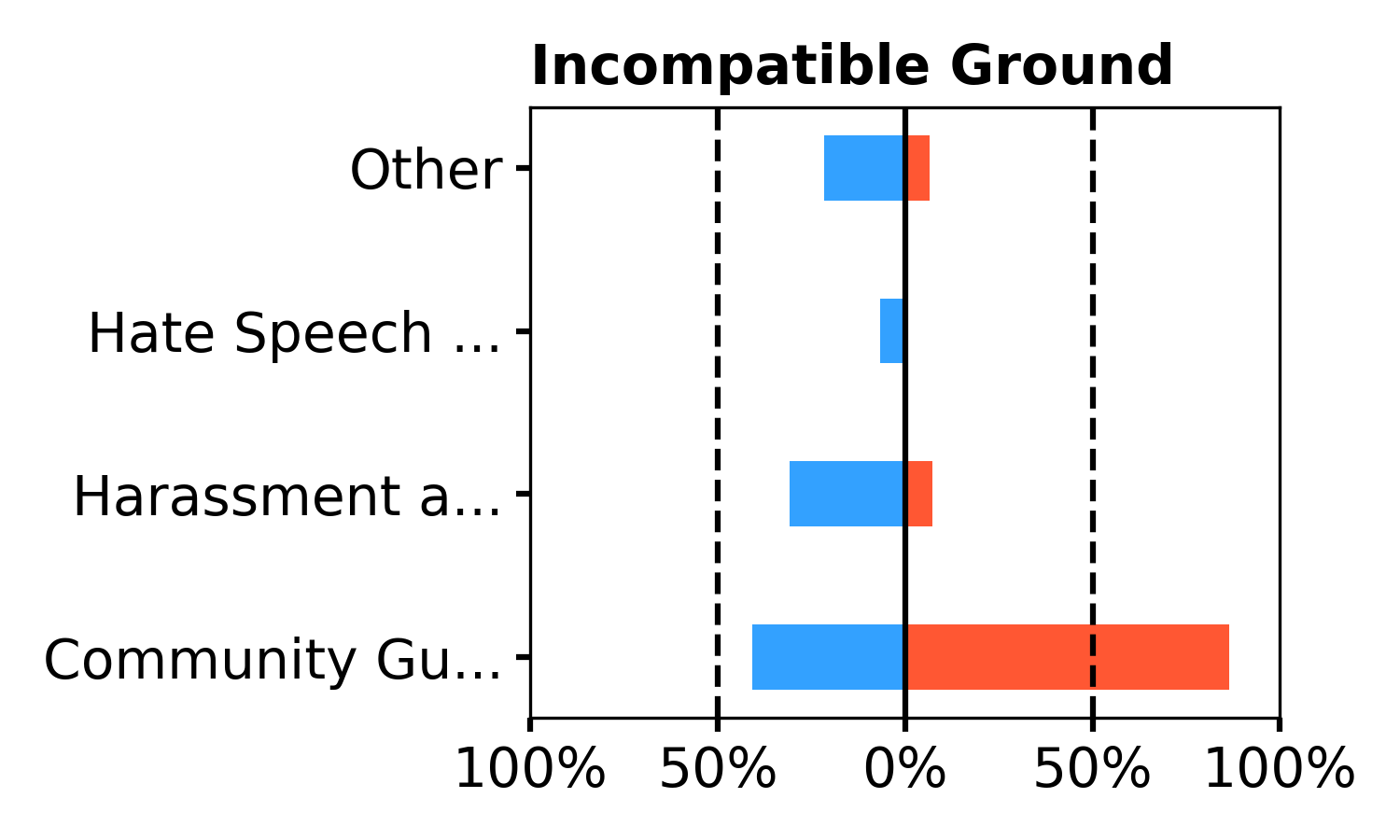}%
    \end{subfigure}%
    \begin{subfigure}{0.45\textwidth}%
        \includegraphics[width=\textwidth]{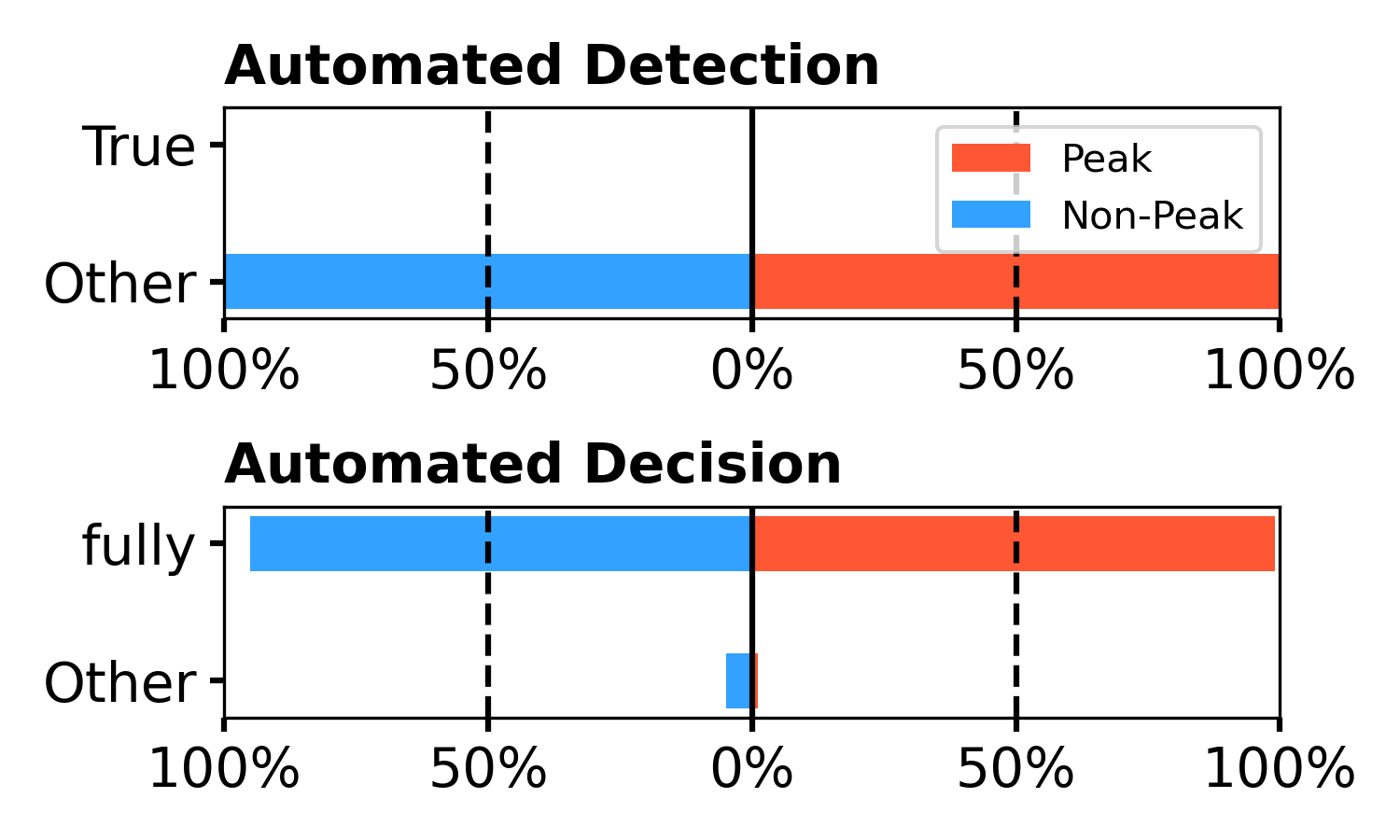}%
    \end{subfigure}%
    \caption{Comparison between the SoRs (right-aligned, red-colored) that caused the moderation anomaly reported for \faTiktok~\textbf{TikTok} on \textbf{May 10, 2024} and the SoRs (left-aligned, blue-colored) by the same platform from the surrounding routine days.}
    \label{anomaly-tiktok-volume}
\end{figure*}

\noindent\textbf{\faTiktok~TikTok---May 10, 2024.} %
Figure~\ref{anomaly-tiktok-volume} reveals that this anomaly is driven by the mass suspension of accounts  operating outside TikTok’s intended use and in violation of its community guidelines.
However, the information carried by these SoRs remains highly generic, preventing an in-depth understanding of the specific drivers of this moderation surge or of any direct connection to the elections.

\begin{figure*}[h!]
    \centering
    \begin{subfigure}{0.45\textwidth}%
        \includegraphics[width=\textwidth]{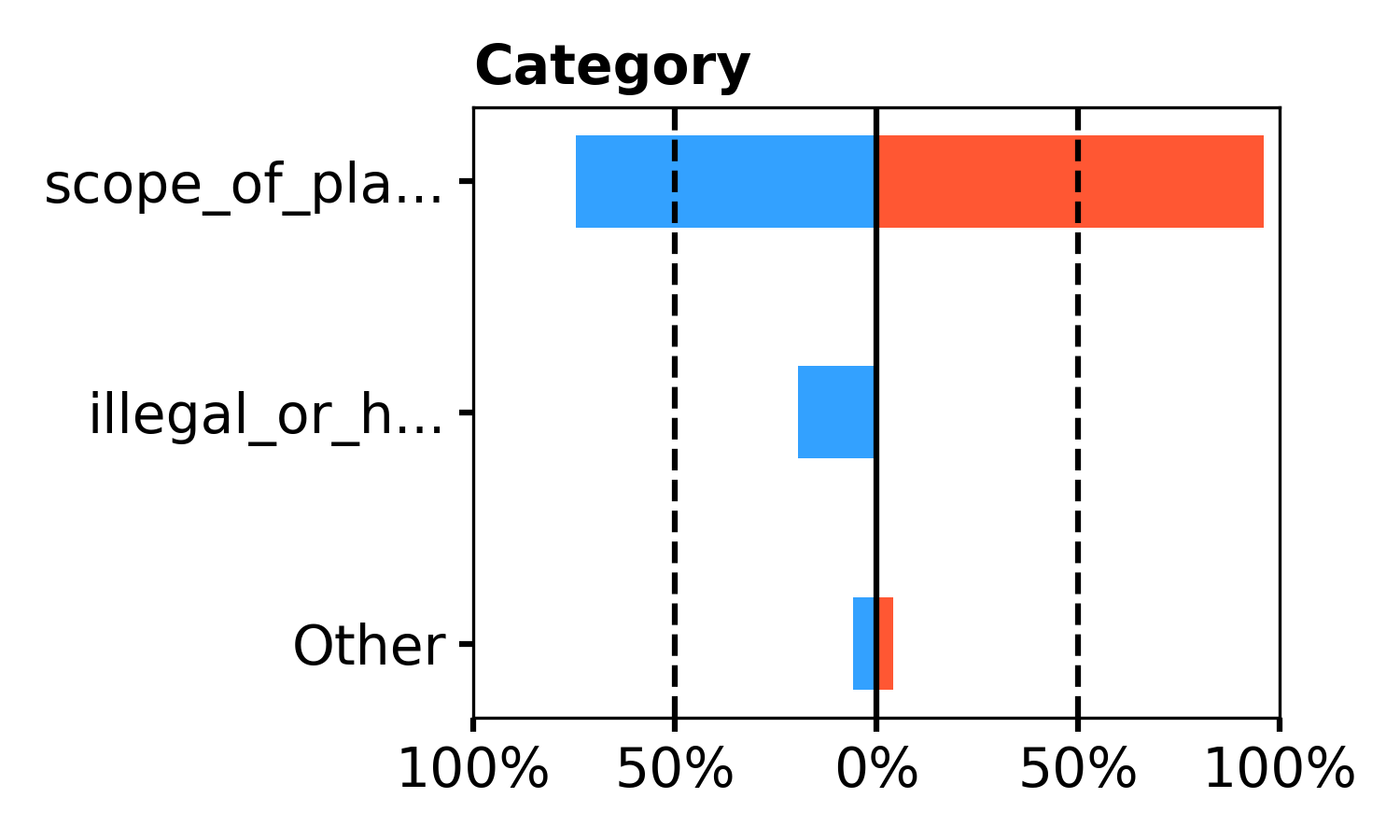}%
    \end{subfigure}%
    \begin{subfigure}{0.45\textwidth}%
        \includegraphics[width=\textwidth]{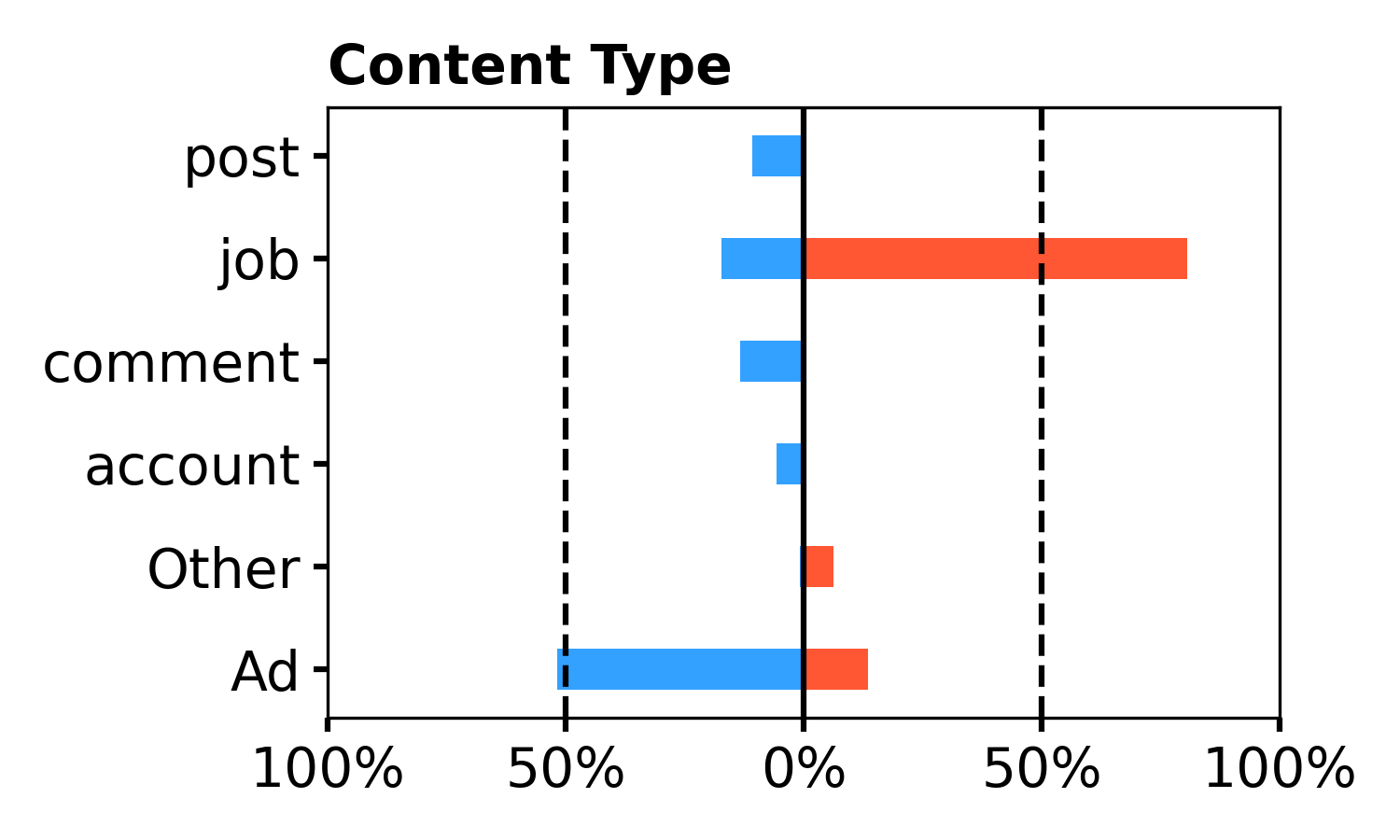}%
    \end{subfigure}%
    \\
    \begin{subfigure}{0.45\textwidth}%
        \includegraphics[width=\textwidth]{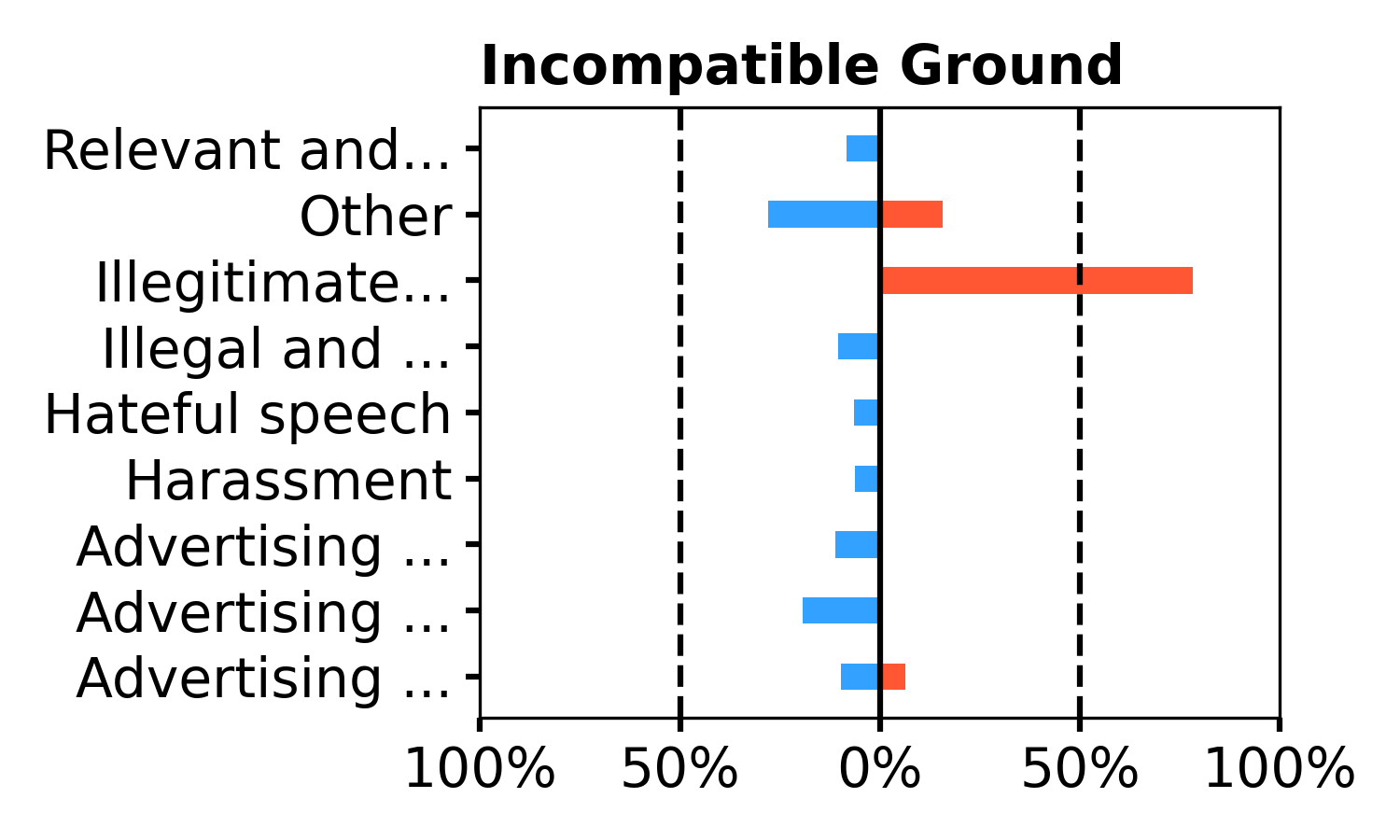}%
    \end{subfigure}%
    \begin{subfigure}{0.45\textwidth}%
        \includegraphics[width=\textwidth]{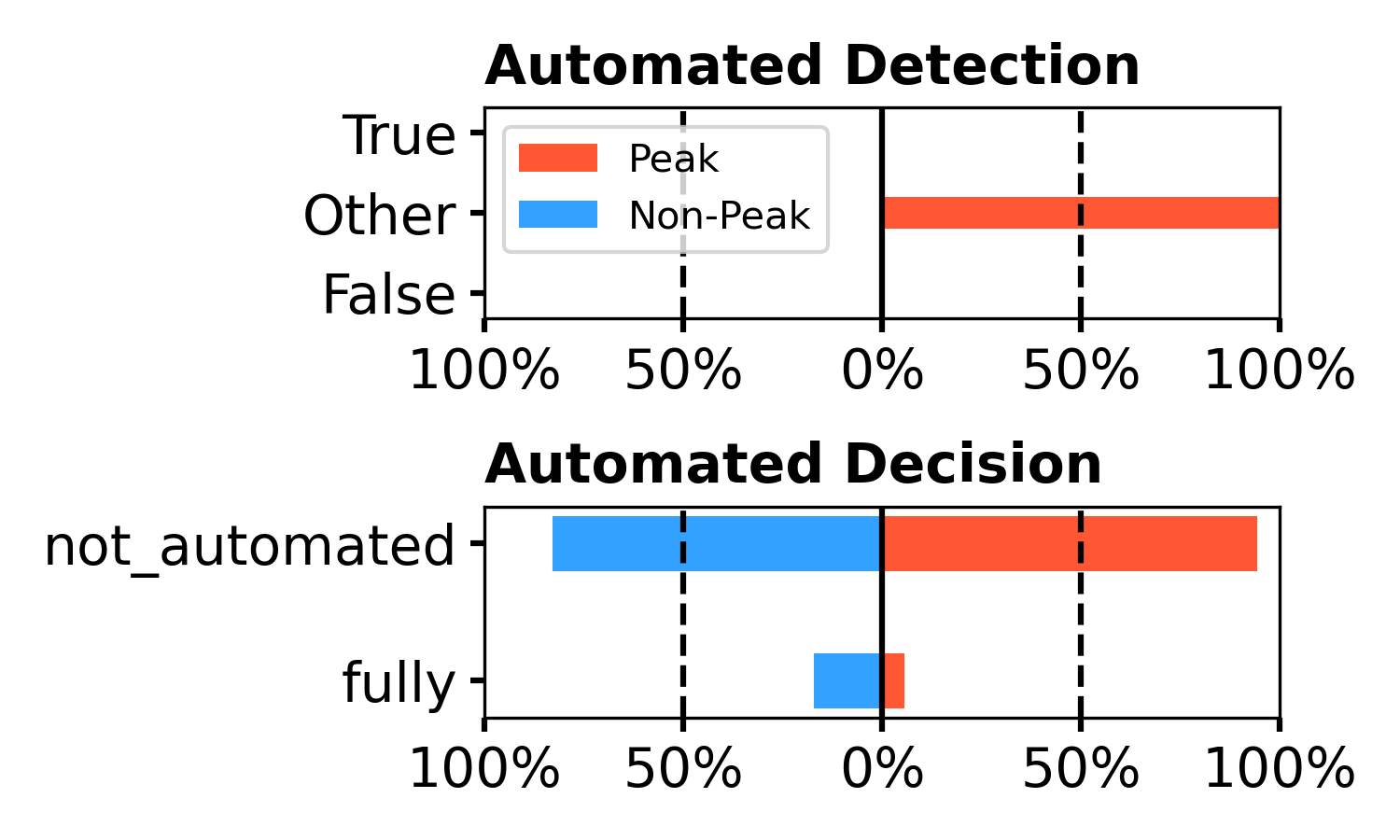}%
    \end{subfigure}%
    \caption{Comparison between the SoRs (right-aligned, red-colored) that caused the moderation anomaly reported for \faLinkedin~\textbf{LinkedIn} on \textbf{May 8, 2024} and the SoRs (left-aligned, blue-colored) by the same platform from the surrounding routine days.}
    \label{anomaly-linkedin-volume}
\end{figure*}

\noindent\textbf{\faLinkedin~LinkedIn---May 8, 2024.} %
Figure~\ref{anomaly-linkedin-volume} shows how the SoRs submitted during the anomaly day are significantly different from the ones submitted during the surrounding days.
While LinkedIn primarily moderated advertisements, during the peak day the platform mostly moderated illegitimate job offers, thus focusing on different targets.
However, the rationale behind the moderation of the job offers cannot be deduced by current data, nor it is possible to determine possible correlations with the elections.

\begin{figure*}[h!]
\centering
    \begin{subfigure}{0.45\textwidth}%
        \includegraphics[width=\textwidth]{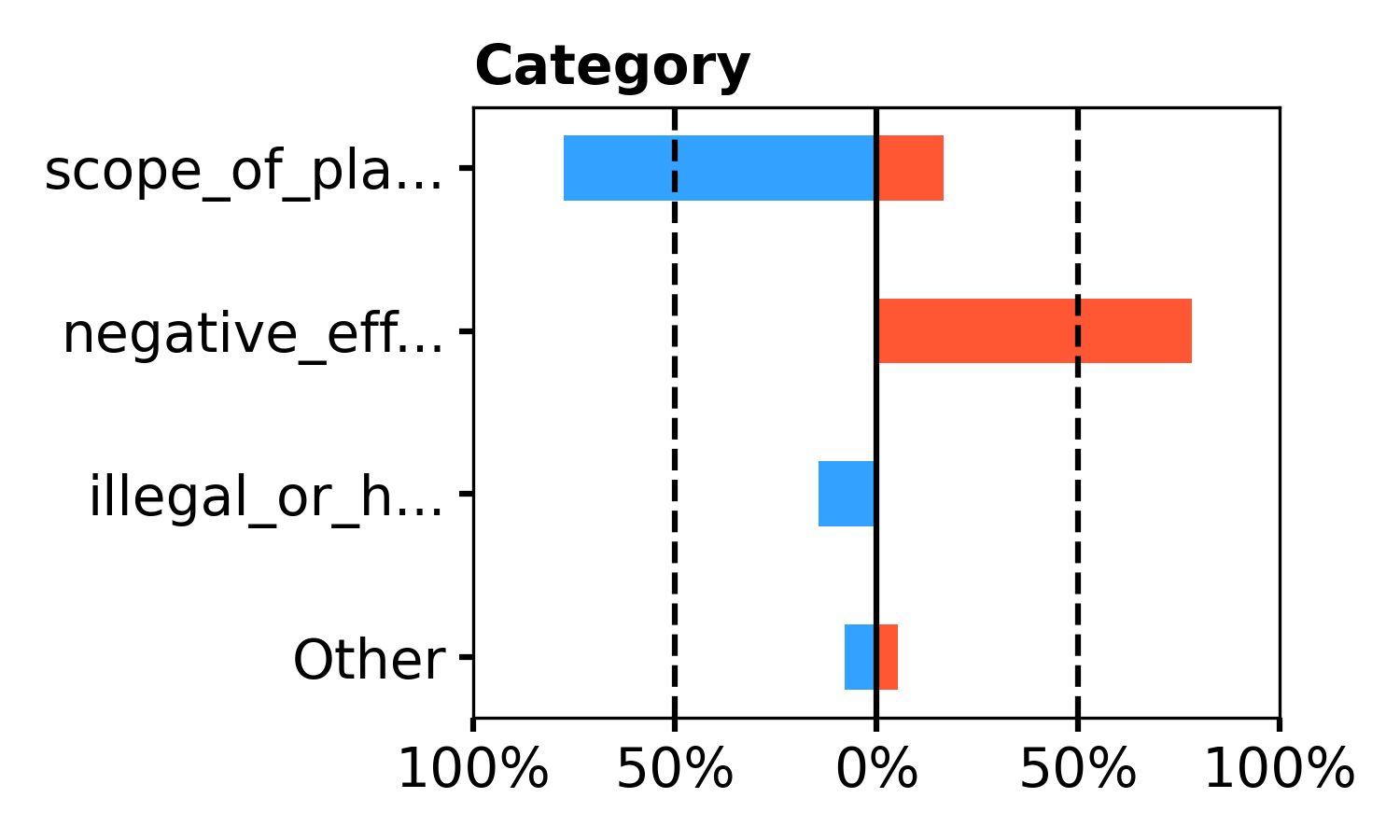}%
    \end{subfigure}%
    \begin{subfigure}{0.45\textwidth}%
        \includegraphics[width=\textwidth]{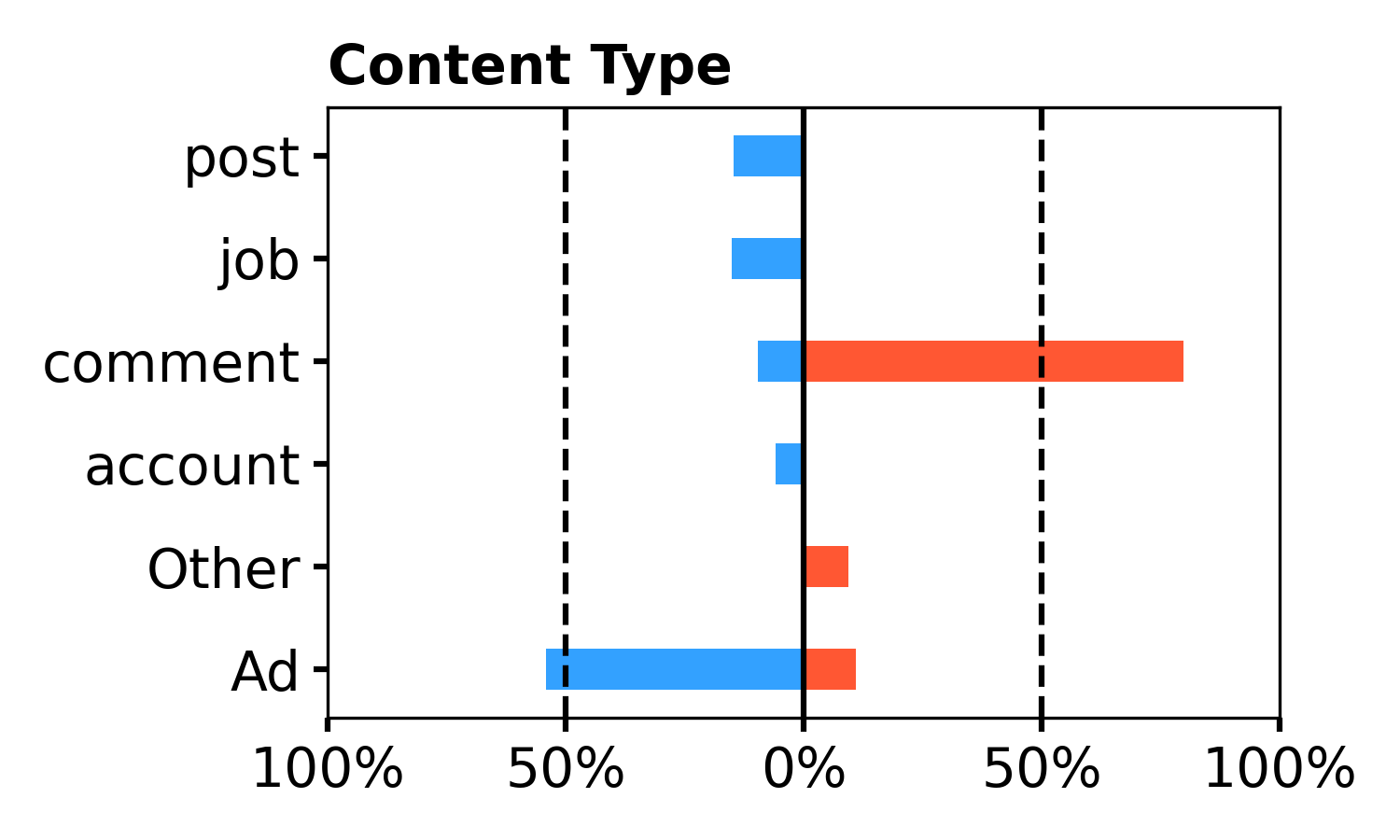}%
    \end{subfigure}%
    \\
    \begin{subfigure}{0.45\textwidth}%
    \includegraphics[width=\textwidth]{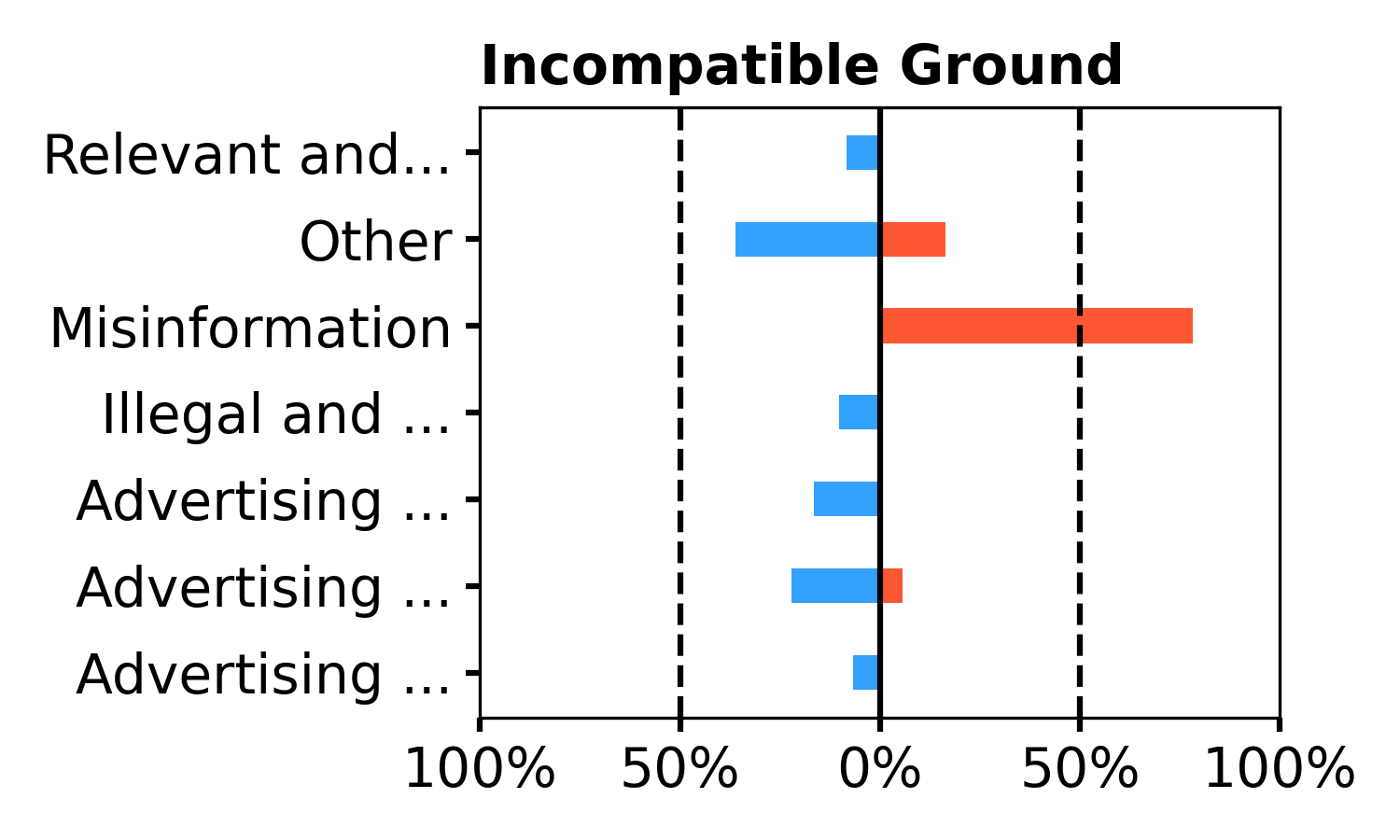}%
    \end{subfigure}%
    \begin{subfigure}{0.45\textwidth}%
        \includegraphics[width=\textwidth]{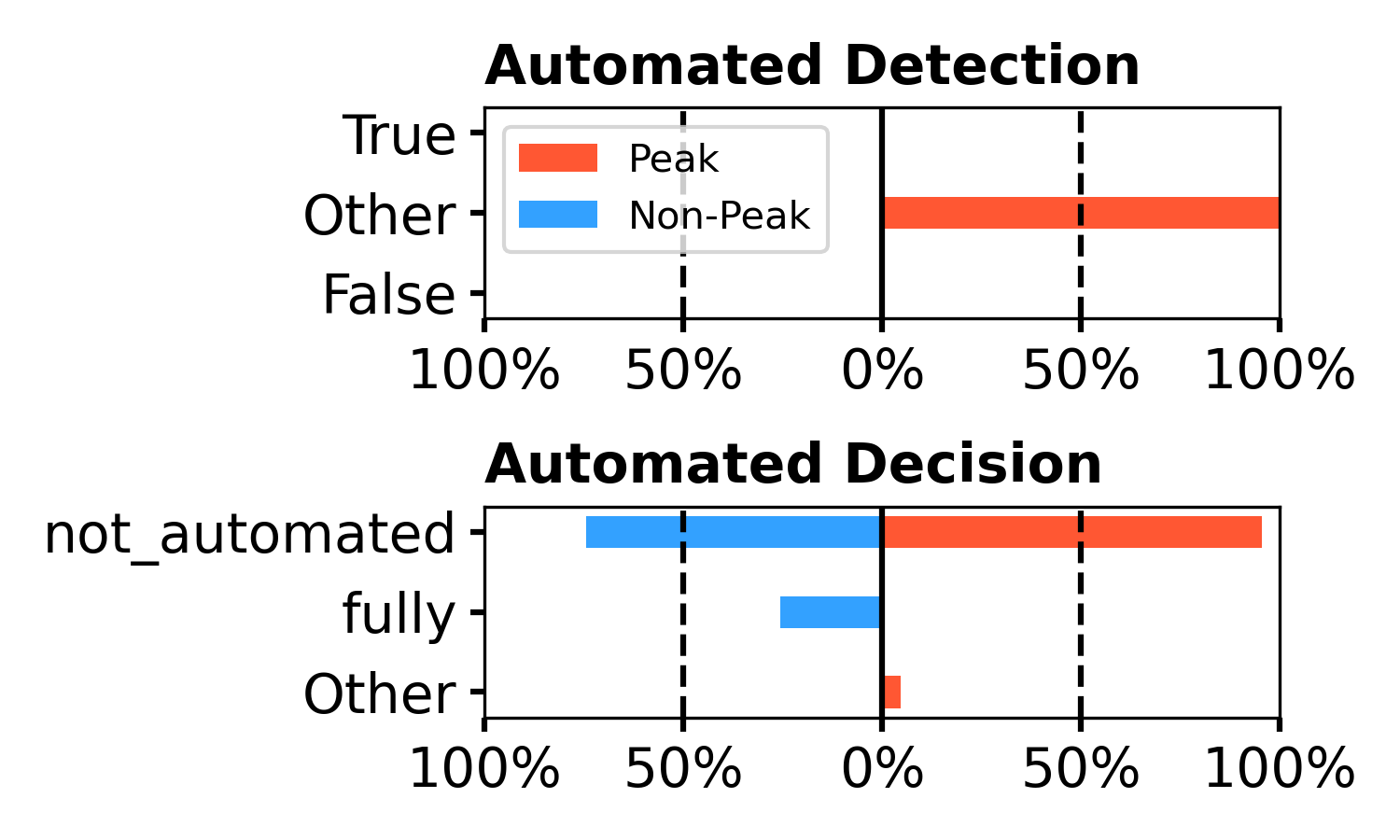}%
    \end{subfigure}%
    \caption{Comparison between the SoRs (right-aligned, red-colored) that caused the moderation \textit{delay} anomaly reported by \faLinkedin~\textbf{LinkedIn} on \textbf{September 11, 2024} and the SoRs (left-aligned, blue-colored) by the same platform from the surrounding routine days.}
    \label{fig:anomaly-linkedin-delay}
\end{figure*}

\noindent\textbf{\faLinkedin~LinkedIn---September 11, 2024.} %
The pattern displayed in Figure~\ref{fig:anomaly-linkedin-delay} emerges as the most significant peak in our analysis, with an increase in moderation mainly caused by a marked rise in comments moderated for election-related misinformation.
Its relevance is highlighted by the explicit use of the \dbField{negative\_effects on\_civic\_discourse\_or\_elections} field, a notably specific designation compared the generic moderation categories usually employed. 
Moreover, the decision to moderate is almost fully manual, signaling a more intentional review process.
We also point out that this surge emerged also in the delay analysis, and is clearly visible in Figure~\ref{fig:avg_delay_time_series} and Figure~\ref{fig:delay_analysis}.
This indicates that the moderated content was not recent, but had been published several weeks earlier. Specifically, the delay amounts to approximately 45 days, as the content was posted in mid-July but moderated no earlier than September. This timing is relevant, as it falls within the inter-electoral phase.
The distinctive features of this anomaly--both in terms of its timing and the specificity of the moderation labels employed--provide strong indications that the observed spike was related to the electoral context.
This case underscores the value of granular labeling within the \texttt{DSA-TDB}, as LinkedIn’s use of precise categories rather than broader classifications such as \dbField{scope\_of\_platform\_service},allowed us to confidently reach this conclusion.

\begin{figure*}[h!]
\centering
    \begin{subfigure}{0.45\textwidth}%
        \includegraphics[width=\textwidth]{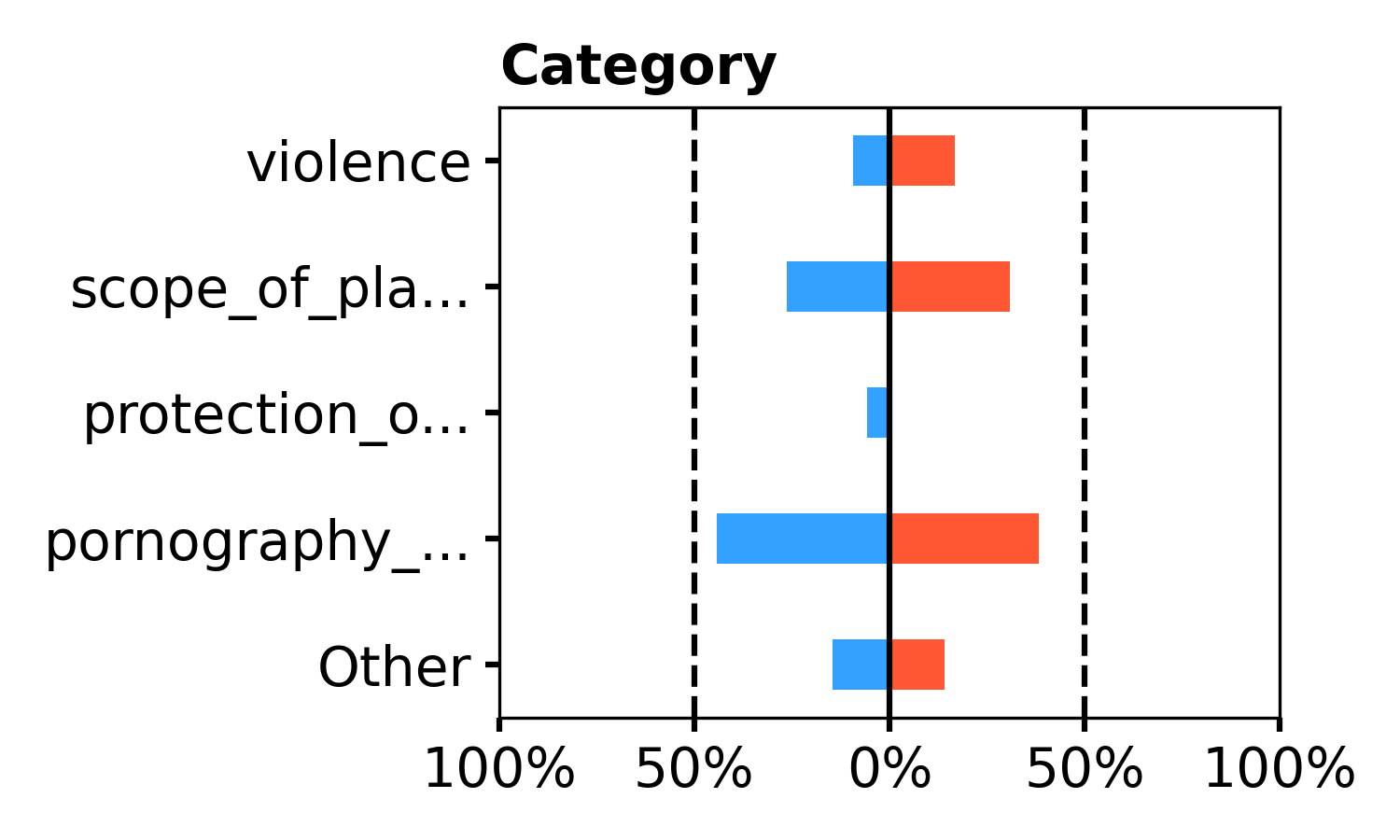}%
    \end{subfigure}%
    \begin{subfigure}{0.45\textwidth}%
        \includegraphics[width=\textwidth]{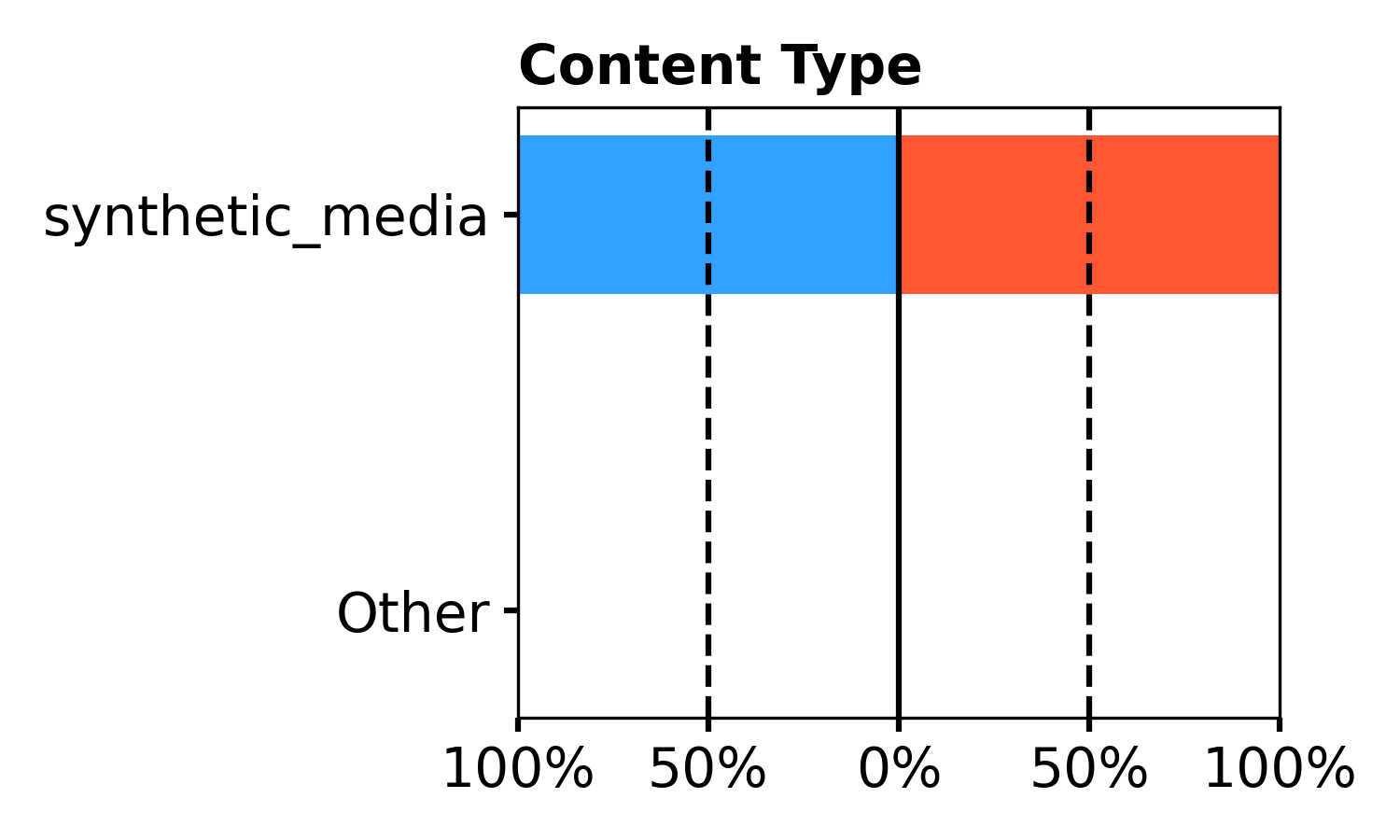}%
    \end{subfigure}%
    \\
    \begin{subfigure}{0.45\textwidth}%
        \includegraphics[width=\textwidth]{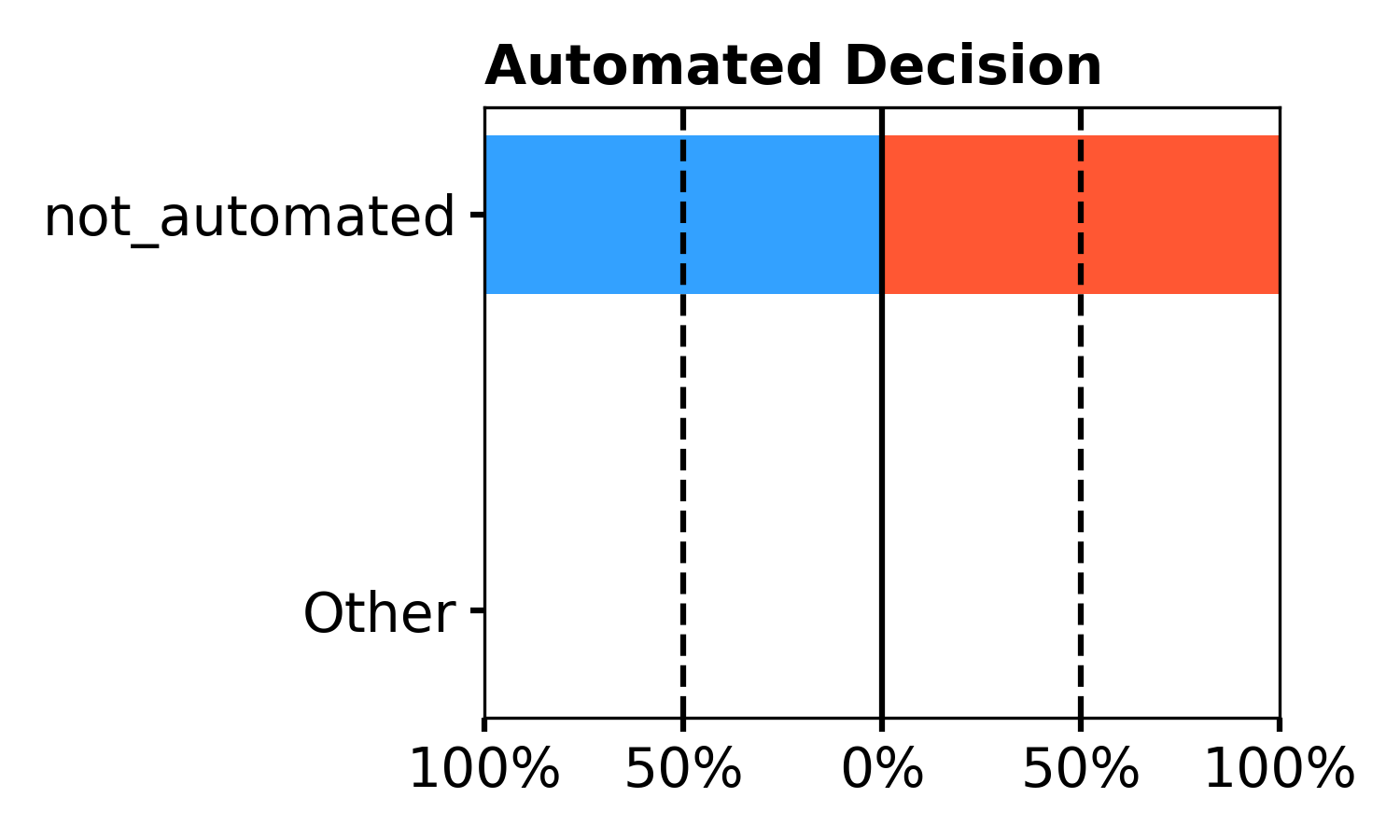}%
    \end{subfigure}
    \begin{subfigure}{0.45\textwidth}%
        \includegraphics[width=\textwidth]{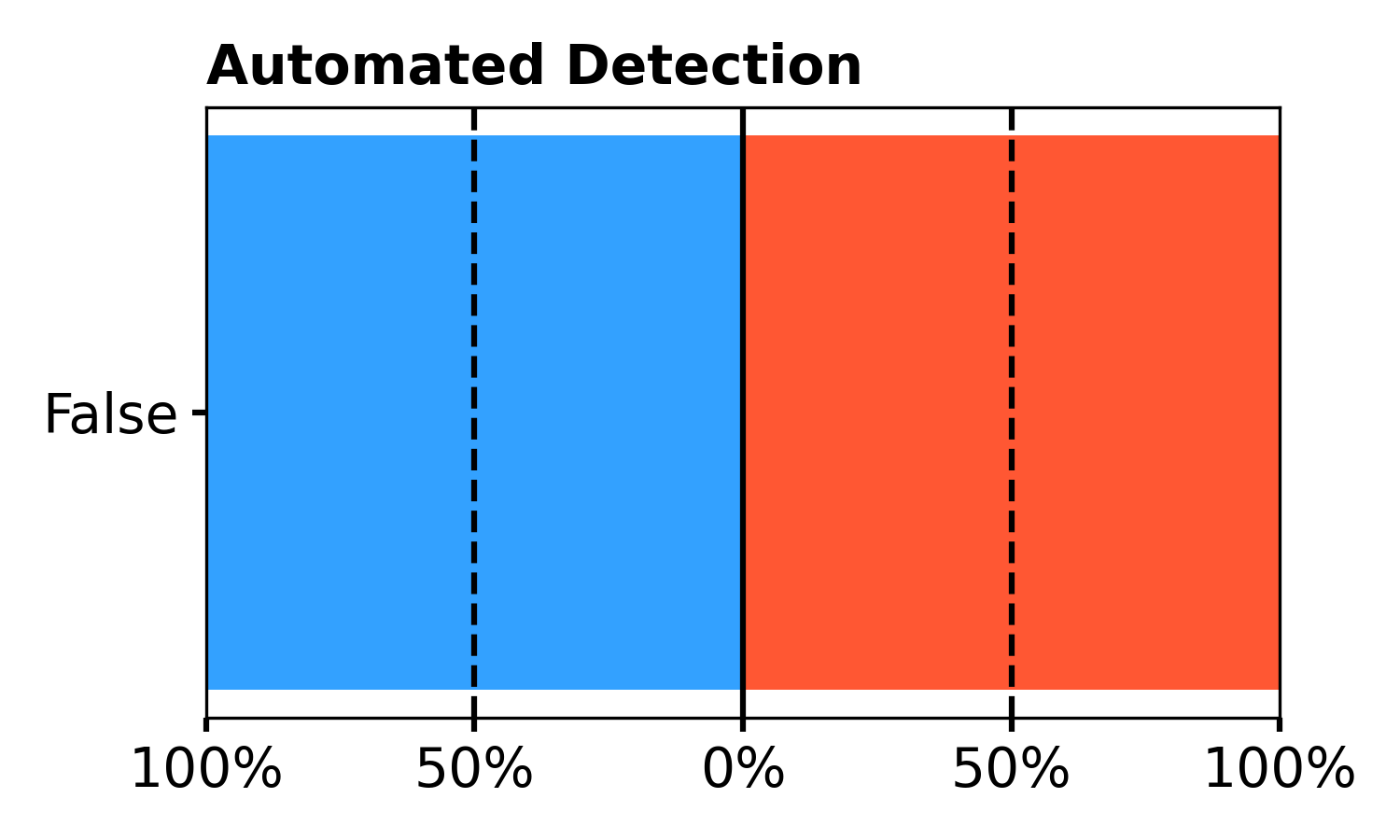}%
    \end{subfigure}%
    \caption{Comparison between the SoRs (right-aligned, red-colored) that caused the moderation anomaly reported for $\mathbb{X}$ during \textbf{June 27, 2024} and the SoRs (left-aligned, blue-colored) from the surrounding routine days.}
    \label{fig:anomaly-x-peak05}
\end{figure*}

\noindent\textbf{$\mathbb{X}$---June 27, 2024.} Figure~\ref{fig:anomaly-x-peak05} shows that, on that day, X moderated synthetic content exclusively through manual review, primarily targeting synthetic content due to pornography and due to it being outside the scope of platform services. However, this pattern does not differ from the surrounding routine days. Moreover, no connection to the electoral period emerges from the data. We further note that, for X, we do not report the \emph{incompatible ground} category, as this is not used by the platform.

\begin{figure*}[h!]
\centering
    \begin{subfigure}{0.45\textwidth}%
        \includegraphics[width=\textwidth]{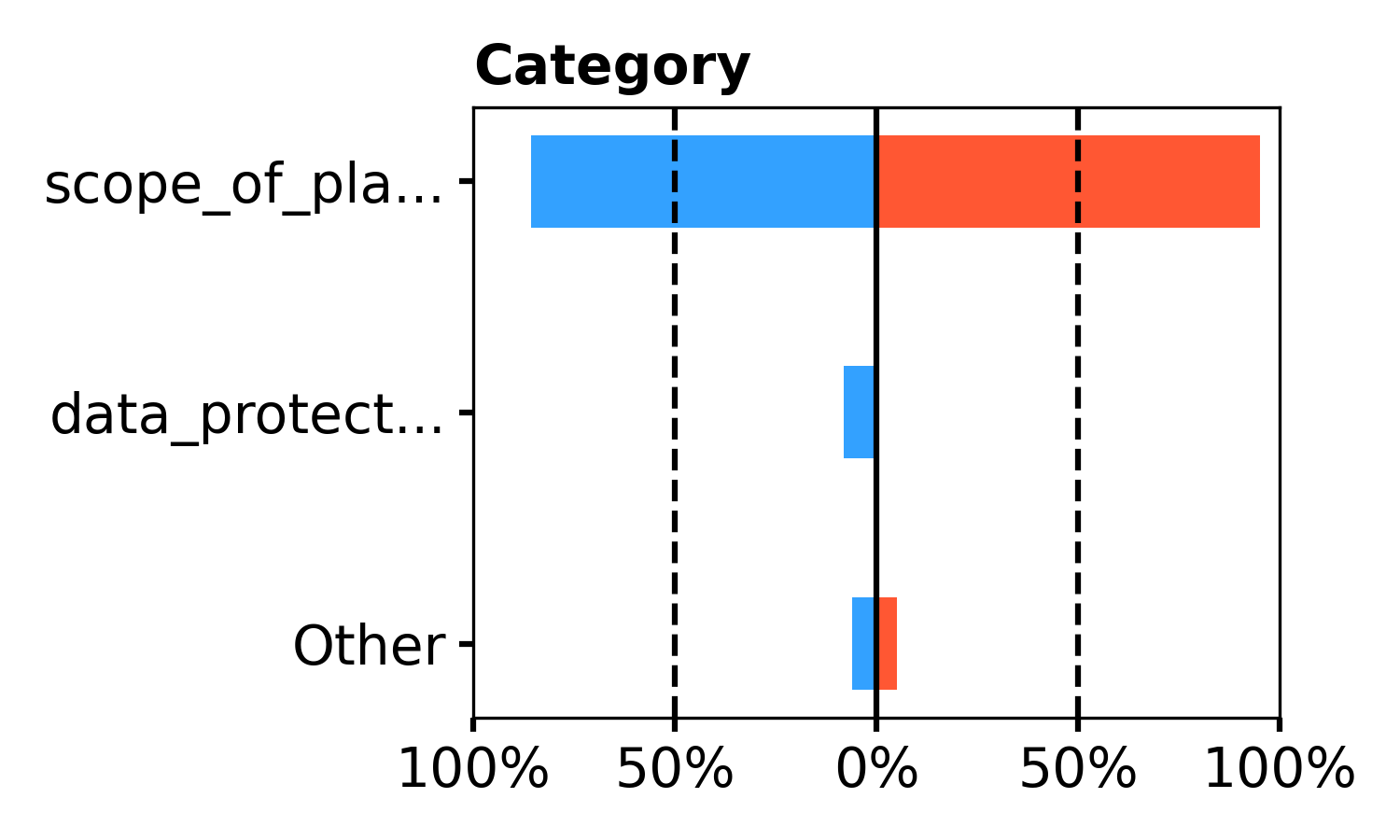}%
    \end{subfigure}%
    \begin{subfigure}{0.45\textwidth}%
        \includegraphics[width=\textwidth]{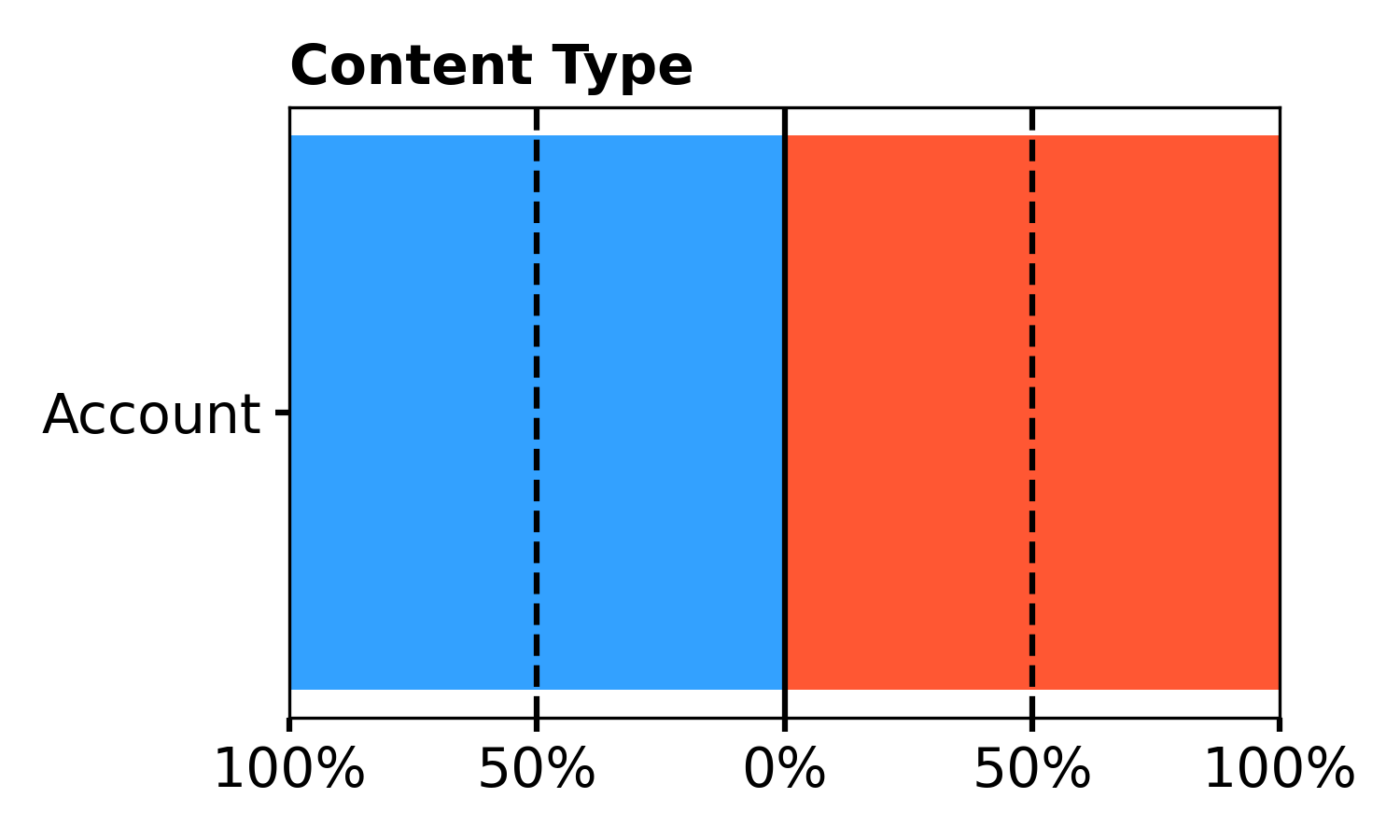}%
    \end{subfigure}%
    \\
    \begin{subfigure}{0.45\textwidth}%
        \includegraphics[width=\textwidth]{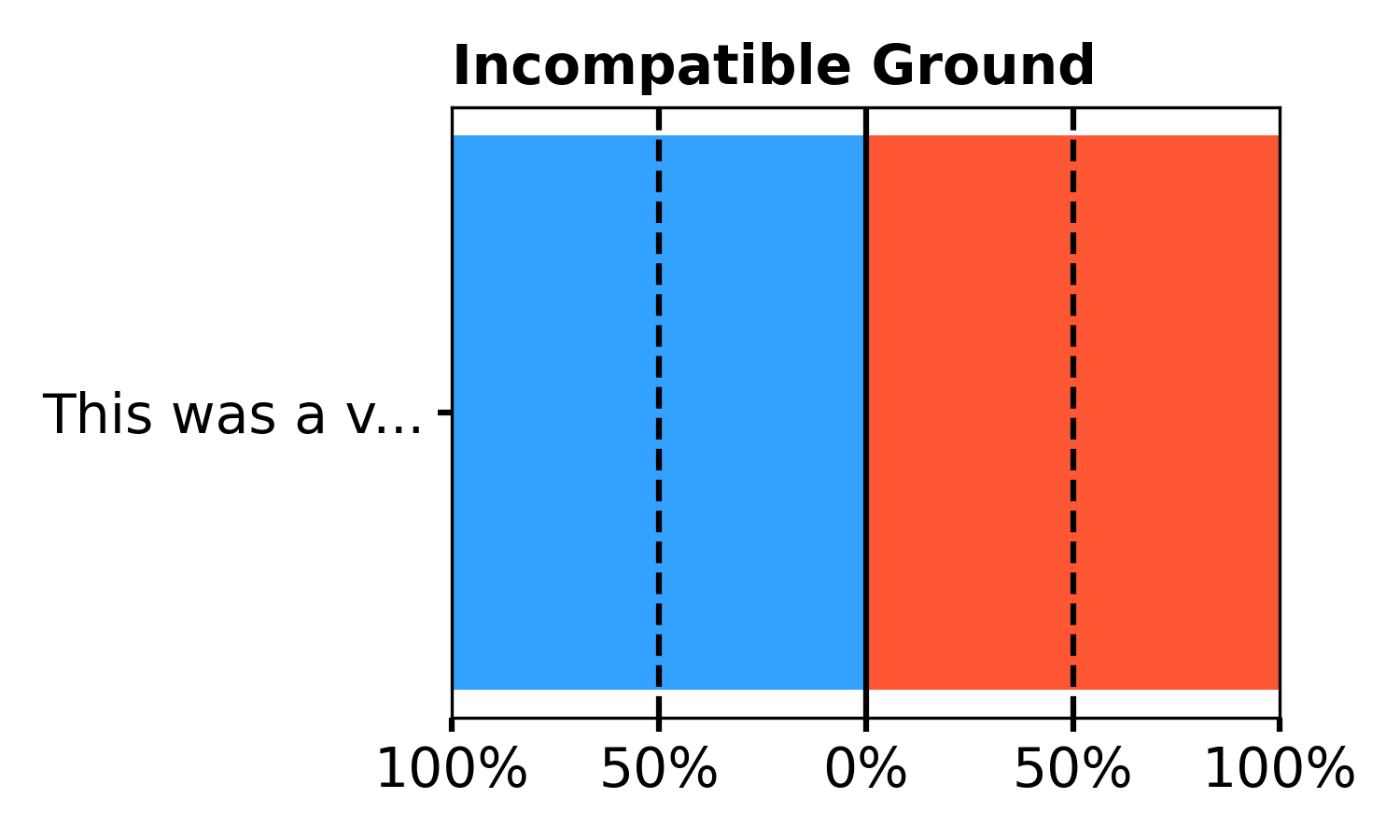}%
    \end{subfigure}%
    \begin{subfigure}{0.45\textwidth}%
        \includegraphics[width=\textwidth]{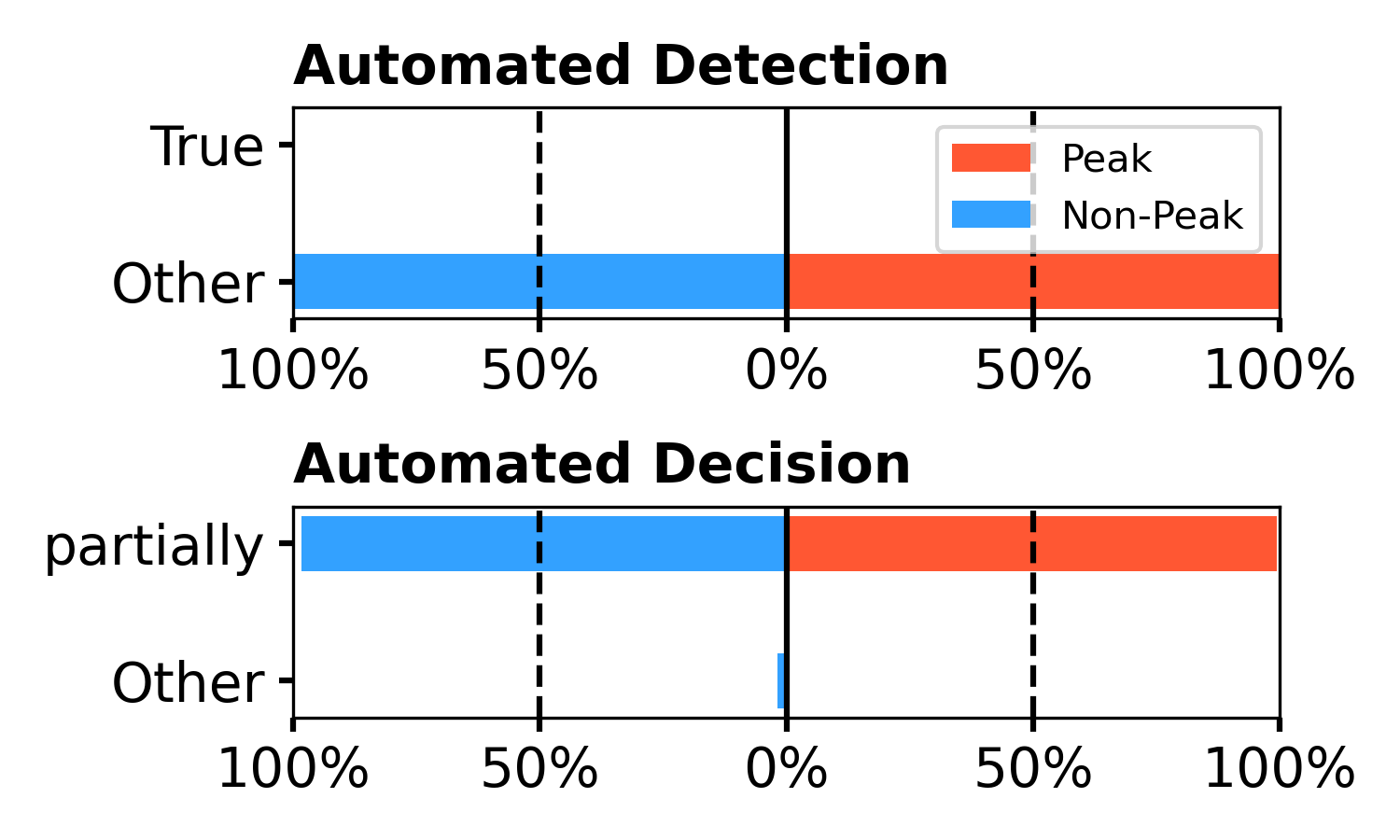}%
    \end{subfigure}%
    \caption{Comparison between the SoRs (right-aligned, red-colored) that caused the moderation anomaly reported for \faFacebook~\textbf{Facebook} during \textbf{September 29, 2024} and the SoRs (left-aligned, blue-colored) from the surrounding routine days.}
    \label{fig:anomaly-facebook-peak06}
\end{figure*}

\noindent\textbf{\faFacebook~Facebook---September 29, 2024.} Figure~\ref{fig:anomaly-facebook-peak06} shows that there are no differences between the peak day and the surrounding routine days. In all cases, Facebook moderated accounts for violations of Section 3.2 of Meta's Terms of Service,\footnote{\url{https://www.facebook.com/terms?section_id=section_3}} which mandates the shareable content and authorized conduct on Meta Products. The explanations provided for these actions are very vague, making it impossible to determine whether they are related to election-related content. There is the possibility that the observed anomaly is more driven by internal algorithmic processes rather than specific content moderation decisions.

\begin{figure*}[h!]
\centering
    \begin{subfigure}{0.45\textwidth}%
        \includegraphics[width=\textwidth]{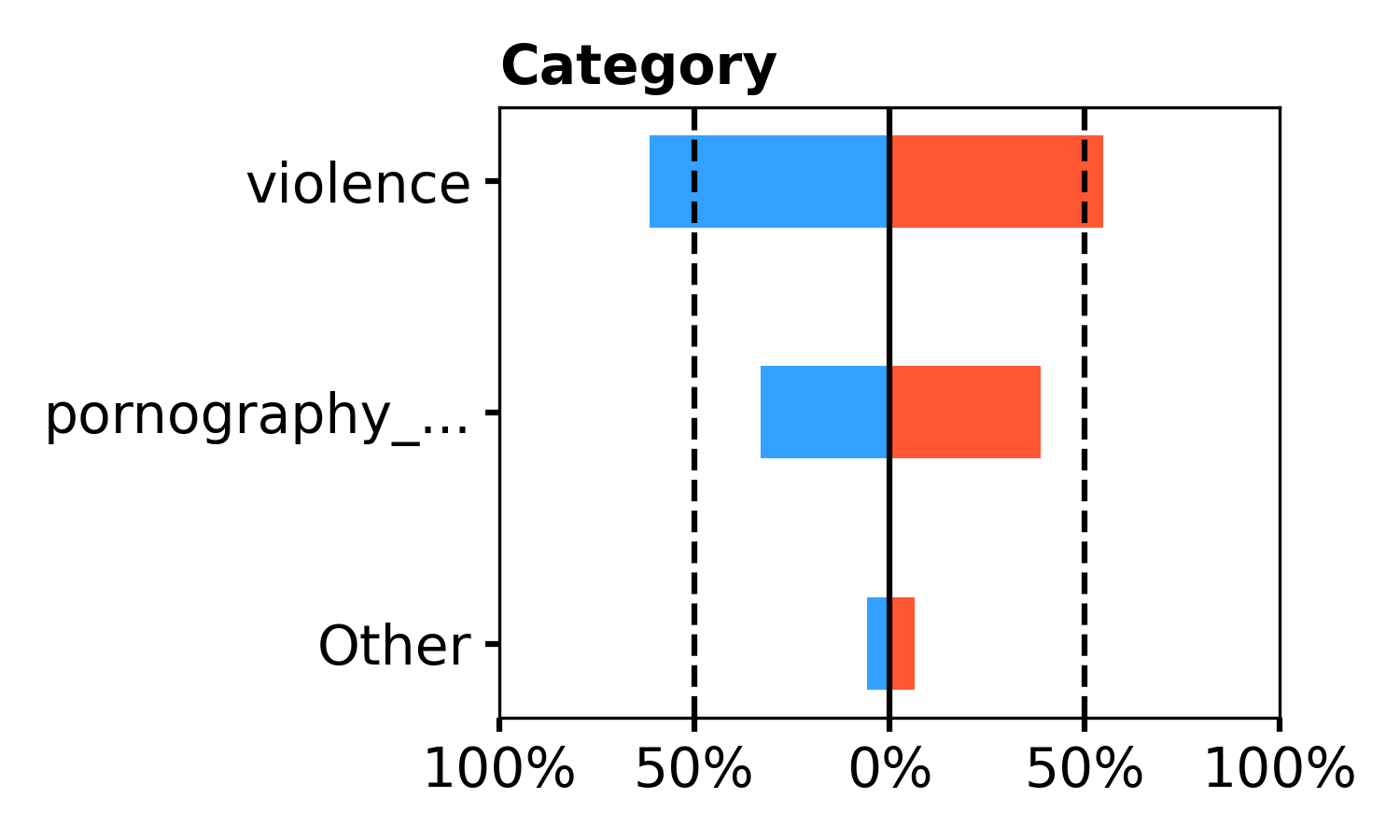}%
    \end{subfigure}%
    \begin{subfigure}{0.45\textwidth}%
        \includegraphics[width=\textwidth]{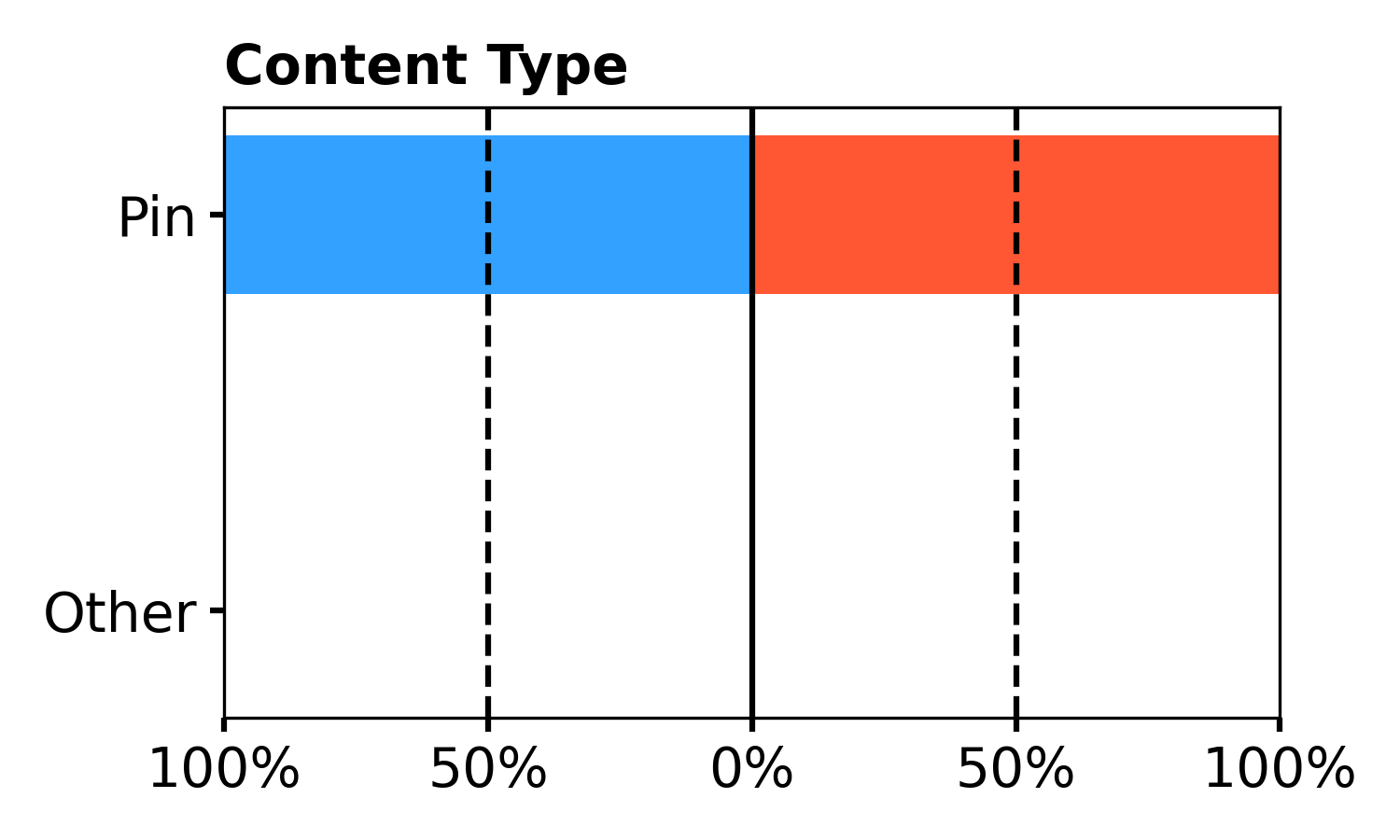}%
    \end{subfigure}%
    \\
    \begin{subfigure}{0.45\textwidth}%
        \includegraphics[width=\textwidth]{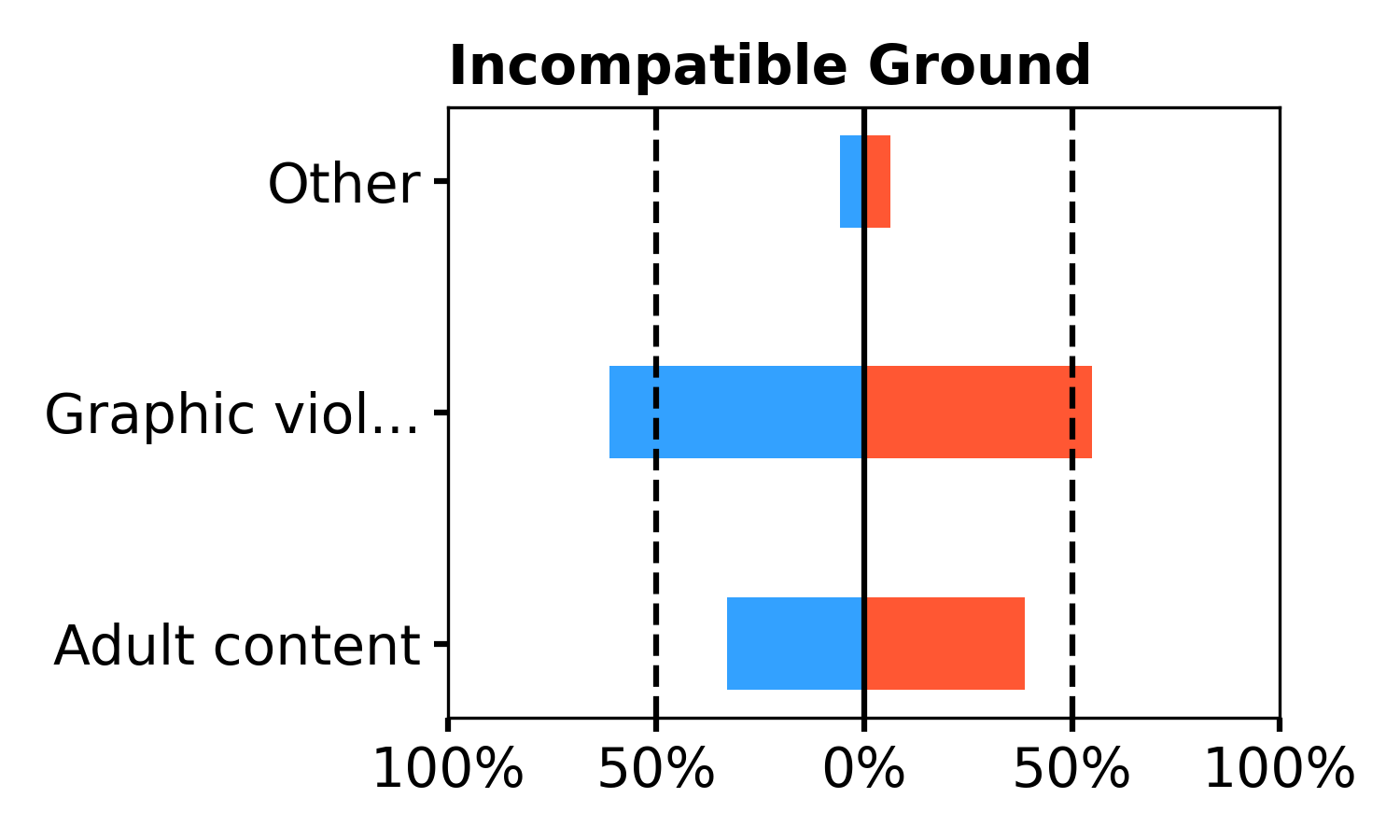}%
    \end{subfigure}%
    \begin{subfigure}{0.45\textwidth}%
        \includegraphics[width=\textwidth]{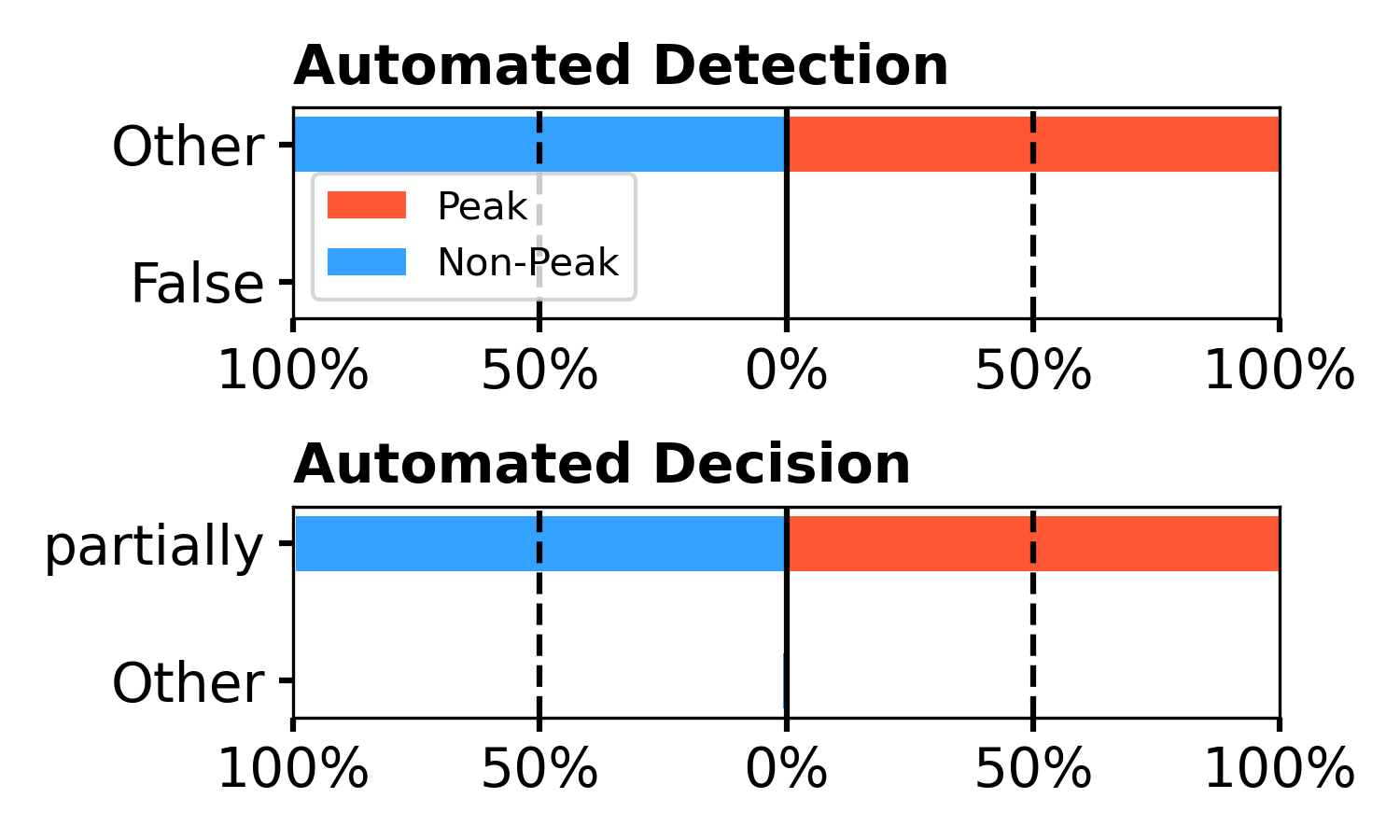}%
    \end{subfigure}%
    \caption{Comparison between the SoRs (right-aligned, red-colored) that caused the moderation anomaly reported for \faPinterest~\textbf{Pinterest} on \textbf{August 21, 2024} and the SoRs (left-aligned, blue-colored) by the same platform from the surrounding routine days.}
    \label{fig:anomaly-pinterest-volume}
\end{figure*}

\noindent\textbf{\faPinterest~Pinterest---August 21, 2024.} %
Figure~\ref{fig:anomaly-pinterest-volume} reveals only minimal differences between the SoRs associated with the anomaly and those from routine days. The limited changes points to a higher share of actions targeting graphic violence rather than pornographic content. Overall, these SoRs are not related to the electoral context.

\begin{figure*}[h!]
\centering
   \begin{subfigure}{0.45\textwidth}%
        \includegraphics[width=\textwidth]{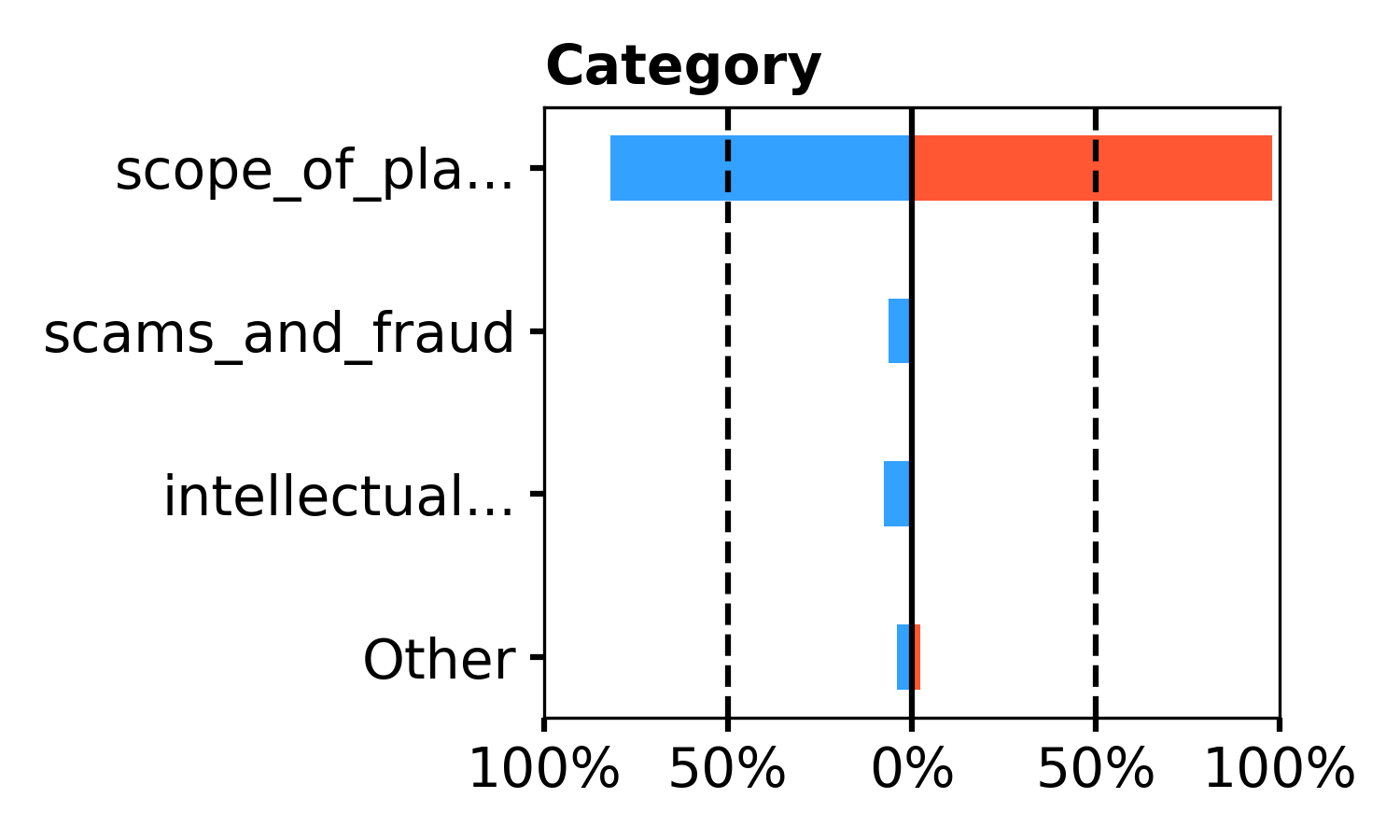}%
    \end{subfigure}%
    \begin{subfigure}{0.45\textwidth}%
        \includegraphics[width=\textwidth]{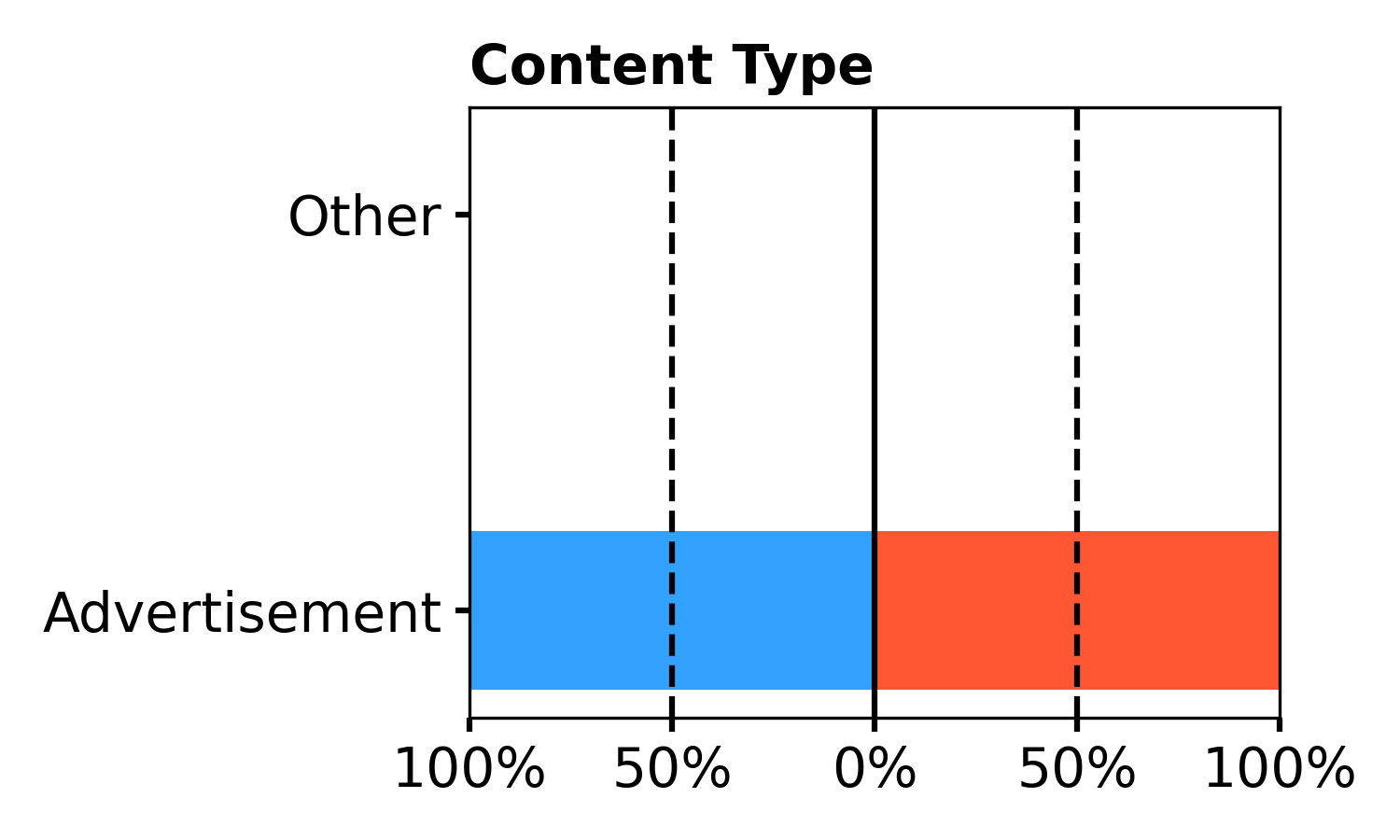}%
    \end{subfigure}%
    \\
    \begin{subfigure}{0.45\textwidth}%
        \includegraphics[width=\textwidth]{new_peak_images/peak_08_bidirectional_category_percentage_youtube.png}%
    \end{subfigure}%
    \begin{subfigure}{0.45\textwidth}%
        \includegraphics[width=\textwidth]{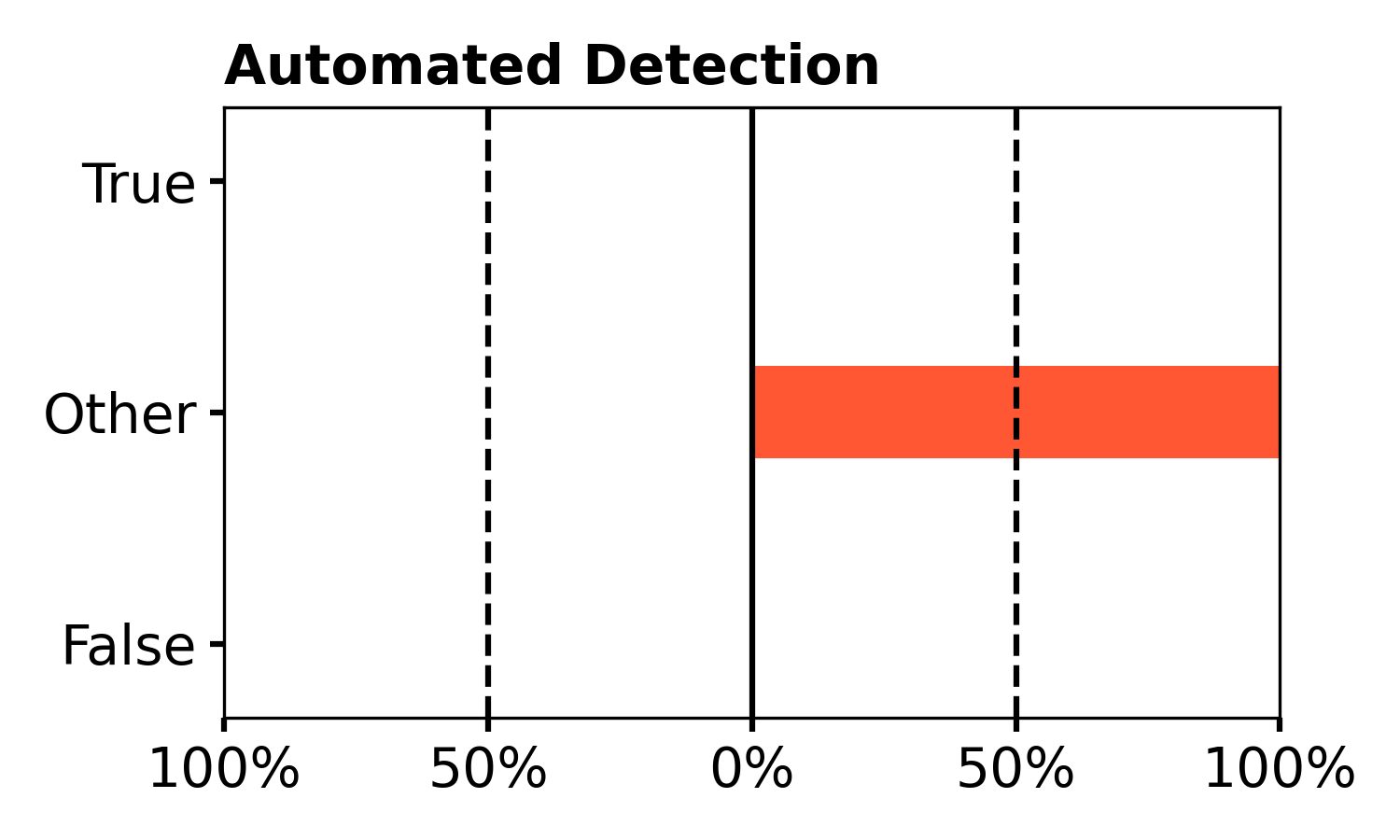}%
    \end{subfigure}%
    \caption{Comparison between the SoRs (right-aligned, red-colored) that caused the moderation anomaly reported for \faYoutube~\textbf{YouTube} on \textbf{July 21, 2024} and the SoRs (left-aligned, blue-colored) by the same platform from the surrounding routine days.}
    \label{anomaly-youtube-volume}
\end{figure*}

\noindent\textbf{\faYoutube~YouTube---July 21, 2024.} %
Figure~\ref{anomaly-youtube-volume} indicates that this peak primarily moderated advertisements deemed as being outside YouTube’s scope and, specifically, related to ``other restricted businesses.''
According to their documentation, some of such businessed could be related to the electoral context, \footnote{\url{https://support.google.com/adspolicy/answer/6368711?hl=en}} such as businesses related to ``Government documents and official services.''
However, there is no further indication on the type of business being restricted. For this reason, similarly to the other anomalies, there is not enough context provided to determine any link with the elections.

Altogether, this analysis brings to light various limitations in assessing the cause behind moderation anomalies.
We were mostly unable to determine wether these surges were related to the electoral context or caused by other factors.
This uncertainty primarily arises from the shallow level of detail provided in the SoRs. This is reflected by: \textit{(i)} mandatory fields frequently filled with generic values, and \textit{(ii)} optional fields, which could provide crucial context, being severely underutilized~\cite{shahi2025year}.

\section{Database Reliability And Consistency}
\label{sec:results-rq2}
As anticipated, various works have brought up concerns regarding the quality and consistency of the SoRs in the \texttt{DSA-TDB}~\cite{trujillo2023dsa,kaushal2024automated,drolsbach2024content,papaevangelou2024content}. Such concerns can threaten the database’s promises, as it was designed to promote transparency and accountability. Now, one year after its launch, we reassess these findings by analyzing our more recent dataset to address RQ2 and understand whether the situation has improved. If the same inconsistencies persist, this would raise concerns not only about the quality of the reported data but also about the broader effectiveness of the database as a regulatory oversight tool~\cite{tessa2025improving}.

\begin{figure}[t]
    \centering
    \includegraphics[width=0.5\columnwidth]{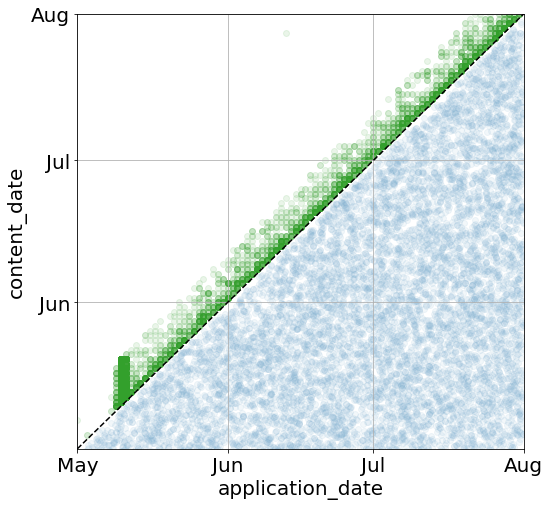}
    \caption{Highlight of 12,010 SoRs from TikTok that erroneously report content being moderated before it was published. The erroneous SoRs are green-colored while the correct ones are blue-colored.}
    \label{fig:tiktok_err}
\end{figure}

\subsection{Erroneous Database Records}
To evaluate the quality of the submitted data, we began by identifying obvious errors. We uncovered two main issues: \textit{(i)} duplicate records, and \textit{(ii)} records indicating that content was moderated before it was even published. The first issue is quite negligible as it affected only a small fraction of SoRs submitted by Facebook and Pinterest. Specifically, it concerns SoRs with identical Universally Unique Identifiers (UUIDs). In all detected duplicates, the SoRs were fully identical across every field. However, the second issue is far more concerning, as TikTok, Facebook, Snapchat, and LinkedIn submitted several SoRs reporting to have moderated content before their publication. Figure~\ref{fig:tiktok_err} displays this issue in a subset of TikTok data from May to August 2024, where the green-colored records incorrectly indicate moderation occurring before publication. While these errors represent only a small fraction of all SoRs---TikTok submitted more than 12k erroneous SoRs on May 10, 2024, representing roughly $\sim$0.1\% of its total for that day---care is nonetheless required when analyzing the data. Depending on the filtering strategy, these flawed records may be disproportionately retained, turning them into a non-negligible share of the analyzed subset. While the duplication issue could stem from mechanisms internal to the \texttt{DSA-TDB} itself, the other issue points to reporting errors on the platforms’ side. In fact, these errors could be easily detected and solved by the platform, possibly suggesting lack of adequate quality checks. Therefore, researchers and policymakers need to perform integrity checks to avoid drawing misleading conclusions~\cite{tessa2025improving}.

\begin{figure*}[t]
    \centering
    \includegraphics[width=\textwidth]{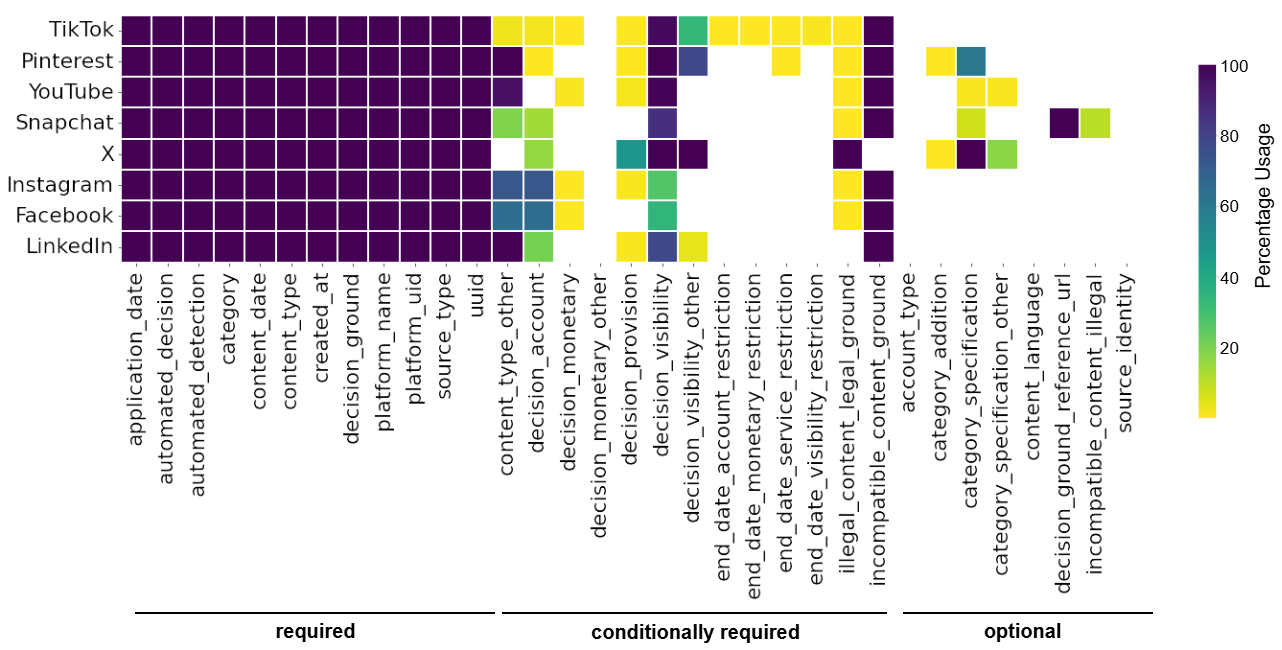}
    \caption{Heatmap showing how frequently each platform populates the attributes of the database. Attributes are grouped into required, conditionally required, and optional fields. Platforms are ordered by overall attribute utilization (from highest to lowest).}
    \label{fig:field_usage}
\end{figure*}

\subsection{Fields Usage And Uninformative Reporting}
The \texttt{DSA-TDB} fields fall into three categories---required, conditionally required, and optional---offering a reporting framework that is both standardized and flexible. Moreover, the value of certain fields can either be selected from a predefined list of values or be free text. However, some of the values in the predefined lists are very specific, while the others are broad and generic. This is why the informativeness and completeness of the data strongly depends on how the platforms populate these fields.

To evaluate the informativeness of the submitted SoRs, we analyzed how frequently each database attribute was used by the different platforms, distinguishing between required, conditionally required, and optional fields. Our findings are presented in Figure~\ref{fig:field_usage}. All platforms consistently fill in the required attributes, ensuring formal compliance with the DSA's requirements. In contrast, the use of conditionally required attributes varies considerably: TikTok makes an extensive use of them, unlike the other platforms. Interestingly, certain conditionally required attributes such as \dbField{decision\_visibility} and \dbField{incompatible\_content\_ground} are consistently filled by all platforms. Others such as \dbField{decision\_monetary\_other}, are completely unused. Moreover, optional fields are severely underused. For example, TikTok, Instagram, Facebook, and LinkedIn never populated any of these optional fields. Pinterest, YouTube, Snapchat, and X filled only a few, in a small fraction of SoRs. This issue was already noted in early assessments of the \texttt{DSA-TDB}~\cite{trujillo2023dsa}, and our analysis confirms that little progress has been made in addressing it.

\begin{figure}[t]
    \centering
    \includegraphics[width=0.75\columnwidth]{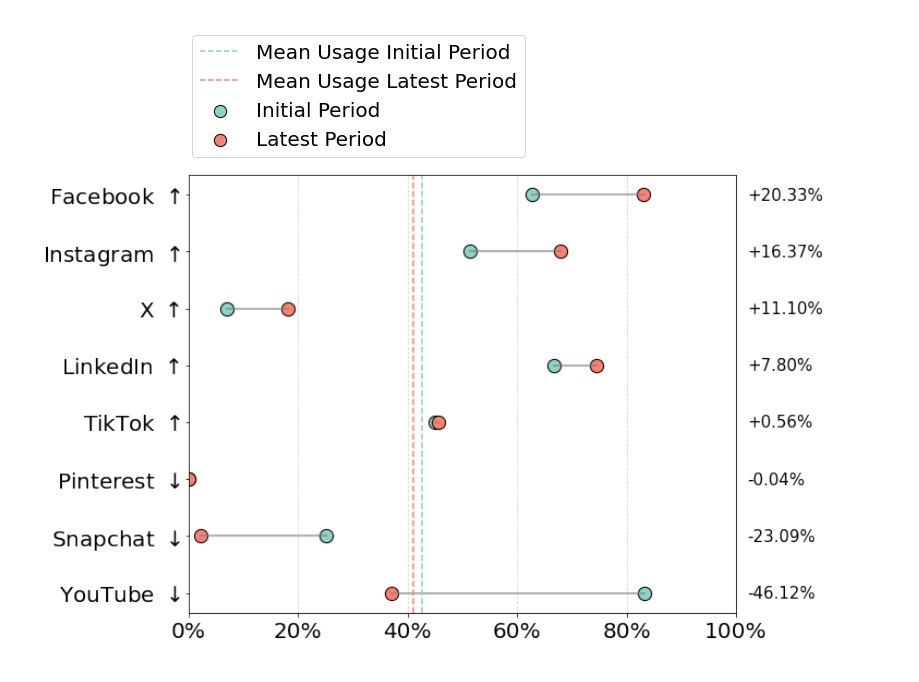}
    \caption{Frequency of platforms’ use of the generic category \dbField{scope\_of\_platform\_service} in their SoRs. Teal dots represent usage in the initial period, and red dots represent usage in the latest period. Horizontal bars indicate the change between periods. Platforms are ordered by decreasing difference between periods. Arrows next to platform names denote increases ($\uparrow$) or decreases ($\downarrow$) in usage. Vertical dashed lines mark the mean value for each period.}
    \label{fig:scope_service}
\end{figure}

One of the required attributes obliges platforms to indicate the reason for each moderation action by selecting a value from a predefined list. While some of the values are fairly specific (e.g., pornography, harmful speech), others, such as \dbField{scope\_of\_platform\_service}, are more broad and vague, covering a wide range of restrictions related to age, geography, language, disallowed goods and services, and nudity.\footnote{\url{https://transparency.dsa.ec.europa.eu/page/documentation\#16-category-specification-category-category-addition-category-specification}} According to previous studies, the reliance on this generic category is heavy and hinders platforms' clarity and transparency~\cite{trujillo2023dsa,kaushal2024automated}. To evaluate whether platforms have become more specific, we compared the use of the \dbField{scope\_of\_platform\_service} category in the initial and latest periods, defined in Section~\ref{sec:datasets}. Figure~\ref{fig:scope_service} shows the results of this comparison. Overall, the average use of this category remain almost unchanged as it slightly decreased from 42.68\% in the initial period to 41.04\% in the latest. However, platform-specific usage varied. Facebook, Instagram, LinkedIn, and X have increased their use of this generic label, whereas YouTube and Snapchat have reduced it considerably. TikTok and Pinterest, in contrast, show little change. These findings confirm that platforms still rely heavily on vague reporting, reducing the informativeness of mandatory fields. Combined with the infrequent use of optional fields, this indicates that although platforms comply with the formal DSA requirements, the transparency and practical usefulness of their reporting remain limited.

\begin{figure}[t]
    \centering
    \begin{subfigure}[b]{0.45\columnwidth}
        \centering
        \includegraphics[width=\linewidth]{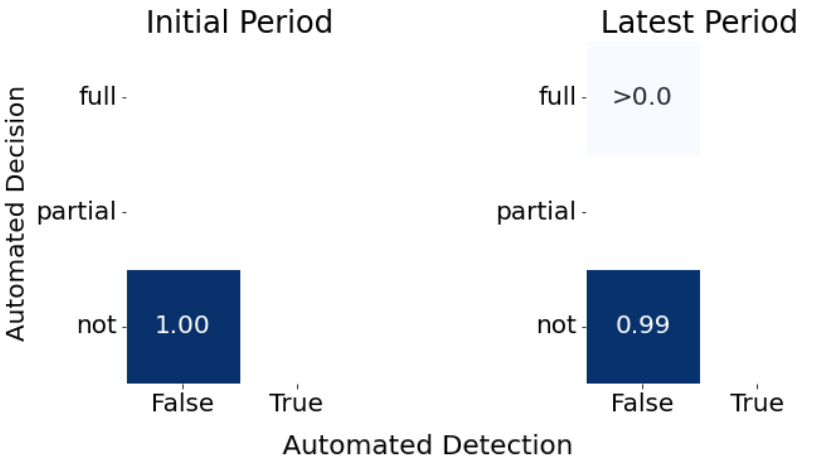}
        \caption{Use of automation.}
        \label{fig:X_automation}
    \end{subfigure}
    \begin{subfigure}[b]{0.45\columnwidth}
        \centering
        \includegraphics[width=\linewidth]{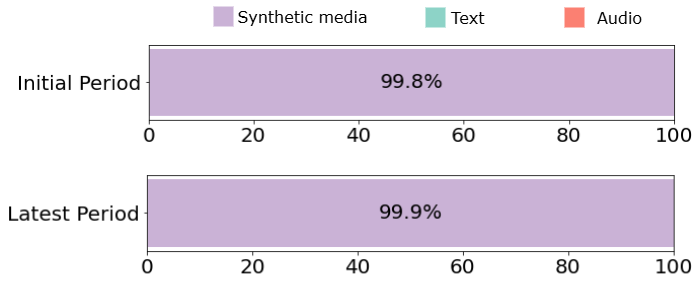}
        \caption{Type of moderated content.}
        \label{fig:X_content_type}
    \end{subfigure}
    \caption{%
    Comparison between X’s reported use of automation (a) and the categories of moderated content (b) across the initial and most recent observation periods.}
    \label{fig:X_comparison}
\end{figure}

\subsection{Unresolved Issues in X’s Reporting}
On top of identifying general issues with the database, previous studies also brought to light platform-specific limitations. In particular, X emerged as the platform with the most inconsistencies in its reporting~\cite{trujillo2023dsa,kaushal2024automated,drolsbach2024content}. For almost all platforms, greater reliance on automated means in moderation correspond to lower moderation delays~\cite{papaevangelou2024content}. However, X continues to report zero moderation delay despite stating that no automation is used (i.e., all moderation is manual). Since deepfakes are X’s primary focus for moderation and are inherently difficult to detect manually at scale~\cite{mirsky2021creation}, this reporting pattern appears implausible. To make matters worse, X’s \texttt{DSA-TDB} submissions have been shown to contradict its own transparency reports~\cite{trujillo2023dsa}. Notably, this lack of transparency was among the factors that prompted the European Commission to initiate formal proceedings against X in December 2023~\cite{eu_proceedings_ag_x}. To evaluate whether X has addressed these inconsistencies, we compared its reporting from the initial and latest periods. Figures~\ref{fig:timeline} and~\ref{fig:delay_analysis} show how X still continues to report zero moderation delay. Furthermore, Figure~\ref{fig:X_automation} shows that 99\% of X’s recent SoRs still report purely manual moderation and Figure~\ref{fig:X_content_type} confirms its focus on synthetic media. The small differences between the two periods are statistically non-significant ($p = 0.99$, $\chi^2$ test), indicating no meaningful changes in X’s reporting practices. These results reinforce earlier concerns about X’s data reliability and show that it has not improved over time.

\section{Discussion}
\label{sec:conclusion}
We examined 1.58 billion moderation actions taken by major social media platforms during the 2024 European Parliament elections, highlighting how such platforms reacted to heightened systemic risks during a major political event. \rev{Our analyses revealed no major shifts in moderation before, during, or after the elections, and no meaningful improvement in reporting quality compared to prior analyses, raising doubts on the effectiveness of the \texttt{DSA-TDB} as co-regulatory mechanism}.

\subsection{Changes in Moderation Practices}
In RQ1 we aimed to determine whether the \texttt{DSA-TDB} revealed any changes in content moderation practices before, during, or after the electoral period. Apart from LinkedIn’s delayed actions against election-related misinformation comments, we found no evidence of significant shifts in moderation behavior across the eight social media platforms analyzed. This is consistently reflected across all time series. Both moderation volumes and moderation delays display largely stable trends over time, and the limited variations observed cannot be traced back to moderation of election-related content. Indeed, while we identified localized spikes and short-term anomalies, these did not align with the elections themselves and instead coincided with other major political or geopolitical events not necessarily limited within the EU.

There are several possible explanations for these findings. For instance, platforms may not have adjusted their moderation practices to face such a politically sensitive period. This could be because they considered their existing practices sufficient or chose not to make adjustments. This observation, however, contrasts with what platforms reported in their DSA Risk Assessment reports, where they explicitly claimed to have implemented election-specific measures, such as increased monitoring, dedicated teams, or the introduction of automated tools to mitigate risks to civic discourse and electoral integrity. Moreover, this interpretation is also misaligned with independent monitoring efforts and external assessments of the 2024 European elections. For instance, post-election analyses have documented the persistence and amplification of disinformation narratives related to electoral fraud, conspiracy theories, and foreign information manipulation and interference (FIMI) across major platforms~\cite{GDI2024,casero2025spreading}. The absence of corresponding signals in the \texttt{DSA-TDB} suggests a mismatch between heightened election integrity risks, declared mitigation efforts, and the actual moderation activity that platforms reported, which was already found in earlier works~\cite{trujillo2023dsa,shahi2025year}. However, the absence of observable changes in the \texttt{DSA-TDB} should not be interpreted as a definitive evidence that platforms took no action. Platforms may have implemented preventive or internal mitigation measures that do not directly translate into an increased number of SoRs and therefore remain hidden in the database.

Another possibility is that the database itself, due to self-reporting limitations and structural shortcomings~\cite{trujillo2023dsa,kaushal2024automated}, may have concealed any changes the platforms implemented. Because the \texttt{DSA-TDB} submissions are self-reported, platforms may have left out certain details about their enforcement actions, either unintentionally due to internal reporting gaps, or deliberately to preserve some opacity in their moderation practices. This is particularly relevant in a high-stakes context such as elections, where platforms may deploy internal or temporary mitigation measures that are not fully reflected in standardized reporting pipelines.
At the same time, even if platforms had fully reported their moderation actions, the database’s design may have inherently limited the ability to detect shifts in moderation~\cite{trujillo2023dsa}. For instance, the predefined categories provided by the DSA are in some cases too broad to capture meaningful changes in how platforms respond to election-related threats. As a result, potentially important adaptations in moderation practices may remain concealed when analyzed through aggregate counts or coarse-grained labels, limiting the database’s capacity to answer research questions about systemic risks.

\subsection{Quality of Moderation Reports}
In RQ2 we reassessed the quality of the \texttt{DSA-TDB} submissions, revisiting earlier analyses that had highlighted several issues in order to determine whether they have since been resolved.
Our findings indicate that many of these problems persist. Incomplete and vague reporting remains widespread, and several submissions contain unreliable or implausible values, with particularly severe issues still observed for X~\cite{trujillo2023dsa}. The consistent presence of these issues suggests that, despite increased regulatory attention and the development of the reporting framework, platforms have not substantially improved the quality of their reporting. This raises concerns not only about the database’s ability to function as a meaningful transparency instrument, but also about the broader effectiveness of regulatory efforts aimed at strengthening accountability in platform governance~\cite{kausche2024platform}.

Transparency mechanisms such as the \texttt{DSA-TDB} can only fulfill their intended role when the data they provide is accurate, comprehensive, and comparable across platforms. When platforms rely on vague classifications, underuse optional fields, or submit data that lacks internal consistency, the analytical value of the database is significantly reduced. In these cases, platforms technically comply with reporting obligations, but the resulting data offers limited insight into how moderation is actually carried out in practice.
As a result, the \texttt{DSA-TDB} may fall short of supporting one of its core objectives: enabling regulators, researchers, and the public to successfully scrutinize how platforms respond to major systemic risks. The persistence of low-quality and opaque reporting also has implications for public trust. Prior research shows that users are more likely to accept and support moderation decisions when platforms provide clear, consistent, and well-documented explanations for their actions~\cite{cai2024content,jhaver2019does}. Continued opacity in platform disclosures may therefore undermine confidence in content moderation systems, reinforcing skepticism about both the fairness of enforcement and the sincerity of platforms’ commitments to transparency.

\subsection{Regulatory Implications}

\rev{In terms of practical implications, our results indicate that even when platforms claimed to implement election-focused mitigation strategies, the effects of such measures were either limited, applied unevenly across platforms, or not fully captured by the \texttt{DSA-TDB}. This limits the database’s effectiveness as a tool for detecting nuanced responses to systemic risks, such as those related to elections. It also underscores the challenges of relying solely on self-reported data for transparency: without additional mechanisms to validate or enrich reported data, subtle but important shifts in moderation practices may remain invisible~\cite{tessa2025improving}.} \rev{These findings also point to a clear practical priority for public managers and regulators: improving the quality, comparability, and auditability of the data submitted to the \texttt{DSA-TDB}. In practice, this would require clearer reporting standards, stricter validation rules, and more systematic checks for missing, vague, inconsistent, or implausible values. Regulators could also identify a smaller set of high-value fields that platforms should be required to complete consistently, especially when moderation actions relate to systemic risks such as electoral integrity.
Beyond improving reporting quality, public authorities could use the \texttt{DSA-TDB} more actively to request clarifications from platforms when reported trends do not align with declared mitigation efforts or with external evidence of heightened risks. Implementing these steps would require technical capacity for large-scale data processing and anomaly detection, legal expertise to assess compliance with DSA obligations, and domain expertise on electoral integrity, disinformation, and platform governance. Therefore, the DSA-TDB should not be treated only as a passive repository of platform disclosures, but as an operational tool to support risk-based oversight and targeted auditing.}

\rev{
From a theoretical point of view, our results call attention to some of the limits of the DSA's co-regulatory framework, particularly in terms of platform self-regulation.
The EC legally mandates transparency and creates a common reporting infrastructure, but platforms retain significant control over the data, classifications, and explanations through which their moderation practices become visible.
Our findings thus challenge the co-regulatory assumption that mandatory disclosure necessarily transforms self-regulation into effective public accountability. Instead, they show that self-regulatory power may persist inside a co-regulatory framework through control over regulatory information, such as the self-assessment of systemic risks and their mitigation~\cite{palumbo2026systemic}. Consequently, its effectiveness  depends on whether platform disclosures are not merely available, but also comparable, independently verifiable, and capable of producing contestation and corrective action.
As a matter of fact, the European Commission has initiated formal proceedings against 
Facebook, Instagram, and X due to shortcomings in mitigating threats to civic discourse and electoral integrity, specifically in relation to the 2024 European Parliament elections~\cite{eu_proceedings_ag_fb_insta,eu_proceedings_ag_x}.
Again, these shortcomings were not visible on the \texttt{DSA-TDB}, however, as we initially expected.
Rather, in its current state the database seems to be a mechanism of \emph{formal} transparency in which platforms merely publish the required data to comply with the law, instead of being a mechanism of \emph{actionable} transparency in which regulators, researchers, and other organizations can make effective use of those data to achieve a safer online environment.}

\subsection{Limitations and Future Work}
Our work relies on the data submitted to the \texttt{DSA-TDB}, which is self-reported by nature and whose accuracy and reliability may be inconsistent or misleading, as previously stated.
Moreover, our analysis was limited to a relatively small number of platforms. Although we focused on the eight most prominent social media in the EU, our findings may not generalize to all platforms.
Another limitation stems from the lack of contextual information in the data. For example, the database does not allow to link the SoRs to the actual piece of content that was moderated.
While necessary to ensure privacy, this limits the depth of our finings. Additionally, it remains unclear whether the lack of observable changes reflects limited discourse around the European Parliament elections, resulting in little content requiring moderation.

One of the main limitations of our study includes the lack of identifiers in the SoRs which cause the moderated content to be inaccessible.
For this purpose, ongoing initiatives by the European Commission to develop operational procedures for data access under Article 40(4) of the DSA could help address this limitation.\footnote{\url{https://www.eu-digital-services-act.com/Digital_Services_Act_Article_40.html}}
The new data access portal\footnote{\url{https://data-access.dsa.ec.europa.eu/home}} represents a fundamental step forward in this direction, as it provides the opportunity to combine and cross-check self-reported data to the corresponding platform data---including non-public data---allowing to gain more detailed data on quality and moderation practices. 
For future work, it would also be insightful to reevaluate the data submitted by TikTok, Facebook, Instagram, and X after their proceedings have ended.
Moreover, maintaining high data quality in the \texttt{DSA-TDB} will be essential for future transparency efforts. Our analysis could also be extended to assess how platform moderation evolves during other major events, whether political or otherwise.

\section{Conclusions}
{\color{black}
We analyzed 1.58 billion moderation actions from the Digital Services Act Transparency Database (\texttt{DSA-TDB}), a tool created to improve transparency and oversight in online moderation. Our extensive analysis aimed to investigate how major social media platforms managed content moderation during the 2024 European Parliament elections---a large-scale, multi-country political event.
First, we assessed whether the \texttt{DSA-TDB} reveals any shifts in how platforms moderated content before, during, and after the election. Our results reveal that, with the exception of LinkedIn’s delayed interventions against election-related misinformation, platforms exhibited largely stable moderation patterns across the entire period. Both the number of moderation actions and the associated delays showed only limited or non-meaningful variations, with localized spikes coinciding mostly with other geopolitical events rather than the elections themselves. %
Then, we evaluated the quality and reliability of the data submitted to the database, assessing whether earlier reporting issues have been resolved or persist over time. We found that the quality of submitted data remains a concern. Platforms frequently rely on broad or generic categories, underuse optional fields, and occasionally report implausible or inconsistent data, particularly in X’s submissions. This persistent opacity limits the utility of the database as a tool for transparency and regulatory oversight and suggests that self-reported data may not fully reflect the enforcement actions undertaken by platforms, even during high-stakes events like elections. Taken together, our results highlight a clear mismatch between the heightened integrity risks, the declared mitigation efforts in platforms' risk assessment and risk mitigation reports, and the observable patterns in the \texttt{DSA-TDB}. \rev{All this points to both reporting and structural limitations within the database, as well as practical and theoretical limitations of the current co-regulatory framework of the DSA}.

For future work, it will be useful to monitor how platforms adjust their reporting practices following regulatory proceedings and to extend the analysis to other major events. Moreover, the development of new data access mechanisms, enabling cross-checks between self-reported and non-public platform data, also offers an opportunity to gain more insights and better understand the reliability of the \texttt{DSA-TDB}.
}

\section*{Declarations}

\textbf{Ethics approval and consent to participate} 
Not Applicable

\noindent \textbf{Consent for publication} All co-authors have approved the content of the manuscript. All authors have given explicit consent to publish this manuscript. %

\noindent \textbf{Availability of data and materials} This study uses the dataset from the DSA Transparency Database (\texttt{DSA-TDB})\footnote{https://transparency.dsa.ec.europa.eu/} an open and centralized repository hosted by the European Commission. %

\noindent \textbf{Competing interests} The authors have no competing interests to declare that are relevant to the content of this article. 

\noindent \textbf{Funding}
This work is partly supported by the ERC project DEDUCE under grant \#101113826, and by the European Union -- Next Generation EU, Mission 4 Component 1, for project PIANO (CUP B53D23013290006).

\noindent \textbf{Authors' contributions} B.T. and G.K.S. contributed equally to this work. B.T. and G.K.S. carried out the analysis and visualisation. S.C. led the project, contributed to the conceptualisation and planning of the study, and revised the manuscript. A.T. and S.C. supervised the work, provided feedback, and proofread the manuscript. All authors contributed to the writing and approved the final manuscript.%

\bibliography{sn-bibliography}
\begin{appendix}

\section{Systemic risk assessment reports}
\label{sec:appendix-dsa-systemic-risk-reports}

Table~\ref{tab:dsa-systemic-risk-reports} contains the URLs to the repositories of the DSA Systemic Risk Assessment Reports described in Section~\ref{sec:self-assessed-risk}. The release time frame of the reviewed reports is between August 2024 and January 2025. It should be noted that there is no regulatory guidance on the systemic risk assessment methodology to follow. Hence, besides Facebook and Instagram (both from Meta), each platform follows a different approach to report risk.

\begin{table}[h]
\small
    \begin{tabular}{ll}
    \toprule
    \textbf{platform} & \textbf{URL of repository} \\ 
    \midrule\addlinespace[2.5pt]
    Facebook & \url{https://transparency.meta.com/reports/regulatory-transparency-reports/} \\ 
    Instagram & \url{https://transparency.meta.com/reports/regulatory-transparency-reports/} \\ 
    LinkedIn & \url{https://www.linkedin.com/help/linkedin/answer/a1678508} \\ 
    Pinterest & \url{https://policy.pinterest.com/en/transparency} \\ 
    Snapchat & \url{https://values.snap.com/privacy/transparency} \\ 
    TikTok & \url{https://www.tiktok.com/transparency/it-it/dsa-transparency} \\ 
    X & \url{https://transparency.x.com/en/reports/dsa-transparency-report} \\ 
    YouTube & \url{https://transparencyreport.google.com/?lu=regulatory-reporting&hl=en} \\ 
    \bottomrule
    \end{tabular}
\caption{Sources of the DSA Systemic Risk Assessment Reports.}
\label{tab:dsa-systemic-risk-reports}
\end{table}

\section{Attributes description}
\label{sec:appendix-description}
\begin{table*}[h]
    \footnotesize
    \centering
    \renewcommand{\arraystretch}{1.2}
    \adjustbox{max width=\textwidth}{
    \begin{tabular}{p{0.25\textwidth}p{0.25\textwidth}p{0.45\textwidth}}
        \toprule
        \textbf{field} & \textbf{reference} & \textbf{description} \\
        \midrule
        \texttt{application date} & \S 4.1 Application Date & Indicates when a content moderation decision was applied \\
        \midrule
        \texttt{automated decision} & \S10. Automated Decision & Indicates whether the decision to moderate a content was automatic or not \\
        \midrule
        \texttt{automated detection} & \S9. Automated Detection & Indicates whether moderated content was detected automatically or not \\
        \midrule
        \texttt{category} & \S16. Category \& Specification & Indicates the type of illegality or incompatibility with the platform's terms of services that led to a content being moderated \\
        \midrule
        \texttt{content date} & \S2.3. Date on which the content was created on the online platform & Indicates when moderated content was created \\
        \midrule
        \texttt{content type} & \S 2.1. Type of content affected & Type of the moderated content (e.g. audio, video, image, etc.) \\
        \midrule
        \texttt{content type other} & \S2.2. Specification of Content Type ``Other'' & Specification required when content type is ``other'' \\
        \midrule
        \texttt{decision ground} & \S11. Decision Grounds & Indicates whether the moderated content was deemed allegedly illegal or incompatible with the platform’s terms of service \\
        \midrule
        \texttt{incompatible content illegal} & \S12.2. Explanation of the applicability of the legal ground & Explains why a specific content has been deemed illegal according to Article 17(3)(d) of DSA \\
        \midrule
        \texttt{incompatible content ground} & \S13.1. Incompatible Content Grounds & Explains why a specific content has been deemed incompatible with the platform's terms of service \\
        \bottomrule
    \end{tabular}}
    \caption{Complete list of the DSA Transparency Database (\texttt{DSA-TDB}) fields analyzed in this study. For each field, we report its name, reference to the official documentation, and brief description \cite{shahi2025year}.}
    \label{tab:dsa-fields}
\end{table*}

Although the DSA Transparency Database provides a wide range of attributes and information, we focus our analysis on a subset of the most relevant attributes for our study. The complete list of analyzed attributes, along with references to the corresponding sections of the official documentation, and brief descriptions, is provided in Table~\ref{tab:dsa-fields}.

\section{Attributes values}
\label{sec:appendix-values}
Platforms can assign a predefined set of values to the attributes \dbField{category}, \dbField{automated\_decision}, and \dbField{automated\_detection}. Therefore, in our analysis of moderation anomalies, we report all possible values for the automation-related attributes, as there are only two for \dbField{automated\_decision} and three for \dbField{automated\_detection}. Conversely, for the \dbField{category} attribute---which allows up to 14 values---we only include those used at least once, for the sake of brevity and clarity. The \dbField{content\_type} attribute also requires the use of a predefined set of values, including the value \dbField{other}. When a platform only used \dbField{other} as the content type, we examined the \dbField{content\_type\_other} attribute, which is a free-text attribute allowing platforms to specify content that did not fit within the predefined set of values. The specifications used in the \dbField{content\_type\_other} attribute differ from platform to platform. Consequently, we merged \dbField{content\_type} and \dbField{content\_type\_other} into a single category and reported only the values that were used at least once. The same reasoning applies to \dbField{incompatible\_content\_ground}, whose content is fully up to the platforms and not predefined.

\section{Inter-platform temporal patterns}
\label{sec:appendix-dtw}
\begin{figure*}[t]
    \centering
    \begin{subfigure}[b]{0.33\linewidth}
        \includegraphics[width=\linewidth]{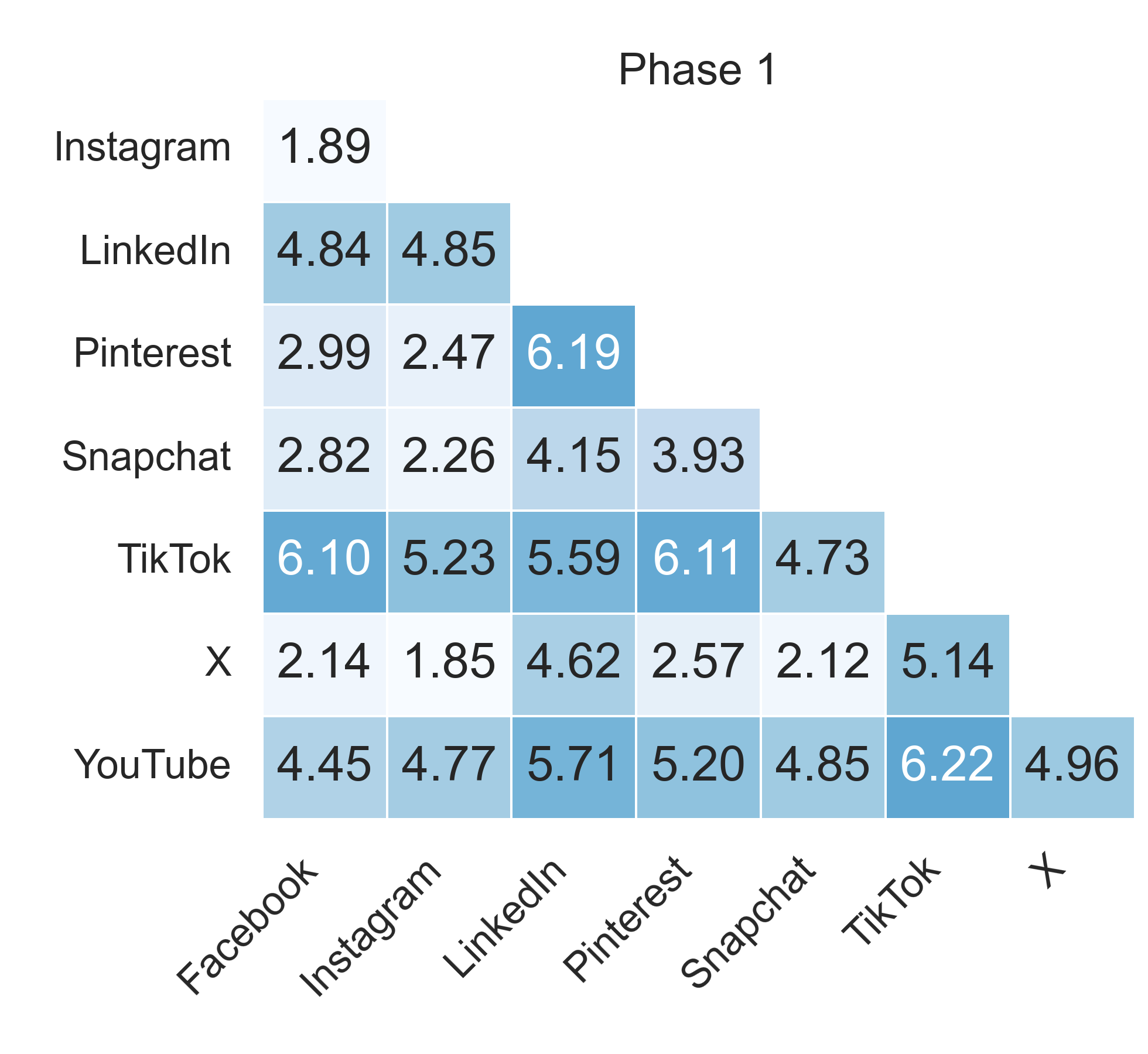}
    \end{subfigure} %
    \begin{subfigure}[b]{0.28\linewidth}
        \includegraphics[width=\linewidth]{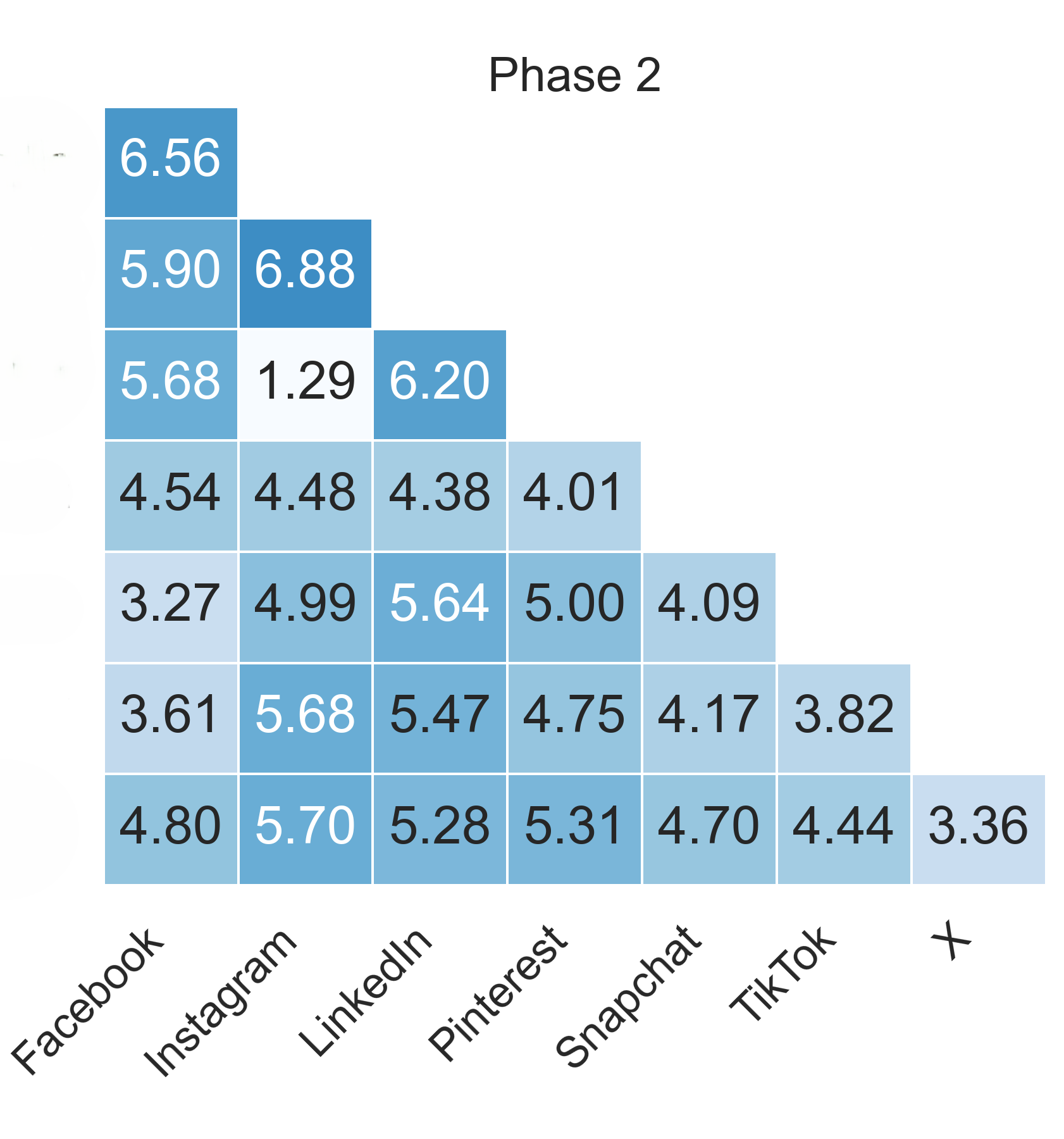}
    \end{subfigure}%
    \begin{subfigure}[b]{0.33\linewidth}
        \includegraphics[width=\linewidth]{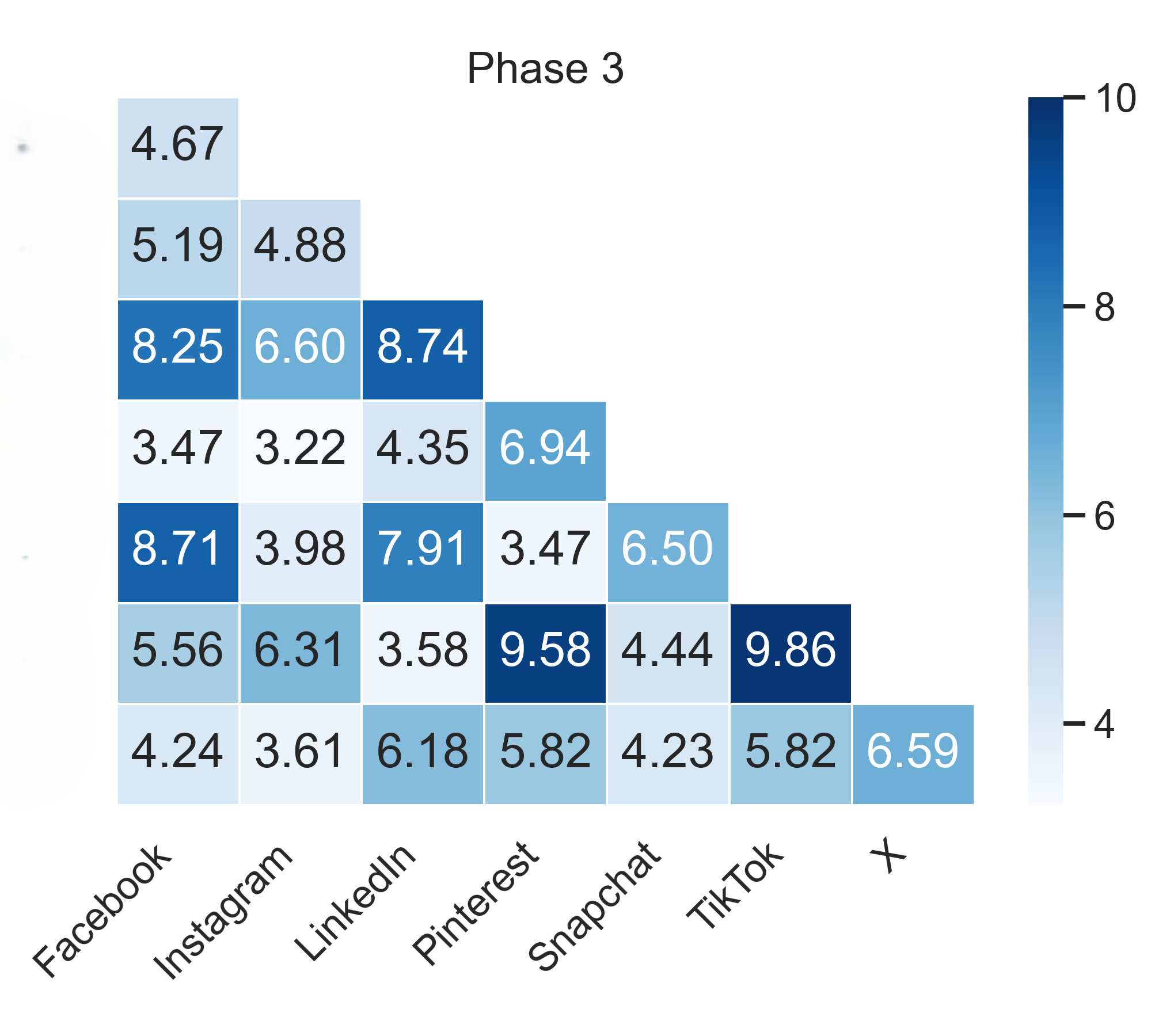}
    \end{subfigure}
    \caption{Dynamic Time Warping Distances across platforms for the number of SoRs.}
    \label{fig:dtw-sors}
\end{figure*}

\begin{figure*}[t]
    \centering
    \begin{subfigure}[b]{0.33\linewidth}%
        \includegraphics[width=\linewidth]{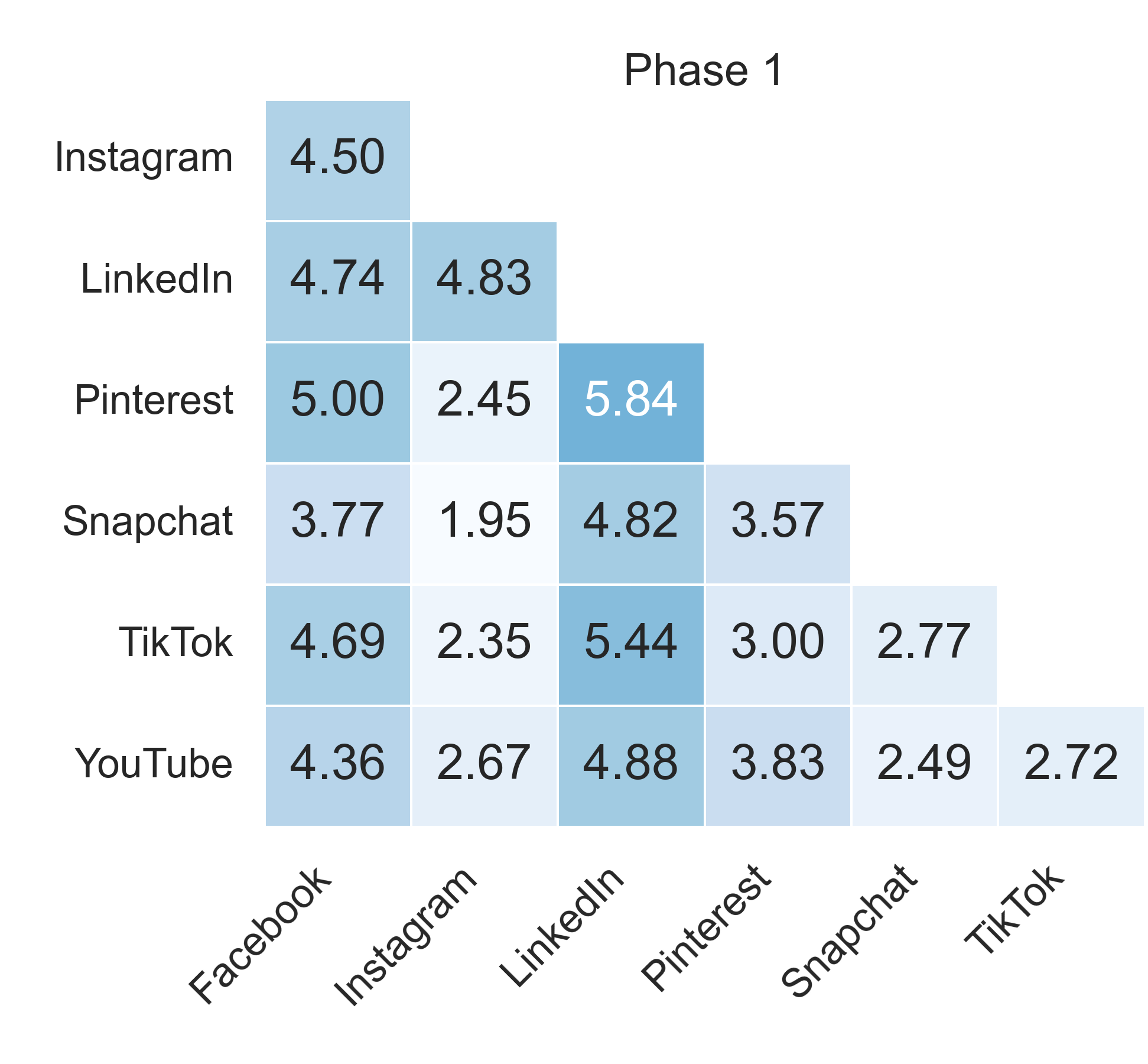}%
    \end{subfigure}
    \begin{subfigure}[b]{0.28\linewidth}%
        \includegraphics[width=\linewidth]{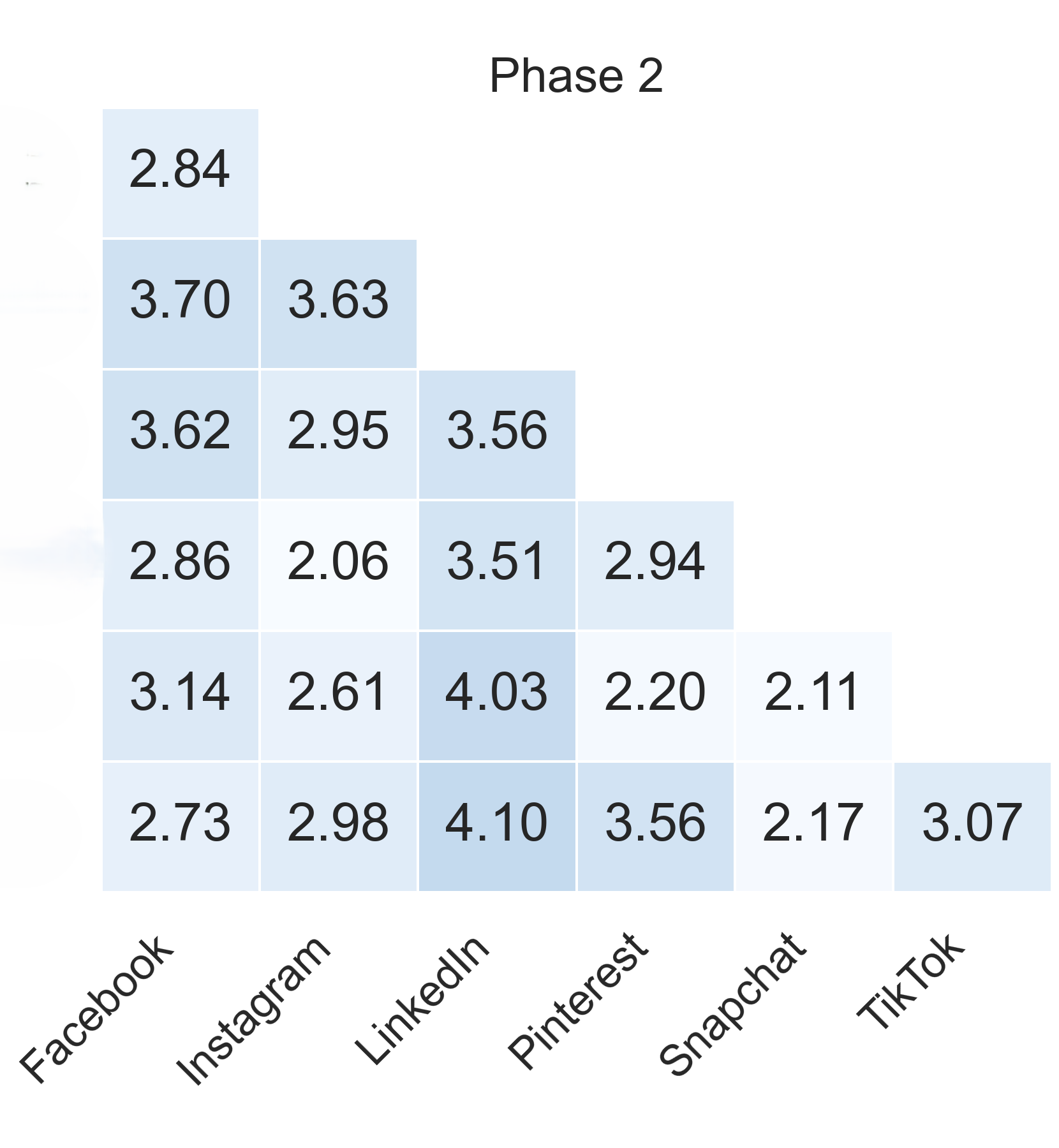}
    \end{subfigure}
    \begin{subfigure}[b]{0.33\linewidth}%
        \includegraphics[width=\linewidth]{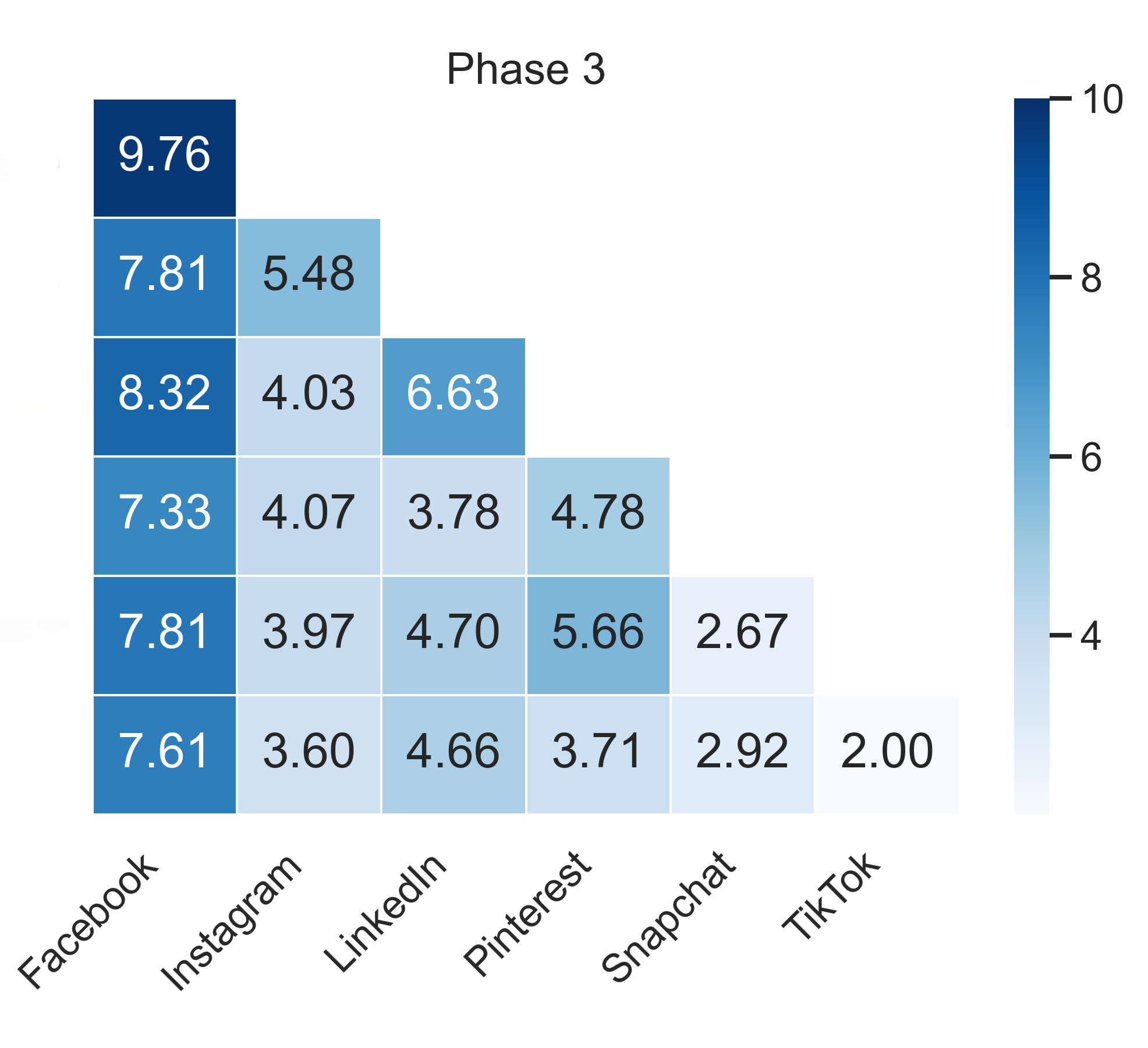}
    \end{subfigure}
    \caption{Dynamic Time Warping Distances across platforms for the average delay.}
    \label{fig:dtw-delay}
\end{figure*}

We computed Dynamic Time Warping (DTW) distances for each phase and for each platform on the clean and standardized time series. Figure~\ref{fig:dtw-sors} shows the DTW distances for the SoRs volume, while Figure~\ref{fig:dtw-delay} for the average delay.

\section{Structural changes in moderation activity}
\label{sec:appendix-change-points}
\begin{figure}
    \centering
    \includegraphics[width=\linewidth]{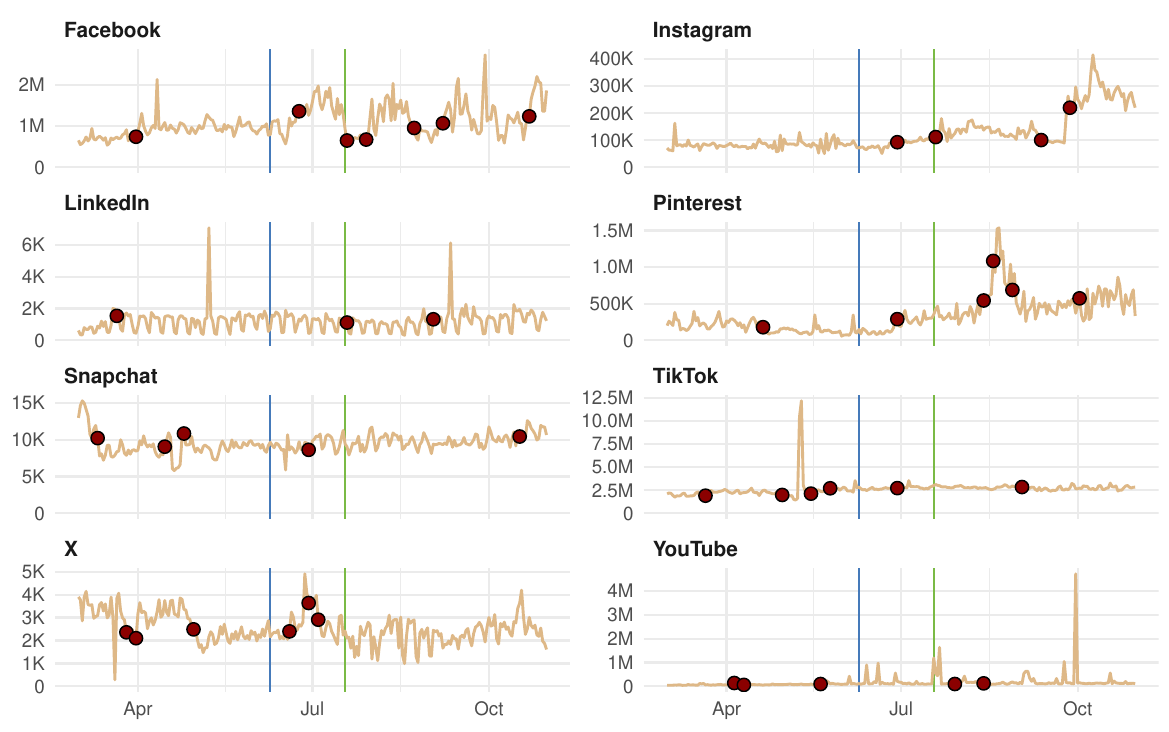}
    \caption{Change points for the number of SoRs time series.}
    \label{fig:cp-sors}
\end{figure}

\begin{figure}
    \centering
    \includegraphics[width=\linewidth]{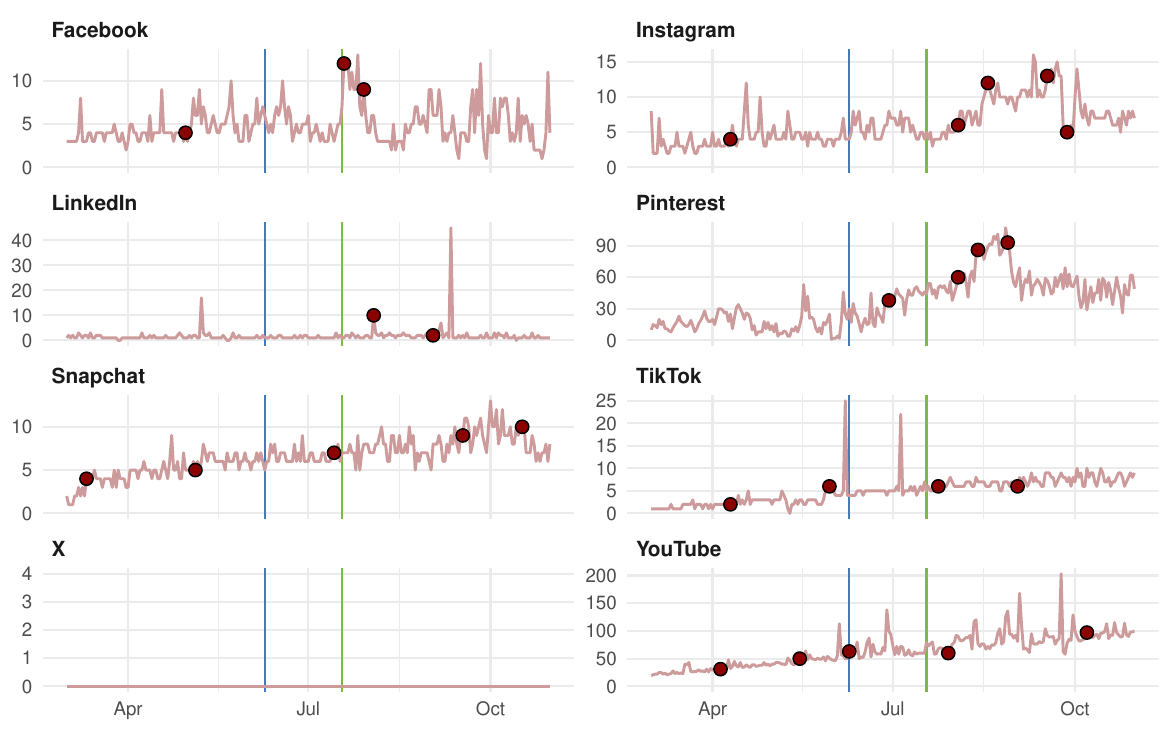}
    \caption{Change points for the delay time series.}
    \label{fig:cp-delays}
\end{figure}

We investigate potential structural changes in moderation activity by carrying out a change point detection analysis to the time series. Figures~\ref{fig:cp-sors} (orange) and~\ref{fig:cp-delays} (red) report the time series of the number of SoRs and the average moderation delay, respectively, with detected change points highlighted by red circles. The vertical blue line marks the parliamentary election days, while the green line denotes the presidential election day. This visualization allows us to assess whether any identified shifts in moderation activity or response times coincide with relevant external events.

\end{appendix}

\begin{appendices}

\end{appendices}

\end{document}